\documentclass[11pt]{article}
\usepackage{amsfonts}
\usepackage{amssymb}
\usepackage{amscd}
\usepackage{amsmath}
\usepackage{epsfig}
\usepackage{graphicx, subfigure}
\usepackage{latexsym}

\newlength{\dinwidth}
\newlength{\dinmargin}
\newtheorem{theorem}{Theorem}
\newtheorem{proposition}{Proposition}
\newtheorem{corollary}{Corollary}
\newtheorem{remark}{Remark}
\newtheorem{lemma}{Lemma}

\def\be{\begin{equation}}
\def\ee{\end{equation}}
\def\ben{\begin{displaymath}}
\def\een{\end{displaymath}}
\def\baa{\begin{eqnarray}}
\def\eaa{\end{eqnarray}}

\def\ba{\begin{array}}
\def\ea{\end{array}}

\def\itwo{{\scriptscriptstyle{2}}}
\def\ithree{{\scriptscriptstyle{3}}}

\def\iN{{\scriptscriptstyle{d}}}
\def\iT{{\scriptscriptstyle{T}}}
\def\iM{{\scriptscriptstyle{\M}}}

\def \surf {{\cal L}}
\def\i{{\rm i}}
\def\E{{\rm {\bf {E}}}}
\def\e{{\rm {\bf {e}}}}
\def\cp{\mathbb {CP}^1}
\def\sig{\sigma}
\def\q{q}

\def\B{\mathbb{B}}

\def\T{{\bf {T}}}
\def\Id{{\bf{1}}}

\def\hom{{\bf h}}

\def\X{{\cal X}}

\def\mid{\Big|}

\makeatletter
\@addtoreset{equation}{section}
\makeatother

\def\CP1{\mathbb {CP}^1}
\def\C{{\mathbb C}}
\def\Z{{\mathbb Z}}

\def\a{\alpha}
\def\g{\gamma}

\def\b{\beta}

\def\l{\lambda}

\def\phi{\varphi}

\def\M{N}
\def\A{{\cal A}}
\def\f{\frac}
\def\la{\label}
\def\H{{\cal H}}
\def\L{{\cal L}}
\def\p{\partial}

\def\con{{\bf s}}
\def\Scont{{\cal S}}

\def\map{{\bf f^{-1}}}
\def\symm{{\bf S}_d}

\def\Gr{{\cal{M}}}
\def\br{\theta}
\def\MA{{\cal{A}}}
\def\MB{{\cal{B}}}
\def\F{T}

\def\N{d}
\def\K{m}

\begin{document}
\title{Riemann-Hilbert problem for Hurwitz Frobenius manifolds: regular singularities}
%\shorttitle{}

%\author{
%D. Korotkin\footnote{e-mail: korotkin@mathstat.concordia.ca;
%address: Department of Mathematics and Statistics, Concordia University
%1455 de Maisonneuve Blvd. West, Montreal H3G 1M8  Qu\'ebec,  Canada}  \;and
%V. Shramchenko\footnote{e-mail: vasilisa.shramchenko@usherbrooke.ca; address: D\'epartement de math\'ematiques, Universit\'e de %Sherbrooke; 2500 boul. de l'Universit\'e, Sherbrooke J1K 2R1 Qu\'ebec, Canada  }  }
\author{D. Korotkin$^1$  \; and V. Shramchenko$^2$ }
\maketitle

\footnotetext[1]{e-mail: korotkin@mathstat.concordia.ca;
address: Department of Mathematics and Statistics, Concordia University
1455 de Maisonneuve Blvd. West, Montreal H3G 1M8  Quebec,  Canada}

\footnotetext[2]{e-mail: vasilisa.shramchenko@usherbrooke.ca; address: D\'epartement de Math\'ematiques, 
Universit\'e de Sherbrooke; 2500 boul. de l'Universit\'e, Sherbrooke J1K 2R1 Qu\'ebec, Canada  }

{\bf Abstract.} In this paper we  study  the Fuchsian Riemann-Hilbert
(inverse monodromy) problem corresponding to Frobenius
structures on Hurwitz spaces.
We find a solution to this Riemann-Hilbert problem in terms of
integrals of certain meromorphic differentials over a basis of an
appropriate relative homology space over a Riemann surface,
study the corresponding monodromy group and compute the monodromy matrices explicitly for
various special cases.

%Although various aspects of this
%Riemann-Hilbert
%problem were studied by Dubrovin himself \cite{2D,DubrovinPainleve}, a

\tableofcontents

\section{Introduction}

The matrix Riemann-Hilbert problems (or inverse monodromy problems)
naturally arise in the context of systems of linear differential equations with meromorphic coefficients (we consider here the simplest
Fuchsian case, when all the poles of the coefficients in the right-hand side are simple):
\be
\f{\p\Phi}{\p\l}=\sum_{j=1}^{\M}\f{A_j}{\l-\l_j}\Phi\;,
\label{lsintr}
\ee
where $A_j$ are $n\times n$ matrices independent of $\l$; $\Phi(\l)$ is $n\times n$ solution matrix.
The solution $\Phi$ is single-valued on the universal cover of the Riemann sphere with punctures at
$\l_1,\dots,\l_\M$ and $\infty$. If one starts at a point $\l_0$ on some sheet of the universal cover,
and analytically continues $\Phi$ along an element $\gamma$ of the fundamental group of the punctured sphere,
one gets a new solution, $\Phi_\gamma$ of the same system; therefore, $\Phi_\gamma$ is related to $\Phi$ by a right multiplier,
$M_\gamma$, which is called the monodromy matrix: $\Phi_\gamma=\Phi M_\gamma$. The monodromy matrix corresponding to the product
$\gamma_1\gamma_2$ is given by $M_{\gamma_1\gamma_2}= M_{\gamma_2} M_{\gamma_1}$; therefore, in this way one gets the
group anti-homomorphism of the fundamental group of Riemann sphere with $\M+1$ punctures to $GL(n)$. The image of this anti-homomorphism is called the monodromy group
of the system (\ref{lsintr}).
The Riemann-Hilbert (or inverse monodromy) problem is the problem of finding the matrix-valued function $\Phi$ (and, therefore, also the
coefficients $A_j$) knowing the positions of singularities $\lambda_j$ and the corresponding monodromy matrices.

It is natural to deform the whole picture by changing infinitesimally the positions of singularities $\lambda_j$ in such a way that
the monodromy matrices remain unchanged.  Such a deformation (called isomonodromic deformation) implies a set of non-linear differential equations
for matrices $A_j$ as functions of $\{\lambda_k\}$; these equations are called {\it the Schlesinger equations}.

Therefore, the solution of the non-linear Schlesinger equations reduces to solution of the complex-analytic inverse monodromy problem.
Typically, one starts with a set of monodromy matrices, finds a solution $\Phi$ of the corresponding Riemann-Hilbert problem, and, finally, gets a
solution $\{A_j\}$ of the Schlesinger system.  In particular, a class of Riemann-Hilbert problems whose monodromy groups are
subgroups of the torus normalizer was solved in \cite{DimaRH}; this allowed to find a class of solutions of the Schlesinger system
associated to the Hurwitz spaces. Another class of solutions of the Schlesinger system related to the Hurwitz spaces was discussed in \cite{Doran}.

The  Riemann-Hilbert
problems of some special type and the corresponding Schlesinger system play an important role in the theory of Frobenius manifolds \cite{2D,DubrovinPainleve}.
In this context the  monodromy groups provide a way of classification  of
Frobenius manifolds; the Schlesinger systems are equivalent to equations for rotation coefficients of the
Darboux-Egoroff metric corresponding to the Frobenius manifold.

To each Frobenius manifold one can naturally associate two systems of linear differential
equations: a Fuchsian system
and a non-Fuchsian one. In the case of the non-Fuchsian system the coefficients have both first and second order poles.
These two systems are related by a formal Laplace transform. For the class of
Frobenius manifolds associated to the Hurwitz spaces (Hurwitz Frobenius manifolds), the non-Fuchsian systems were recently solved in
\cite{V3} (although many essential elements of this construction were already
given by Dubrovin in \cite{2D,DubrovinPainleve}); the corresponding Stokes and monodromy
matrices  were also computed in \cite{V3}. In principle, one can  apply the formal Laplace transform to the solution
from \cite{V3} and get solutions to the corresponding Fuchsian systems, however, this does not give a
satisfactory final result due to a non-trivial superposition of various direct and inverse Laplace
transforms. 
%(see Remark \ref{fuchs-nefuchs} below for more details).

The goal of this paper is to present a different approach (not involving the Laplace transform of solutions to the dual problem) to
constructing solutions to the Fuchsian Riemann-Hilbert  problems of  Hurwitz Frobenius manifolds and study the
corresponding
 monodromy group.
% ; these solutions  are not  related in an obvious way to the solutions to the non-Fuchsian systems found in  \cite{V3}.

The coefficients of the system of Fuchsian linear ODE's 
corresponding to a given Frobenius manifold are written in terms of rotation
coefficients
$\Gamma_{ij}$ of the Darboux-Egoroff metric on the manifold. In the context of Frobenius manifolds the number of singularities
$\M$ in (\ref{lsintr}) is the dimension of the Frobenius
manifold;
 $\{\l_i\}_{i=1}^\M$ are the canonical coordinates on
the Frobenius  manifold. In the case under consideration, the number of singularities  $\l_j$
coincides with the matrix dimension of the system (\ref{lsintr}).

The residues $A_j$ are given by
\be
A_j=-E_j (V+\q I),
\la{AjV}\ee
where $E_j={\rm diag}(0,\dots,1,\dots,0)$ is the diagonal $\M\times \M$ matrix with $1$ on $j$th place; $\q\in \C$ is an arbitrary
constant. The matrix $V$ is defined as follows:
\be
V:=[\Gamma, U]\;,
\la{VGU}\ee
 where $\Gamma$ is the
matrix  of rotation coefficients:
$(\Gamma)_{jk}:=\Gamma_{jk}$ if $j\neq k$ and $(\Gamma)_{jj}:=0$;
 $U:={\rm diag}(\l_1,\dots,\l_{\iM})$.
Thus each matrix $A_j$ in (\ref{lsintr}) has only one non-trivial row (the $j$th row).

Dubrovin in \cite{2D, DubrovinPainleve} studies the linear system with $\q=1/2$. In this paper
we focus on the case $\q=-1/2$;
 the relationship between systems (\ref{lsintr}), (\ref{AjV})
 with the values of $\q$ different by an integer is discussed in Remark \ref{rmk_alphas} below.

In the  context of the Fuchsian system (\ref{lsintr}), (\ref{AjV}), (\ref{VGU})
  the   solution of the Schlesinger system is given by  the rotation coefficients, which  were found earlier in \cite{2D,KokKor}.
Moreover, in \cite{KokKorB} the corresponding
Jimbo-Miwa isomonodromic tau-function was explicitly computed. This tau-function turned out to be an object of fundamental importance:
it appears in various contexts from the large N limit of Hermitian matrix models to the determinants of Laplacians on Riemann surfaces \cite{KokKorJDG} and geometry of Hurwitz spaces \cite{KKZ}.

 However, a solution to the
corresponding Fuchsian Riemann-Hilbert problem (which coincides with the solution of the Fuchsian system
(\ref{lsintr})) was missing so far. It is this gap which we fill in this  paper: we solve this  Fuchsian system, compute the
 monodromy matrices and describe the corresponding monodromy group. Thus, the logic of this paper is different from the logic of
the paper \cite{DimaRH}, where the
Riemann-Hilbert problem was solved first, and the solution of the corresponding Schlesinger system was found
as a corollary.

We  also discuss the transformation of the solution $\Phi$ under
the action of the braid group on the set of singularities $\l_j$, by introducing the notion of {\it  the braid monodromy group}.
In particular, we discuss the action of the braid group on the set of monodromy matrices of the system (\ref{lsintr}) following the
ideas of the work by Dubrovin and Mazzocco \cite{DubMaz} where such an action was considered in the context of algebraic solutions of the Painlev\'e VI equation.

Let us now describe the settings and our results in more details.

The Hurwitz space $\H_{g,\N}(k_1,\dots,k_m)$ is the space of equivalence classes of pairs $\X:=(\L,f)$, where $\L$ is a compact
Riemann surface of genus $g$, and $f$ is a meromorphic function of
degree $\N$ on $\L$ with simple critical points and $\K$ poles of multiplicities $k_1,\dots,k_\K$ ($k_1+\dots+k_\K=\N$); two pairs $\X_1:=(\L_1,f_1)$ and $\X_2:=(\L_2,f_2)$ are
equivalent if there exists a biholomorphic
map  $h:\L_1\to\L_2$, such that $f_1=f_2\circ h$.  Using the function $f$ we can realize the Riemann surface $\L$ as
 a $\N$-sheeted branched covering of the Riemann sphere; the branch points of this covering are given by the critical
values of the function $f$. Therefore, the Hurwitz space can be viewed as a space of branched covering of the Riemann sphere with the
fixed number of sheets and the fixed branching structure.

 The Frobenius structures
can be defined on each space $\H_{g,\N}(k_1,\dots,k_m)$.
The branch points, which we denote  by
$\l_1,\dots,\l_{\iM}$ (the corresponding ramification points on $\L$ are the critical points of the function $f$; they
are denoted by $P_1,\dots,P_{\iM}$, i.e., we have $\lambda_i=f(P_i)$), can be used as local coordinates on the Hurwitz space; they also play the role of canonical coordinates on the
corresponding Frobenius manifold.

The main tool in our construction is the canonical meromorphic bidifferential $W(P,Q)$ on
the Riemann surface $\L$. To define this bidifferential we have to choose some weak marking of the Riemann surface $\L$, i.e.,
a canonical basis $({\bf a}_\a, {\bf b}_\a)$ ($\a=1,\dots,g$) of the homologies $H_1(\L)$ with coefficients in $\Z$.

The bidifferential $W$ is  symmetric, has a quadratic
pole on the diagonal $P=Q$ with biresidue $1$ and is normalized by the condition of vanishing of all periods along cycles
${\bf a}_\a$ with respect
to both $P$ and $Q$. Therefore, in fact, $W$ depends only on the choice of a Lagrangian subspace (spanned by the $a$-cycles) in  $H_1(\L)$. Therefore it is natural to introduce the space
$\H_{g,\N}^{\{\bf a\}}(k_1,\dots,k_m)$ which is the space of pairs
$(\X,\{{\bf a}\})$,
where $\X=(\L,f)\in\H_{g,\N}(k_1,\dots,k_m)$ and $\{{\bf a}\}$ is a Lagrangian subspace of $H_1(\L)$.

The rotation coefficients (the matrix $\Gamma$ in (\ref{VGU})) of the Frobenius structures on the
Hurwitz spaces are written in terms of the bidifferential $W$ \cite{KokKor}:
\be
\Gamma_{jk}=\f{1}{2}W(P_j,P_k):=\f{1}{2}\f{W(P,Q)}{d x_j(P)\;d x_k(Q)}\Big|_{P=P_j,\,Q=P_k}\;.
\label{Gammaijintr}
\ee
Here $x_j(P)$ in a local parameter on $\L$ near the branch point $P_j$ defined by the equations $2 x_j dx_j =df$ and $x_j(P_j)=0$. These two conditions imply
that
$x_j(P):=\pm\sqrt{f(P)-\l_j}$; these local parameters  near the ramification points $P_j$ are called {\it distinguished}.
Different choices of the signs of $x_j(P)$  in (\ref{Gammaijintr})  lead to different sets of rotation coefficients. If one considers
$\Gamma_{jk}^2$, the freedom of choosing different signs disappears and one can write the following invariant
 expression:
$$
\Gamma_{jk}^2= {\rm Res}\big|_{P=P_j}\,{\rm Res}\big|_{Q=P_k}\left\{\frac{W^2(P,Q)}{df(P) \,df(Q)}\right\}
$$
For a given covering $(\L,f)$ and a Lagrangian subspace $\{{\bf a\}}$
we therefore get $2^N$ different sets of rotation coefficients. Each set gives rise to a family of $N$ Frobenius manifolds of dimension $N$.

To construct a solution of the Fuchsian linear system (\ref{lsintr}) we introduce, for any $\l\in\C\setminus\{ \lambda_1,
\dots, \lambda_\M \},$ the relative homology group
$H_1(\L\setminus f^{-1}(\infty)\,;\,f^{-1}(\l))$ with coefficients in $\Z$. This is the homology group  of the Riemann surface $\L$
punctured at the poles of the function $f$, relative to the set of  $\N$ points  on $\L$ where the value of $f$ equals $\l$.
The dimension of this relative homology space equals $\M=2g+\N+\K-2$,
where $\K$ is the number of poles of the function $f$, i.e.,  the number of points  in the set $f^{-1}(\infty)$.

Our first main result is that for any contour $\con\in H_1(\L\setminus f^{-1}(\infty)\,;\,f^{-1}(\l))$ the vector function with the components
\begin{equation}
\Phi^{(\con)}_j(\l):= \lambda\int_{\con} W(P,P_j) - \int_{\con} f(P)W(P,P_j)\;,
\label{defPhiintr}
\end{equation}
where $j=1,\dots,\M$, satisfies the linear system (\ref{lsintr})-(\ref{VGU}) with $\q=-1/2$. In (\ref{defPhiintr})
\begin{equation}
W(P,P_j):=\f{W(P,Q)}{d x_j(Q)}\Big|_{Q=P_j}\; ,
\la{WPPjint}
\end{equation}
and the signs of the distinguished local parameters $x_j(Q):=\sqrt{f(Q)-\l_j}$ in (\ref{WPPjint}) have to be
 chosen in the same way as in
(\ref{Gammaijintr}). The square, $W^2(P,P_j)$, is defined by the formula
\begin{equation*}
W^2(P,P_j)=2\,{\rm Res}\big|_{Q=P_j}\left\{\frac{W^2(P,Q)}{df(Q)}\right\}\; .
\end{equation*}

Our second main result is that choosing $\con$ to run through a basis in  $H_1(\L\setminus
f^{-1}(\infty)\,;\,f^{-1}(\l))$, we get the complete set of $2g+\N+\K-2$
independent solutions to (\ref{lsintr}); the proof of this
independence is a tedious exercise involving  analysis of the behaviour
of the bidifferential $W(P,Q)$ at the boundary of the Hurwitz space.

Let us choose a neighbourhood $D$ of a point $\l_0\in\mathbb{C}$ which
contains no branch points $\l_k$.

A basis of  contours in  $H_1(\L\setminus
f^{-1}(\infty)\,;\,f^{-1}(\l))$ for $\lambda\in D$ can be chosen as follows. First we choose a canonical
basis of $2g$ cycles in  $H_1(\L)$ (this canonical basis
does not
necessarily coincide with the set of cycles on $\L$  which enter the
definition of the bidifferential $W$). To this set we add the set of  small contours around  $\K-1$
points which can be arbitrarily chosen from the set $f^{-1}(\infty)$
consisting of $\K$ points. Finally, the remaining $\N-1$ contours can be taken
to connect pairwise the $\N$ points from $f^{-1}(\l)$; for the linear
independence of these contours one has to require connectedness of  the graph whose
edges are given by these contours and the
vertices are the $\N$ points from $f^{-1}(\l)$. 
We assume that the chosen bases of cycles in the spaces $H_1(\L\setminus f^{-1}(\infty)\,;\,f^{-1}(\l))$
for any two values of $\l\in D$ can be smoothly deformed one into another on the Riemann surface $\L$.
In this way we get a non-degenerate matrix-valued function $\Phi(\l)$
solving (\ref{lsintr}), which is analytic for $\l\in D$.

The function $\Phi=\Phi(\{\lambda_i\};\lambda)$
depends on (i) the  choice of a Lagrangian subspace in $H_1(\surf)$ generated by the $a$-cycles
(which enters the definition of the canonical bidifferential $W$); 
(ii) the choice of a basis $\con_1,\dots,\con_\M$ in  $H_1(\L\setminus
f^{-1}(\infty)\,;\,f^{-1}(\l_0))$, where $\lambda_0\in\mathbb{C}\setminus\{\lambda_k\}$ is some base point, and (iii) the choice of the signs of the distinguished local
parameters $x_j$ near the ramification points.

If one preserves the integration contours  $\con_1,\dots,\con_\M$, but changes the Lagrangian subspace $\{{\bf a}\}$ used for normalization of $W$,
 the new function $\Phi$
turns out to be related to the old one by a Schlesinger transformation (multiplication of $\Phi$ from the left by a rational function of $\l$ of a special form), which we find explicitly.  Therefore, a change of normalization of $W$ does not
influence the monodromy matrices of the function $\Phi$ (i.e. the new and the old functions have the same set of monodromies although they
satisfy the linear system (\ref{lsintr}) with different coefficients).

If, on the other hand, we preserve the normalization of $W$ but change the set of the  integration contours
$\con_1,\dots,\con_\M$, the new function $\Phi$ can be  obtained from the old one by  multiplication from the right
with some constant matrix: this corresponds to a linear transformation in the space of solutions of the linear system (\ref{lsintr}).

Finally, if one changes the sign of some of distinguished local parameters $x_j$: $x_j\to \epsilon_j x_j$ with  $\epsilon_j^2=1$, the new function $\Phi$ differs from the old one by multiplication from the left by the matrix ${\rm diag}(\epsilon_1,\dots,\epsilon_{\M})$.

Let us define  the following covering $\widehat{\H}_{g,d}(k_1,\dots,k_m)$ of the Hurwitz space. An element of the
space  $\widehat{\H}_{g,d}(k_1,\dots,k_m)$ is a quadruple $(\L, f, \{{\bf a}\}, \{\epsilon_j\})$ i.e. the
covering  $\X=(\L,f)\in \H_{g,d}(k_1,\dots,k_m)$ with the chosen Lagrangian subspace $\{{\bf a}\}$ in the homology space of $\L$ and
the choice of signs of distinguished local parameters at the critical points of the function $f$.

Any  solution  vector of the Fuchsian system (\ref{lsintr}) is
a section of a vector bundle on the punctured sphere $\mathbb{C}\setminus\{\lambda_1,\dots,\lambda_N\}$. On the other hand, for fixed $\lambda$,
the  same solution vector is a section of a vector bundle over  the space  $\widehat{\H}_{g,d}(k_1,\dots,k_m)$.
Each of these vector bundles corresponds to a monodromy group.
Let us discuss these two monodromy groups in more detail.
\begin{itemize}
\item

A solution to the system (\ref{lsintr}) is non-singlevalued in the complex plane. Upon analytical continuation with respect to $\lambda\in\mathbb{C}$ along the generators of the fundamental group $\pi_1(\mathbb{C} \setminus\{\l_1,\dots,\l_{\M},\infty\})$, the function  $\Phi$ is multiplied from the right by monodromy matrices $M_k$, $k=1,\dots,\M,\infty$.

Since the only non-linear dependence on $\lambda$ of our solution is due to the $\lambda$-dependence of the contours of integration,
the monodromy matrices describe the transformation of a chosen basis
in  $H_1(\L\setminus f^{-1}(\infty)\,;\,f^{-1}(\l))$ under the natural action of an element
of $\pi_1(\mathbb{C}\setminus\{\l_1,\dots,\l_{\M},\infty\}, \lambda)$. Therefore all
entries of the monodromy matrices are integer numbers.

If a basis in  $H_1(\L\setminus
f^{-1}(\infty)\,;\,f^{-1}(\l))$ is chosen as described above, the monodromy matrices  possess the following structure:
\begin{equation}
M_k= \left(\begin{array}{cc} I & S_k\\
                             0 & \Sigma_k \end{array}\right),
\label{monintr}
\end{equation}
where $\Sigma_k$ are square  $(\N-1)\times (\N-1)$ matrices; they generate a
subgroup of $GL(\N-1,\Z)$ given by the image  in $GL(\N-1,\Z)$ of the
monodromy group of the covering $\L$  under the natural group homomorphism. The unit matrices in
the upper diagonal block are of the size $(2g+\K-1)\times (2g+\K-1)$; the matrices $S_k$ of the
size $(2g+\K-1)\times (\N-1)$ depend on the choice of a basis in the subspace 
$H_1(\L\setminus f^{-1}(\infty))$ of space $H_1(\L\setminus f^{-1}(\infty)\,;\, f^{-1}(\l))$. However,  the change of a basis
in $H_1(\L\setminus f^{-1}(\infty))$ results in a simultaneous
conjugation of all monodromy matrices $M_k$ by the
same matrix; thus the monodromy group is in fact independent of the
choice of this basis.

The monodromy group formed by the matrices (\ref{monintr}) can be described as a semidirect product of the
free group $\mathbb{Z}^{(2g+\N-1)\times(\N-1)}$ and the symmetric group $\symm$, the monodromy group of the covering.
Thus the monodromy group of our Fuchsian system coincides with the Weyl group of the algebra of formal power series in $2g+\N-1$ variables with coefficients in $A_{\N-1}$.

A  Schlesinger system corresponding to a block-diagonal structure of monodromy matrices as in (\ref{monintr}), was
called {\it reducible} in \cite{DubMaz1}; this means that its solution can in principle be expressed in terms of solutions of two
Schlesinger systems of lower dimension ( $(\N-1)\times(\N-1)$ and $(2g+m-1)\times(2g+m-1)$ in our case).

%there exists a Schlesinger transformation
%(which does not necessarily can be effectively found) which brings the solution $\Phi$ to a block-diagonal form with two blocks
%of the size $(\N-1)\times(\N-1)$ and $(2g+m-1)\times(2g+m-1)$. At the moment we don't know how to white this
%Schlesinger transformation explicitly.

\item
The second type of monodromy transformation is the transformation of the solution (\ref{defPhiintr}) under analytical continuation with respect to the arguments $\lambda_k$ along the generators of the fundamental group of the covering $\widehat{\H}_{g,\N}$ of the Hurwitz space $\H_{g,\N}$. This fundamental group is a subgroup of the plane braid group on $\M$ strands acting on the set of local coordinates $\{\lambda_k\}_{k=1}^\M$ on the Hurwitz space. Such an analytical continuation also results in the multiplication of the solution $\Phi$ from the right by some monodromy matrices with integer entries. We called the arising group of transformations  {\it the braid monodromy group} of the  solution to the Fuchsian system.

We note that the action of the braid group on solutions to the Schlesinger system was used to classify the algebraic solutions of
the Painlev\'e-VI equation in \cite{DubMaz}. In the context of Knizhnik-Zamolodchikov equations (which can be considered as a special case of equations of isomonodromic deformations \cite{Reshet,Harnad}), the braid monodromies were studies starting from Drinfeld's paper \cite{Drinfeld}; these monodromies play a fundamental role in the theory of
quantum groups  \cite{Kassel}. The action of the braid group on the coverings with $\Z_d$ symmetry was recently studied in \cite{McMullen}.

\end{itemize}

The central object associated to  any Riemann-Hilbert problem and
the corresponding equations of isomonodromic deformations (the Schlesinger
system) is the  isomonodromic
Jimbo-Miwa tau-function, a function of $\{\l_k\}$. The divisor of
zeros of  the tau-function consists of those  configurations of poles
$\{\l_k\}$ where the Riemann-Hilbert problem loses its solvability
(see \cite{Bolibruch}).
 In the context of the Frobenius manifold structures on
Hurwitz spaces, the tau-function determines the $G$-function of the
Frobenius manifold, which is the genus one free energy of the
corresponding topological field theory.
The isomonodromic tau-function associated to the solutions
(\ref{AjV}),(\ref{VGU}), (\ref{Gammaijintr}) of the
Schlesinger system coincides with the so-called Bergman tau-function on the Hurwitz space \cite{KokKorG}.
The Bergman tau-function plays a key role in the computation of the determinant of the
Laplacian in flat metrics on Riemann surfaces \cite{KokKorJDG} and of the genus one
free energy in the Hermitian two-matrix models \cite{EKK}. In \cite{KKZ} it was constructed a line
bundle on compactified Hurwitz spaces (spaces of admissible covers proposed by Harris and Mumford)
whose holomorphic section is given by the Bergman tau-function; this line bundle is closely related to the
Hodge line bundle on the moduli space of Riemann surfaces.

The paper is organized as follows. Section \ref{sect_Frobenius} contains a few basic facts about the Fuchsian and non-Fuchsian
Riemann-Hilbert problems appearing in the theory of Frobenius manifolds. In Section \ref{sect_solution} we construct  a solution to the
Fuchsian system and discuss its dependence on the choice of normalization for the main building block of the solution, the bidifferential $W$. In Section \ref{sect_monodromy} we describe the monodromy group of the solution. In Section \ref{sect_braid} we discuss the braid group action on our solution, i.e., the behaviour of the solution under the analytical continuation along nontrivial loops in the Hurwitz space; we compute the generators of the braid monodromy group for the case of Hurwitz space of two-fold genus one coverings. Technical details of the computation of monodromy matrices and of the proof of the non-degeneracy of our solution are given in the Appendices \ref{app_monodromy} and \ref{app_completeness}, respectively.

\section{ The Fuchsian  Riemann-Hilbert problem in Frobenius manifolds theory}
\label {sect_Frobenius}

For the reader's convenience and to set up the notations we shall review here the
connections between
solutions to systems of linear differential equations with meromorphic coefficients,
matrix Riemann-Hilbert (inverse monodromy) problems, and Frobenius manifolds.

Consider a matrix linear differential equation (\ref{lsintr});
depending on the context we shall understand $\Phi$ as either a vector
solution to this equation, or a square $\M\times\M$ matrix of
linearly independent vector solutions to this equation.
%with meromorphic coefficients with
%simple poles:
%\be
%\f{\p\Phi}{\p\l}=\sum_{j=1}^{\M}\f{A_j}{\l-\l_j}\Phi
%\label{ls0}
%\ee
%where $\Phi$ is a matrix-valued function  of $\l$;
%$A_j$ are matrices
%independent
%of $\l$.
Generically, a solution to  equation (\ref{lsintr}) has non-trivial monodromy under the
analytical continuation around singularities $\{\l_i\}$ and around the point
$\l=\infty$.
Let us choose a set of generators $\g_1,\dots,\g_{\iM},\g_{\infty}$ of the fundamental
group
of the punctured sphere $\CP1\setminus\{\l_1,\dots,
\l_{\iM},\infty\}$ such that each generator $\g_j$ encloses only the point $\l_j,$
the generator
$\g_{\infty}$ goes around the point at infinity, and the following relation is fulfilled:
\begin{equation}
\label{genpi1}
\g_1\dots\g_{\iM}\g_{\infty}=id \,.
\end{equation}
Suppose that the matrix solution $\Phi$, being analytically continued along $\g_j$, gains
the right multiplier
$M_j$ (which is called the monodromy matrix). Being  analytically continued along $\g_{\infty},$
the solution $\Phi$ gains the right multiplier
$M_{\infty}$.
As a corollary of relation (\ref{genpi1}) the monodromy matrices satisfy the relation
\begin{equation}
\label{relationmonodromy}
M_{\infty}M_{\M}\dots M_1=I\;.
\end{equation}
%
%i.e., they give  an anti-representation of the fundamental group.

At the poles $\l_j$ of the coefficients of the system (\ref{lsintr}), the function $\Phi$ has
regular singularities
(i.e., $\Phi(\lambda)$ grows at these points not faster than some power of
$\l-\l_j$). If the matrices $A_j$ are diagonalizable (this is the only case considered
in this paper),
the
behaviour of $\Phi$ in a neighbourhood of $\l_i$ looks as follows:
\begin{equation}
\Phi(\l)= G(\l)(\l-\l_j)^{T_j}  C_j \,,
\label{locli}
\end{equation}
where $T_i$ is a diagonal matrix,  $G(\l)=G_j+O(\l-\l_i)$ is a function holomorphic in
a neighbourhood of $\l_j$. If some matrix $A_j$ is non-diagonalizable, the asymptotics
of $\Phi$ near $\l_j$  contains logarithmic terms.

The monodromy matrices can be expressed in terms of $C_j$ and $T_j$ as follows:
\be
M_j=C_j^{-1}e^{2\pi \i T_j} C_j \,.
\label{MCT}
\ee

The Riemann-Hilbert (or inverse monodromy) problem is the problem of
reconstructing the function $\Phi$ knowing its monodromy matrices $\{M_j\}$
and the positions of singularities $\{\l_j\}$. Obviously, a solution to
the Riemann-Hilbert problem is not unique: multiplying such a solution
from the left with an arbitrary matrix-valued rational
function of $\l$, we again get a solution to the same Riemann-Hilbert
problem.
On the other hand, assuming that $\Phi$ has at $\{\l_j\}$ regular
singularities of the form
(\ref{locli}) with the given $\{T_j,\,C_j\}$, and has no other
singularities (including zeros of ${\rm det} \Phi$),
 the solution of the Riemann-Hilbert problem is unique.

Let us now impose the isomonodromy condition, i.e., the condition of independence of the
monodromy data  $\{T_j,\,C_j\}$  of the positions of
singularities $\{\l_j\}$. The isomonodromy condition implies a
system of differential equations, called the Schlesinger equations, for the residues $A_j$ as functions
of $\{\l_j\}$.
The Schlesinger equations of a special type together with the corresponding Riemann-Hilbert problem
play a significant role in the theory of Frobenius manifolds.

We shall now briefly outline the way the equations of the type (\ref{lsintr}) appear in the Frobenius manifold theory.
We skip the complete description of the notion of a Frobenius manifold
and associated objects,
referring the reader to \cite{2D,DubrovinPainleve}.  We recall only that to each Frobenius
manifold
 one can associate a Darboux-Egoroff (i.e., diagonal flat potential) metric.
The poles  $\l_j$, $j=1,\dots,\M,$ of the coefficients in equation (\ref{lsintr})
coincide with the canonical coordinates on the Frobenius manifold. The following two differential operators are also associated to a Frobenius manifold structure:
${\bf e}=\sum_{j=1}^{\iM} \f{\p}{\p\l_j}$,
called the unit vector field, and
${\bf E}=\sum_{j=1}^{\iM} \l_j\f{\p}{\p\l_j}$,
called the Euler vector field.

For the Darboux-Egoroff metrics appearing in the theory of Frobenius manifolds the
rotation coefficients $\Gamma_{ij}$ satisfy the following system of equations:
\be
\f{\p\Gamma_{ij}}{\p\l_k}=\Gamma_{ik}\Gamma_{jk},
\label{ggg}
\ee
where all $i,j,k$ are  distinct, and %the following two equations:
\be
{\bf e}(\Gamma_{ij})=0\;,\hskip0.7cm {\bf E}(\Gamma_{ij})=-\Gamma_{ij}\;.
\label{unitg}
\ee

The non-linear system (\ref{ggg}), (\ref{unitg}) is the
compatibility condition for the
following system of linear differential equations \cite{2D,DubrovinPainleve}:
\be
\f{d\Phi}{d\l}=-\sum_{j=1}^{\M} \f{E_j(V+\q I)}{\l-\l_j}\Phi,
\label{ls1}
\ee
\be
\f{d\Phi}{d\l_j}=\left(\f{E_j(V+\q I)}{\l-\l_j}+[\Gamma,E_j]\right)\Phi,
\label{ls2}
\ee
where $\Phi$ is an $\M\times \M$ matrix-valued function of $\l$ and $\{\l_j\}$; $\q \in\C$
is an arbitrary
constant;
matrices $V$, $\Gamma$ and $E_j$ are defined after (\ref{lsintr}).

% and
%$V:= [\Gamma, U]$,
%where $U$ is the diagonal matrix $U={\rm diag}(\l_1,\dots,\l_{\M})$;
%$\Gamma$ is the matrix of rotation coefficients defined after
%(\ref{lsintr}); $E_j$ is the
%diagonal matrix
%with the only non-vanishing entry on the diagonal standing on the
%$j$th place: $E_j={\rm diag}(0,\dots,1,\dots,0)$.

The system (\ref{ls2}) provides the isomonodromy condition for the Fuchsian system (\ref{ls1}).

%From the identity ${\bf tr}\Phi_{\l}\Phi^{-1} = ({\rm det}\Phi)_{\l}({\rm
%det}\Phi)^{-1}$
%we conclude from (\ref{ls1}), (\ref{ls2}) that
%\be
%{\det} \Phi = C\prod_{j=1}^M (\l-\l_j)^{-\a}
%\label{deta}
%\ee
%where $C$ is a constant which is independent of $\l$ and $\{\l_j\}$.

In this way, a family  (\ref{ls1}), (\ref{ls2}) of  isomonodromic linear systems and the corresponding Riemann-Hilbert problems are associated to any semisimple Frobenius manifold.

The Fuchsian linear system introduced in the original papers
\cite{2D,DubrovinPainleve} corresponds to the value $\q=1/2$. In
this paper, we shall study the case $\q=-1/2$; below (see Remark \ref{rmk_alphas}) we discuss the
relationship between linear systems (\ref{ls1}), (\ref{ls2}) with the
values of $\q$ which differ by an integer.

In the sequel we shall  use the following convenient
 alternative formulation of the linear system (\ref{ls1}), (\ref{ls2}).
\begin{proposition}
\label{theuler}
A vector $\Phi:=(\phi_1,\dots,\phi_{\iM})^T$ satisfies the linear system (\ref{ls1}), (\ref{ls2}) if and only if the following equations are fulfilled
\begin{eqnarray}
&&\l\f{\p\phi_j}{\p\l} + {\bf E}(\phi_j)=-\q\phi_j
\label{eulereq} \\
&&\f{\p\phi_j}{\p\l} +{\bf e}(\phi_j)=0
\label{uniteq}\\
&&\f{\p\phi_j}{\p\l_k}=\Gamma_{jk}\phi_k\;,\qquad j\neq k.
\label{maineq}
\end{eqnarray}
\end{proposition}
{\it Proof.} Equation (\ref{ls1}) for the vector
$(\phi_1,\dots,\phi_\M)^T$ reads in the components:
\begin{equation}
\frac{\partial\phi_j}{\partial\l} = - \frac{1}{\l-\l_j} \left( \q\phi_j +
\sum_{k=1,k\neq j}^{\M}  \Gamma_{kj} (\l_k-\l_j) \phi_k  \right).
\label{phi_lambda}
\end{equation}
Similarly, equation (\ref{ls2}) for the vector $(\phi_1,\dots,\phi_\iM)^T$ is
equivalent to
\begin{eqnarray}
\label{rauch1}
&&\frac{\partial\phi_j}{\partial\lambda_k}  = \Gamma_{jk}\phi_k, \qquad j\neq k , \\
%\end{equation}
%
%and
%
%\begin{equation}
&&\frac{\partial\phi_j}{\partial\lambda_j} = \frac{1}{\l-\l_j} \left( \q\phi_j +
\sum_{k=1,k\neq j}^{\M} \Gamma_{kj} (\l_k-\l_j) \phi_k \right) - \sum_{k=1,k\neq j}^{\M}
\Gamma_{kj} \phi_k. \nonumber
%\label{rauch2}
\end{eqnarray}
The latter equation  rewrites due to (\ref{phi_lambda}) as
\begin{equation*}
\frac{\partial\phi_j}{\partial\lambda_j} = -\frac{\partial\phi_j}{\partial\lambda} -
\sum_{k=1,k\neq j}^{\M} \Gamma_{kj} \phi_k,
\end{equation*}
which, by virtue of  (\ref{rauch1}), coincides with (\ref{uniteq}).

We thus need to show the equivalence of equations (\ref{eulereq}) and
(\ref{phi_lambda}) provided (\ref{uniteq}) and (\ref{maineq}) hold. Using
(\ref{maineq}), we rewrite (\ref{phi_lambda}) as follows:
\begin{equation*}
\frac{\partial\phi_j}{\partial\l} = - \frac{1}{\l-\l_j} \left( \q\phi_j +
\sum_{k=1,k\neq j}^{\M}  (\l_k-\l_j) \partial_{\lambda_k}\phi_j \right).
\end{equation*}
Adding and subtracting $\l_j\partial_{\l_j}\phi_j$ in the right hand side and using
the unit and Euler vector fields, we obtain
\begin{equation}
(\l-\l_j)\frac{\partial\phi_j}{\partial\l} = - \q\phi_j -{\bf E}(\phi_j) +\l_j
{\bf e} (\phi_j).
\label{temp1}
\end{equation}
Plugging equation (\ref{uniteq}) into the above relation (\ref{temp1}), we obtain
(\ref{eulereq}).
$\Box$
\begin{remark}
\label{rmk_alphas}
\rm
Using Proposition \ref{theuler} we can easily see that
the solutions to the linear systems (\ref{ls1}), (\ref{ls2}) corresponding to values
 $\q$ and $\q+1$ are related by differentiation in $\lambda$.
Namely, let us indicate explicitly the dependence of a solution to the system (\ref{ls1}),
(\ref{ls2})
on $\q,$ i.e., we denote $\Phi$ by $\Phi^{\q}$. Then
\be
\label{aap1}
\Phi^{\q+1}=\f{\p\Phi^{\q}}{\p\l}\equiv A^{\q}(\l)\Phi^\q(\l),
\ee
where
$A^{\q}(\l)=-\sum_{i=1}^\M \f{E_i(V+\q I)}{\l-\l_i}$ is the matrix of
coefficients in (\ref{ls1}).

In this paper we find a  complete system of
linearly independent solutions to the system (\ref{ls1}), (\ref{ls2})
for the case $\q=-1/2$. Several columns of our solution
$\Phi$ turn out to be independent of $\l$, therefore formula
(\ref{aap1}) cannot be used to generate fundamental solutions to the
system with $\q=-1/2 +n$ for integer $n\geq 1$. However, from our
solution for $\q=-1/2$ we can obtain the complete system of solutions
for any negative half-integer value of $\q$. 
%(see also Remark \ref{otherq} below).
\end{remark}

\section{Solution to the Fuchsian system associated to the Hurwitz Frobenius manifolds}
\label{sect_solution}

\subsection{Preliminaries}
\label{sect_preliminaries}

Let $\L$ be a Riemann surface of genus $g$ and $f$ be a meromorphic
function on $\L$ of degree $\N$. Let us fix the degrees of the poles of $f$ to
be $k_1,\dots,k_\K$ ($k_1+\dots +k_\K=\N$), and assume that all finite critical
points of the function $f$ (i.e., zeros of $df$) are simple; we denote them by
$P_1,\dots,P_{\M}$, where, according to the Riemann-Hurwitz formula, $\M=2g+\N+\K-2$. We denote by
$\H_{g,\N}(k_1,\dots,k_\K)$ the Hurwitz space, i.e., the space of
equivalence classes of  pairs (or branched coverings) $\X:=(\L,f)$ (two coverings $\X_1:=(\L_1,f_1)$ and
$\X_2:=(\L_2,f_2)$ are called equivalent if there exists a
biholomorphic isomorphism  $h:\L_1\to\L_2$ such that
$f_1 = f_2\circ h $).
The critical values of the function $f,$ i.e., the values $\l_k:=f(P_k)$ with $k=1,\dots,\M,$
can be chosen to be the local coordinates on this Hurwitz space.

The branched covering $\X=(\L,f)$ is  a $\N$-sheeted covering of
$\CP1$ ramified at the points $P_1,\dots,P_{\M}$ as well as at those
poles of $f$ whose degrees are higher than 1. The critical values
$\{\l_k\}$ are the finite branch points of the  covering $\X$. In a
neighbourhood of the ramification
point $P_j$ we introduce a local parameter $x_j(P)$ (called {\it distinguished} \cite{Zorich})
satisfying  equations
$$
df= 2 x_j dx_j\;,\hskip0.7cm x_j(P_j)=0\;.
$$
This differential equation has two solutions: $x_j(P)=\pm \sqrt{f(P)-\l_j}$.
Therefore, for each $j$ we have two possible choices (which differ by a sign) of the distinguished local parameter.
Altogether we get $2^N$ sets of distinguished local parameters.

Let us introduce the canonical meromorphic bidifferential $W(P,Q)$, where
$P,Q\in\L$. This bidifferential is symmetric; it has a quadratic pole on the diagonal $P=Q$
with the singular part given by $dx(P) dx(Q)(x(P)-x(Q))^{-2}$ in any
local parameter $x$, and is normalized by the requirement that all of
its $a$-periods with respect to some symplectic basis
$({\bf a}_\a,{\bf b}_\a)$ in $H_1(\L)$ vanish. Let us also introduce the canonical
basis of holomorphic differentials $w_1,\dots,w_g$ on $\L$ normalized
by $\oint_{{\bf a}_\a} w_\b=\delta_{\a\b}$, where $\delta_{\a\b}$ is
  the Kronecker symbol and $\a,\b=1,\dots,g$. Integrals of these
  differentials over the cycles ${\bf b}_\a$ give the Riemann matrix $\B$ of the surface:  $\B_{\a\b}=\oint_{{\bf b}_\a} w_\b$.

We are going to use  the  Rauch variational formulas, which
describe the dependence of $w_\a$, $W$ and $\B$ on the branch points
$\{\l_k\}$ (see \cite{Rauch,KokKorG}):
\be
\f{d }{d\l_j}\{\B_{\a\b}\}=2 {\pi \i}\,{\rm Res}\big|_{Q=P_j}\left\{ \frac{w_{\a}(Q)  w_{\b}(Q)}{df(Q)}\right\}\;;
\label{RauchB0}
\ee
\be
\f{d }{d\l_j}\Big|_{f(P)}\{w_\a(P)\}={\rm Res}\big|_{Q=P_j}\left\{  \frac{w_{\a}(Q)\,W(Q,P)}{df(Q)} \right\};
\label{Rauchw0}
\ee
\be
\f{d}{d\l_j}\Big|_{f(P),\,f(Q)} \{W(P,Q)\}={\rm Res}\big|_{R=P_j}\left\{\frac{ W(P,R)W(Q,R)}{df(R)}\right\}\;\;.
\label{RauchW0}
\ee
Here the derivative with respect to $\l_k$ is taken keeping the
projections $f(P)$ and $f(Q)$ of the points $P$ and $Q$ to $\CP1$
constant.

Note that the variational formulas for normalized holomorphic differentials
 (\ref{Rauchw0}) can  be alternatively stated as horizontality  of the column vector $(w_1(P),\dots,w_g(P))^T$
with respect to the Gauss-Manin connection 
\be
d-\sum_{j=1}^N  \Theta_j d\lambda_j\;,
\ee
where the connection coefficients $\Theta_j$ are  the $g\times g$ diagonal matrices with
$$
(\Theta_j)_{\a\a}:=
{\rm Res}|_{Q=P_j}\left\{\f{w_{\a}(Q)W(P,Q)}{df(Q)\,w_{\a}(P)}\right\}
$$

The formulas (\ref{RauchB0}) - (\ref{RauchW0}) can be alternatively rewritten in
the following less invariant  form which we are going to use below:
\be
\f{d }{d\l_j}\{\B_{\a\b}\}= \pi i  w_{\a}(P_j)w_{\b}(P_j)
\;;
\label{RauchB}
\ee
\be
\f{d }{d\l_j}\Big|_{f(P)}\{w_\a(P)\}=\f{1}{2} w_{\a}(P_j)W(P,P_j);
\label{Rauchw}
\ee
\be
\f{d}{d\l_j}\Big|_{f(P),\,f(Q)} \{W(P,Q)\}=\f{1}{2} W(P,P_j)W(Q,P_j)\;,
\label{RauchW}
\ee
 where
\be
\label{defW(P,P_j)}
w_\a(P_j):=\f{w_\a(P)}{dx_j(P)}\Big|_{P=P_j}\;,\hskip0.7cm
W(P,P_j):=\f{W(P,Q)}{dx_j(Q)}\Big|_{Q=P_j} .
\ee
and $x_j(Q)$ is one of the possible sets of distinguished local parameters.

For the squares, $w_\a(P_j)^2$ and $W^2(P,P_j)$, we have the following invariant expressions:
\be
w_\a^2(P_j)= 2{\rm Res}\big|_{P=P_j}\left\{\frac{w_\a^2(P)}{df(P)}\right\}\;, \hskip0.7cm
W^2(P,P_j)=2{\rm Res}\big|_{Q=P_j}\left\{\frac{W^2(P,Q)}{df(Q)}\right\}\;.
\ee

In the next section we  are going to solve the linear system (\ref{ls1}), (\ref{ls2}),
where the rotation coefficients are given by (\ref{Gammaijintr}) \cite{2D,KokKorG}:
i.e.,
\be
\Gamma_{jk}=\frac{1}{2}W(P_j,P_k):=\frac{1}{2}\f{W(P,Q)}{dx_j(P)dx_k(Q)}\Big|_{P=P_j\,,Q=P_k}\;.
\la{coefGamma}
\ee
where $\{x_j\}$ is some set of distinguished local parameters. By changing signs of some of the $x_j$
 we get $2^N$ different sets  of rotation coefficients.
These coefficients satisfy the system (\ref{ggg}), (\ref{unitg}) as a
simple corollary of the Rauch formulas (\ref{RauchW}).

The squares $\Gamma_{jk}^2$ of rotation coefficients, are defined by the following residue formulas:
$$
\Gamma_{jk}^2= {\rm Res}\big|_{P=P_j}\,{\rm Res}\big|_{Q=P_k}\left\{\frac{W^2(P,Q)}{df(P) \,df(Q)}\right\}.
$$

\subsection{Construction of a solution to the Fuchsian system for $q=-1/2$}

%Notations:  number of $\l_k$: $\M$ (v tekste  ``slash M''); $j,k=1,\dots,\M$. In the text write  as ``slash M'',
%later may change to something also...

%Number of sheets: $N$; $n=1,\dots, N$.

%Genus: $g$; $\a,\b=1,\dots, g$.

%Number of infinities: $\K$; $s=1,\dots,\K$.

%\vskip1.0cm

Let us fix some $\lambda\in \CP1$ which does not coincide with any
of $\l_j,$ i.e., such that its pre-image $f^{-1}(\l)$ consists of $\N$ different
points
$\l^{(k)}$, $k=1,\dots,\N$.
Let us also enumerate in some way the points of  $f^{-1}(\infty)$,
which we denote by $\infty^{(s)}$, $s=1,\dots,\K$ (if some of $\infty^{(s)}$ are
ramification points then $\K<\N$).

Let us introduce the homology group
$H_1(\L\setminus f^{-1}(\infty)\,;\, f^{-1}(\l))$, with coefficients in $\Z,$ of the Riemann
surface $\L$ punctured at $\K$  points  $\infty^{(s)}$, $s=1,\dots,\K$,
relative to the set $f^{-1}(\l)$ of $\N$ points $\l^{(k)}$, $k=1,\dots,\N$.
The dimension of $H_1(\L\setminus f^{-1}(\infty)\,;\,f^{-1}(\l))$ is $2g+\N+\K-2$; this dimension
equals $\M$, the number  of the branch points $\{\l_j\}$.
The set of basis contours $\con_k$, $k=1,\dots, 2g+\N+\K-2$ in $H_1(\L\setminus f^{-1}(\infty)\,;\,f^{-1}(\l))$ can be chosen as follows:
\be
\con_{2\a-1}:= a_\a\;\hskip0.7cm
\con_{2\a}:=b_\a\;,\hskip0.7cm \a=1,\dots,g,
\label{bacyc}
\ee
where $(a_\a,\b_\a)$ is a canonical basis of cycles in the homology
space $H_1(\L,\Z)$;
\be
\con_{2g+s}:= l_s\;,\hskip0.7cm s=1,\dots,\K-1,
\label{lsss}
\ee
where $l_s$ is the closed contour encircling $\infty^{(s)}$ in the
positive direction (in $H_1(\L,\Z)$ the contour $l_s$ is trivial);
\be
\con_{2g+\K-1+n}:= \gamma_{n,n+1}(\l)\;,\hskip0.7cm n=1,\dots,\N-1,
\label{gkkp1}
\ee
where $\gamma_{n,n+1}(\l)$ is some contour connecting the points $\l^{(n)}$ and
$\l^{(n+1)}$ from the pre-image $f^{-1}(\lambda)$.

It is sometimes convenient to choose the symplectic basis $(a_\a,b_\a)$ in $H_1(\L)$
 (\ref{bacyc}) which forms a part of the
basis in the space of relative homologies $H_1(\L\setminus f^{-1}(\infty)\,;\,
f^{-1}(\l))$) independently of the basis $({\bf a}_\a,{\bf b}_\a)$ used
 for normalization  of the bidifferential $W$  and  the holomorphic $1$-forms $w_\alpha$ (see Section \ref{sect_preliminaries}). That's why we denote these two bases of
 $H_1(\L)$ by different letters.

Let us consider the
 meromorphic differential $W(P,P_j)$ on $\L$ (see (\ref{defW(P,P_j)})). This is the Abelian differential of the second
 kind, having a
second order pole at $P_j$ with the singular part $(x_j(P))^{-2}dx_j(P)$ and
 all vanishing periods over the cycles ${\bf a}_\a$, where $\{x_j\}$ is the same set
of distinguished local parameters as in (\ref{coefGamma}).
A change of sign of  $x_j$ implies the change of the sign of $W(P,P_j)$.

The meromorphic differential $f(P)W(P,P_j)$
has a second order pole at $P_j$ and  poles of order $k_s$ at all poles
$\infty^{(s)}$, $s=1,\dots,\K$ of the function $f$. Generically, the differential $f(P)W(P,P_j)$ does not satisfy any normalization conditions.
% (the order of the poles of this differential at $\infty^{(s)}$ depends on the multiplicity of the branching at these points).

Now we are going to construct a solution to the Fuchsian system (\ref{ls1}) and isomonodromy equations (\ref{ls2}) in terms
of integrals of the differentials  $W(P,P_j)$ and
$f(P)W(P,P_j)$ over the basis  (\ref{bacyc})-(\ref{gkkp1})  in the relative
homology space $H_1(\L\setminus f^{-1}(\infty)\,;\, f^{-1}(\l))$.

Consider some point $\l_0\in\C$ which does not coincide with any of $\l_j$.
Consider an open simply-connected  neighbourhood  $D\subset \C$ of $\l_0$ such that $f^{-1}(D)$ consists of $d$ connected components.

 For all $\l\in D$ we choose
 the  basis elements (\ref{bacyc})-(\ref{gkkp1}) of  the space  $H_1(\L\setminus
f^{-1}(\infty)\,;\, f^{-1}(\l))$ to be obtained by a small smooth deformation from the respective elements of  $H_1(\L\setminus
f^{-1}(\infty)\,;\, f^{-1}(\l_0))$
(this concerns in fact only the contours $\gamma_{n,n+1}(\l)$
(\ref{gkkp1}): we require that for all $\l\in D$ these contours differ
from
 $\gamma_{n,n+1}(\l_0)$ only by paths connecting  the endpoints
$[\l_0^{(n)},\l^{(n)}]$ and $[\l_0^{(n+1)},\l^{(n+1)}]$ within
 $f^{-1}(D)$).

For any contour $\con\in H_1(\L\setminus f^{-1}(\infty);\;
f^{-1}(\l))$ we introduce the column vector-function $\Phi^{(\con)}$ with
values in $\C^{\iM}$  whose $j$th component ($j=1,\dots,\M$) is given by:
\begin{equation}
\Phi^{(\con)}_j(\l):= \lambda\int_{\con} W(P,P_j) - \int_{\con}
f(P)W(P,P_j),%\equiv \int_{\con}(\lambda-f(P)))W(P,P_j)\;,
\label{defPhicon}
\end{equation}
where $\lambda\in D$.

Let us choose for a moment the canonical basis of cycles $({\bf a}_\a,{\bf b}_\a),$ with respect to which
 the meromorphic bidifferential $W$ is normalized  (see Section \ref{sect_preliminaries}), to coincide with
the  canonical basis of cycles $(a_\a,b_\a)$  from the basis (\ref{bacyc}) in $H_1(\L\setminus f^{-1}(\infty)\;,\;f^{-1}(\l)).$
%used in the definition of the integration contours $\con_k$ .
Then the
vectors $\Phi^{(a_\a)}$, $\a=1,\dots,g,$ do not depend
on $\l$, since $a$-periods of the differentials $W(P,P_j)$ vanish:
\begin{equation*}
\Phi^{(a_\a)}_j(\l)=-\oint_{a_\a}f(P) W(P,P_j)\;.
\end{equation*}
The vectors  $\Phi^{(b_\a)}$, $\a=1,\dots,g,$ are linear in $\l$; since
$b$-periods of $W$ are given by the holomorphic normalized differentials $\{w_\a\}$:
\begin{equation*}
\Phi^{(b_\a)}_j(\l)= 2\pi \i \, \lambda\, w_{\a}(P_j) - \oint_{b_\a} f(P) W(P,P_j)\;.
\end{equation*}
The columns corresponding to the contours $l_s$  do not depend on
$\l$ either, since the differentials $W(P,P_j)$ are non-singular at
$\infty^{(s)}$:
\begin{equation}
\Phi^{(l_s)}_j(\l)=-2\pi \i \,{\rm res}|_{P=\infty^{(s)}} [f(P) W(P,P_j)]\;,\hskip0.7cm
s=1,\dots,\K-1.
\label{Phils}
\end{equation}
In particular, if all $\infty^{(s)}$ are not ramification points, i.e., $\K=\N$,
the residues in (\ref{Phils}) can be easily computed to give
\begin{equation}
\Phi^{(l_s)}_j(\l)= - 2\pi \i \, W(\infty^{(s)}, P_j):=- 2\pi \i \, \frac{W(Q,P_j)}{dz_s(Q)}\mid_{Q=\infty^{(s)}} \;,\hskip0.7cm
s=1,\dots,\N-1,
\label{Philssp}
\end{equation}
where $z_s=1/\lambda$ is the local parameter at $\infty^{(s)}.$ The columns $\Phi^{(\g_{n,n+1}(\l))}$ depend on $\l$
non-trivially through the dependence of the integration contours
$\g_{n,n+1}(\l)$ on $\l$.
\begin{theorem}
\label{thm_solution}
For any contour $\con\in H_1(\L\setminus f^{-1}(\infty)\,;\, f^{-1}(\l)),$ the vector function  $\Phi^{(\con)}$ defined by (\ref{defPhicon}), satisfies the linear system (\ref{ls1}), (\ref{ls2}) with $q=-1/2$ and $\l\in D$.
\end{theorem}
{\it Proof.}
We shall check that the vector $\Phi^{(\con)}(\l)=(\Phi^{(\con)}_1(\l),\dots, \Phi^{(\con)}_\iM(\l))^\iT$ satisfies the system (\ref{eulereq}), (\ref{uniteq}), (\ref{maineq}) with $q=-1/2$, which is equivalent to the original system (\ref{ls1}), (\ref{ls2}) with the same value of the parameter $q$.

The validity of equations (\ref{maineq}) is an immediate consequence of (\ref{coefGamma}) and the Rauch variational formulas for the bidifferential $W(P,Q)$.

To verify (\ref{uniteq})  we lift the functions
$\Phi^{(\con)}_j(\lambda),$ $\lambda\in D \subset \cp,$ (\ref{defPhicon}) to the
function $\Phi^{(\con)}_j(f(P))$ defined locally on any connected component of $f^{-1}(D)$ on the Riemann surface $\L$.
%We do so by
%letting  $\l=f(P)$ be the projection of the point $P\in\L$ to the base
%of the covering.
%We shall study the behaviour of the functions
%$\Phi^{(\con)}_j(f(P))$ under  biholomorphic transformations of the Riemann surface $\L.$

The equation  (\ref{uniteq}) is an infinitesimal form of the
invariance of the function
$\Phi^{(\con)}(f(P))$ under a simultaneous translation of all
$\lambda_j$ and $\l=f(P)$ by a constant.
Namely, consider a biholomorphic mapping of the Riemann surfaces
$\surf \to \surf^\delta$ which
acts in every sheet of
 $\surf$ by sending the point $P$ with the projection $\l=f(P)$  to
the point $P^\delta$  projecting to
$\l^{\delta}:=f(P^\delta)=f(P)+\delta$ on the base of the covering. The branch points
$\{\l_i\}$ are then mapped to
$\{\l_i+\delta\}.$ Due to the invariance of the local parameters
$x_i(P) = \sqrt{f(P)-\l_i}$
under the mapping and the invariance of the bidifferential $W$ under all biholomorphic mappings of the surfaces, the equality $W(P,P_i) = W^\delta(P^\delta,P_i^\delta)$ holds, where $W^\delta$ is the bidifferential $W$ defined on $\surf^\delta.$ Therefore, for the function $\Phi_j^{(\con)}(f(P))$  we have:
\begin{multline*}
(\Phi^{(\con)}_j)^\delta(f(P^\delta)):= f(P^\delta)\!\!\int_{\con^\delta}\!\! W^\delta(Q,P^\delta_j) - \int_{\con^\delta} \!\! f(Q)W^\delta(Q,P^\delta_j)
%=(\lambda+\delta)\int_{\con} W^\delta(P^\delta,P^\delta_j) - \int_{\con} f(P^\delta)W^\delta(P^\delta,P^\delta_j) \\
\\=(f(P)+\delta)\!\!\int_{\con} \! W(Q,P_j) - \int_{\con} \! (f(Q)+\delta)W(Q,P_j),
\end{multline*}
where the second equality is obtained by changing the variable of
integration $Q\mapsto Q^\delta$ and using the invariance $W(P,P_j) =
W^\delta(P^\delta,P_j^\delta).$  Differentiating the above relation
with respect to $\delta$ at $\delta=0$ we get
$\partial_\l\Phi_j^{(\con)}(\l)+ {\bf e}( \Phi_j^{(\con)}(\l))=0,$
i.e., the first   equation in (\ref{uniteq}).

Finally, the  equation  (\ref{eulereq})  with $\q=-1/2$ can be verified by considering the transformation of the function $\Phi^{(\con)}(f(P))$ under the biholomorphic mapping of the Riemann surfaces $\surf \to \surf^\epsilon$ which maps the point $P$ with the projection $f(P)$  to the point $P^\epsilon$ belonging to the same
sheet and  projecting to $f(P^\epsilon)=(1+\epsilon)f(P)$ on the base. The local parameters $x_j(P)$ get multiplied by $\sqrt{1+\epsilon}$ and the bidifferential $W$ stays invariant, i.e., $W(P,Q)=W^\epsilon(P^\epsilon,Q^\epsilon).$ Thus for the differential  $W(Q,P_j)$ we have  $W^\epsilon(Q^\epsilon,P^\epsilon_j)=W(Q,P_j)/\sqrt{1+\epsilon},$ see (\ref{defW(P,P_j)}). Therefore, for the function $\Phi_j^{(\con)}(f(P))$ (\ref{defPhicon}) we have:
\begin{equation*}
(\Phi^{(\con)}_j)^\epsilon(f(P^\epsilon)):= f(P^\epsilon) \!\! \int_{\con^\epsilon} \!\! W^\epsilon(Q,P^\epsilon_j) - \int_{\con^\epsilon} \!\!f(Q)W^\epsilon(Q,P^\epsilon_j)
\end{equation*}
$$
=\sqrt{1+\epsilon} \left[ f(P) \!\! \int_{\con} \!\! W(Q,P_j) - \!\! \int_{\con} \!\! f(Q)W(Q,P_j) \right]\,,
$$
where the second equality is obtained by changing the variable of  integration $Q\mapsto Q^\epsilon$ and using the relation $W^\epsilon(Q^\epsilon,P^\epsilon_j)=W(Q,P_j)/\sqrt{1+\epsilon}.$ This implies for the function $\Phi^{(\con)}_j(\l(P)):$
\begin{equation*}
(\Phi^{(\con)}_j)^\epsilon(f(P^\epsilon)) = \sqrt{1+\epsilon}\, \Phi^{(\con)}_j(f(P)).
\end{equation*}
Differentiating this relation with respect to $\epsilon$ at $\epsilon=0$ we get
\begin{equation*}
\l\,\partial_\l\Phi^{(\con)}_j(\l)+ {\bf E}( \Phi^{(\con)}_j(\l))=
\partial_\epsilon\mid_{\epsilon=0}(\Phi^{(\con)}_j)^\epsilon(\l^\epsilon)=\frac{1}{2}\Phi^{(\con)}_j(\l).
\end{equation*}
$\Box$

Now from $\M$ vectors $\Phi^{(\con_k)}$, $k=1,\dots,\M,$ corresponding to
the basis (\ref{bacyc}),
(\ref{gkkp1}), (\ref{lsss}) of  $H_1(\L\setminus f^{-1}(\infty)\,;\,
f^{-1}(\l))$, we construct the $\M\times \M$ matrix
\be
\Phi(\l):=(\Phi^{(\con_1)},\Phi^{(\con_2)},\dots,\Phi^{(\con_{\M})}) \qquad \mbox{for } \l\in D.
\label{phisol}
\ee

\begin{theorem}
The matrix $\Phi(\l)$ (\ref{phisol}) gives a complete set of linearly independent
solutions to the Fuchsian linear system (\ref{ls1}) for $\l\in D$ with
$\q=-1/2$. The matrix $\Phi(\l)$
also satisfies the isomonodromy deformation equations (\ref{ls2}).
\end{theorem}
{\it Proof.} The  matrix $\Phi$ satisfies equations (\ref{ls1})  and (\ref{ls2}) since each of its columns satisfies these equations. The proof of linear independence of its columns is rather tedious.   We postpone it to Appendix \ref{app_completeness} which is entirely devoted to this proof.
$\Box$

\vskip0.3cm

\subsubsection{ Solutions for other half-integer $q$.} 
The solution (\ref{defPhicon}) can be formally rewritten in the following form:
\begin{equation}
\Phi^\con(\lambda) = \int_\con df(P)\int^P W(R,P_i),
\la{solmod}
\end{equation}
where $\con$ is again one of the integration contours (\ref{ls1}) - (\ref{gkkp1}) and we assume that the closed contours start and end at one of the points from the set $f^{-1}(\lambda)$ (not necessarily the same for all contours). This can be achieved by deformation of contours.
This solution satisfies our linear system with $\q=-1/2$. Similarly to the proof of (\ref{solmod})  (Theorem \ref{thm_solution}) one can prove that a  solution for the system (\ref{ls1}), (\ref{ls2}) with $\q=-3/2$ can be written in the form
\begin{equation}
\label{q3}
\tilde{\Phi}^\con(\lambda) = \int_\con df(P) \int^P  df(Q)\int^Q W(R,P_i)\;.
\end{equation}
Adding one more integration, we get a solution to the system (\ref{ls1}), (\ref{ls2}) with $\q=-5/2:$
\begin{equation}
\label{q5}
\tilde{\tilde{\Phi}}^\con(\lambda) = \int_\con df(P^\prime) \int^{P^\prime}  df(P) \int^P  df(Q)\int^Q W(R,P_i),
\end{equation}
and so on.

Let us perform integration by parts in (\ref{q3}) and (\ref{q5}). Then the solutions take, respectively, the form:
\begin{equation}
\label{q31}
\tilde{\Phi}^\con(\lambda) = \frac{1}{2} \lambda^2 \int_\con  W(P,P_i) - \lambda  \int_\con f(P) W(P,P_i)  + \frac{1}{2} \int_\con f^2(P) W(P,P_i).
\end{equation}
\begin{equation}
\label{q51}
\tilde{\tilde{\Phi}}^\con(\lambda) = \frac{1}{6} \lambda^3 \int_\con  W(P,P_i) - \frac{1}{2} \lambda^2 \int_\con f(P) W(P,P_i) + \frac{1}{2} \lambda \int_\con f^2(P)  W(P,P_i) - \frac{1}{6} \int_\con f^3(P) W(P,P_i).
\end{equation}
A straightforward differentiation of (\ref{q51}) with respect to $\l$ gives  (\ref{q31}); and differentiation of (\ref{q31}) gives
 (\ref{defPhicon}), in agreement with (\ref{aap1}).

Similarly, we get solutions to systems  (\ref{ls1}), (\ref{ls2}) for any negative half-integer value of $\q$. However, for $\q=1/2,3/2\,\dots$
we do not get a complete system of solutions of  (\ref{ls1}), (\ref{ls2}) since some of columns of (\ref{defPhicon}) do not depend on
$\lambda$ and turn into zero vectors after differentiation.

\subsubsection{ Fuchsian and non-Fuchsian linear systems.}
The same system of equations (\ref{ggg}),  (\ref{unitg})  describes
isomonodromic deformations of the non-Fuchsian equation
\be
\f{d\Psi}{d z}= (U+\f{1}{z}V)\Psi\;.
\label{nfls}
\ee
A solution $\Psi$ to the system (\ref{nfls}) has an irregular singularity of Poincar\'e
rank 1 at
$z=\infty$, and a regular singularity at the origin.

Solutions to the Fuchsian system (\ref{ls1}) and the non-Fuchsian system
(\ref{nfls}) are related by a formal Laplace transform
 (see also \cite{2D}, p. 87, (3.149)).
Namely, suppose $l$ is a contour in the $\l$
-plane satisfying the following conditions: the solution $\Phi^\q$ of the system (\ref{ls1}),
(\ref{ls2}) is analytic on of $l$; the contour $l$ is either closed or its both ends approach the point at infinity in the sector where ${\rm Re}\{\lambda z\}<0, \; \lambda\in l$. Then the function \be
\label{Laplace} \Psi(z) = z^{(1-\q)} \int_l e^{z\l}\Phi^\q(\l) d\l
\ee solves the corresponding non-Fuchsian system (\ref{nfls}). Vice
verse, if one knows $\Psi$, the function $\Phi^\q$ can be obtained
by inversion of the Laplace transform.

In \cite{V3} solutions of the non-Fuchsian system (\ref{nfls}) were
constructed in the form \be \Psi_{ij}(z)= \f{1}{\sqrt{z}}\int_{C_j}
e^{z f(P)} W(P,P_i), \la{irreg} \ee where $\{C_j\}$ is a family of
$N$ contours connecting poles of the function $f$. A comparison of
(\ref{irreg}) and (\ref{Laplace})  (choosing $l=C_1$) shows that the
column vector with components $W(P,P_j)$ satisfies the Fuchsian
system (\ref{ls1}) with $\q=3/2$ (this fact can be easily proved
directly, without using the Laplace transform).  However, to find
the remaining $N-1$ linearly independent vectors satisfying the
system (\ref{ls1}) with $\q=3/2$ one needs to take a superposition
of the  Laplace transforms over  contours $C_2,\dots,C_N$ with the
inversion of the Laplace transform over the contour $C_1$.  We do not
know whether these remaining columns admit a more explicit
representation in the case $q=3/2$. On the other hand, formula (\ref{irreg}) for solution of the non-Fuchsian system (\ref{nfls}) can be obtained by substitution of our solution (\ref{defPhicon}) (with $q=-1/2$ and $s$ being one of the contours (\ref{gkkp1})) into (\ref{Laplace}). Choosing the integration contour $l$ to coincide with the contour $C_j$, we can perform integration by parts in (\ref{Laplace}) leading to (\ref{irreg}).

\subsection{Dependence of the solution on the choice of homology basis}

\label{sect_normalization}

In this section we discuss the dependence of the solution $\Phi$
(\ref{defPhicon}), (\ref{phisol})
%to the linear system (\ref{ls1}), (\ref{ls2})
on the choice of a Lagrangian subspace $\{{\bf a}\}$ generated by the {\it a}-cycles ${\bf a}_1,\dots,{\bf a}_g$ in $H_1(\L)$
with respect to which  $W(P,Q)$ is normalized
on the choice of the integration contours $\con_1,\dots,\con_{\M}$
and on the choice of the signs of distinguished local parameters $x_j$.
%which form a basis in $H_1(\L\setminus f^{-1}(\infty)\;,\;f^{-1}(\l))$.

Let us denote by ${\bf a}$ and ${\bf b}$ the column vectors of basis cycles: ${\bf
  a}:=({\bf a}_1,\dots,{\bf a}_g)^\iT$ and ${\bf b}:=({\bf
  b}_1,\dots,{\bf b}_g)^\iT.$ Consider a new symplectic basis,  $({\bf \hat{a}},{\bf \hat{b}})$,
in $H_1(\L)$
    which is related to the old one by a symplectic transformation:
\begin{equation}
\left(\begin{array}{r}
      {\bf \hat{b}}  \vspace{.2cm} \\
    {\bf  \hat{a}} \\
   \end{array} \right) =
\left(\begin{array}{rr}
      A & B  \vspace{.2cm} \\
      C & D\\
   \end{array} \right)
   \left(\begin{array}{r}
      {\bf b}  \vspace{.2cm} \\
     {\bf a} \\
   \end{array} \right).
   \label{symplectic}
\end{equation}

Then the canonical bidifferentials $\widehat{W}$ and $W$ corresponding to the bases $({\bf \hat{a}},{\bf \hat{b}})$ and $({\bf a}, {\bf b})$, respectively, are related by  (see \cite{Fay92}, p.10):
\begin{equation}
\label{FayW}
\widehat{W}(P,Q) = W(P,Q) - 2 \pi\i \; w^\iT(P) (C\B+D)^{-1}C  w(Q),
\end{equation}
where $w$ is the vector of holomorphic  differentials,
$w:=(w_1,\dots,w_g)^\iT,$ normalized  by $\oint_{{\bf a}_\alpha}
w_\beta=\delta_{\alpha \beta},$
and $\B$ is the  matrix of  ${\bf b}$-periods:  $\B_{\a\b}:=\oint_{{\bf b}_{\a}} w_{\b}.$

Let us denote by $\con$ the row vector whose components are given by the contours $\con_1,\dots,\con_\M$. For another basis
 $\hat{\con}= (\hat{\con}_1,\dots,\hat{\con}_\M)$ in $H_1(\L\setminus f^{-1}(\infty)\;,\;f^{-1}(\l))$ we have $\hat{\con}= \con R$,
where $R$ is a non-degenerate $\M\times \M$ matrix with integer entries.

Then we can form a matrix-function  $\widehat{\Phi}(\lambda)$  defined by the formulas
 (\ref{defPhicon}),
(\ref{phisol}) with the bidifferential $W$ replaced by the transformed bidifferential $\widehat{W}$,
with integration contours $\{\hat{\con}_k\} \in H_1(\L\setminus f^{-1}(\infty)\;,\;f^{-1}(\l))$,
and a new set of distinguished local parameters $\hat{x}_j=\epsilon_j x_j$ with $\epsilon_j^2=1$.
 The function $\widehat{\Phi}(\lambda)$ solves the system (\ref{ls1}),
 (\ref{ls2}) with the matrix $V$ built from the rotation coefficients
 given by the deformed bidifferential:
 \begin{equation*}
 \Gamma_{ij}=\f{\widehat{W}(P,Q)}{d \hat{x}_i(P) \, d\hat{x}_j(Q)}\Big|_{P=P_i\,Q=P_j}\;.
 \end{equation*}
 
\begin{theorem}
\label{thm_Schlesinger}

The matrix-functions  $\widehat{\Phi}$ and $\Phi$ are related as follows:
\begin{equation}
\widehat{\Phi} (\lambda) = Y \left(\Id - \T(\l) \right) \Phi (\lambda) R \;,
\label{Schlesinger}
\end{equation}
where $\Id$ denotes the $\M\times\M$ identity matrix; $\T$ is a symmetric matrix with the entries:
\begin{equation}
%(\T)_{ij} = \pi \i \; \lambda_j \omega^\iT(P_i) (C\B+D)^{-1}C  \omega(P_j),
(\T)_{ij} = \pi \i \; (\lambda_j - \lambda) \sum_{\a,\b=1}^g \left[
    (C\B+D)^{-1}C\right]_{\a\b} w_{\a}(P_i)  w_{\b}(P_j)\;,
\label{T}
\end{equation}
where
$w_{\a}(P_j):=({w_\a(P)}/{d x_j(P)})|_{P=P_j}$;   the constant matrices $C$ and $D$ are
blocks of the symplectic transformation (\ref{symplectic}) between the two canonical homology bases;
$R$ is the transformation matrix between the sets of new and old integration contours: $\hat{\con}=  \con\, R$; the diagonal matrix $Y$ is formed by the factors $\epsilon_j$, i.e., $Y:={\rm diag}(\epsilon_1,\dots,\epsilon_{\M})$.
\end{theorem}
{\it Proof.}
It is sufficient to check the statement of the theorem in three cases:
\begin{enumerate}
\item
The symplectic matrix in (\ref{symplectic}) is the unit matrix, all $\epsilon_j=1$ (i.e. $Y=\Id$), while the transformation matrix $R$ between the bases $\con$ and $\hat{\con}$ in $H_1(\L\setminus f^{-1}(\infty)\;,\;f^{-1}(\l))$ is non-trivial.
Then $W=\widehat{W}$ and the only difference between $\Phi$ and $\widehat{\Phi}$ is the choice of the integration contours; therefore, $\widehat{\Phi}=\Phi\, R$.
\item
Matrix $R$ is the unit matrix (i.e. the contours of integration $\con_j$ remain unchanged), all  $\epsilon_j=1$ while the symplectic transformation
matrix in
(\ref{symplectic}) is non-trivial.

In this case the formula (\ref{Schlesinger}) with $Y=R=\Id$  can be proved by a direct computation as follows. Relation (\ref{Schlesinger}) is equivalent to
\begin{equation}
\label{comp1}
%\int_{\con} {\l(Q)}\widehat{W}(Q,P_i) =\int_{\con} {\rm e}^{z\l(Q)} \sum_{j=1}^n \left( \Id - \frac{1}{z}\T \right)_{ij} W(Q,P_j) ,
\lambda\int_{\con_k} \widehat{W}(P,P_i) - \int_{\con_k} f(P)\widehat{W}(P,P_i) = \sum_{j=1}^\M \left( \Id - \T \right)_{ij} \left(\lambda\int_{\con_k} W(P,P_j) - \int_{\con_k} f(P)W(P,P_j) \right),
\end{equation}
for any $i=1,\dots,\M$ and $k=1,\dots,\M.$
Using the definition (\ref{T}) of the matrix $\T$ and the Rauch
variational formula (\ref{Rauchw}) for the
holomorphic differentials $w_\a$, we obtain: %rewrite the right-hand side of (\ref{comp1}) as follows:
\begin{multline}
\label{comp2}
\sum_{j=1}^\M \left( \Id - \T \right)_{ij} \int_{\con_k} f(P)W(P,P_j)
%
%= \int_{\con} f(P)W(P,P_i) - \pi \i \sum_{k,l=1}^g \left[
%(C\B+D)^{-1}C\right]_{kl} \omega_k(P_i)
%\sum_{j=1}^\M  (\lambda_j-\lambda)\omega_l(P_j) \int_{\con} %f(P)W(P,P_j)\\
%
%= \int_{\con} f(P)W(P,P_i) - 2\pi \i \sum_{k,l=1}^g \left[
%(C\B+D)^{-1}C\right]_{kl} \omega_k(P_i)
%\sum_{j=1}^\M  (\lambda_j-%\lambda)\partial_{\lambda_j}\int_{\con} f(P)\omega_l(P)\\
= \int_{\con_k} f(P)W(P,P_i) \\ - 2\pi \i \sum_{\a,\b=1}^g \left[
  (C\B+D)^{-1}C\right]_{\a\b} w_\a(P_i)
\left[  \E\left(\int_{\con_k} f(P)w_\b(P)\right) -\lambda \e\left(\int_{\con_k} f(P)w_\b(P)\right)\right],
\end{multline}
where $\E=\sum_{j=1}^\iM \lambda_j\partial_{\lambda_j}$ is the Euler vector
field  and
$\e=\sum_{j=1}^\iM \partial_{\lambda_j}$ is the unit vector field on the Frobenius manifold.
We compute the action of these fields on our integrals using the
invariance of the holomorphic differentials $w_k$ with respect to the biholomorphic mappings of Riemann surfaces
$\surf\to\surf^\epsilon$ and $\surf\to\surf^\delta$ from the proof of Theorem \ref{thm_solution}:
\begin{multline}
\E\left(\int_{\con_k} f(P)w_\b(P)\right) = \frac{d}{d\epsilon}
\mid_{\epsilon=0} \int_{\con_k^\epsilon}
f(P)w^\epsilon_\b(P) = \frac{d}{d\epsilon}\mid_{\epsilon=0} \int_{\con_k} f(P^\epsilon)w^\epsilon_\b(P^\epsilon)  \\
%=\frac{d}{d\epsilon} \mid_{\epsilon=0} \int_{\con} f(P^\epsilon)\omega^\epsilon_l(P^\epsilon)\\
= \frac{d}{d\epsilon} \mid_{\epsilon=0} \int_{\con_k} f(P)(1+\epsilon)w_\b(P)=\int_{\con_k} f(P)w_\b(P).
\label{Euler1}
\end{multline}
\begin{multline}
\label{e1}
\e\left(\int_{\con_k} f(P)w_\b(P)\right) = \frac{d}{d\delta}
\mid_{\delta=0} \int_{\con_k^\delta}
f(P)w^\delta_\b(P)
%=\frac{d}{d\delta} \mid_{\delta=0} \int_{\con} f(P^\delta)\omega^\delta_l(P^\delta)\\
= \frac{d}{d\delta} \mid_{\delta=0} \int_{\con_k} (f(P)+\delta)w_\b(P)=\int_{\con_k}w_\b(P).
\end{multline}
To obtain the second equalities in the above lines we used the
invariance
$w^\epsilon_\b(P^\epsilon)=w_\b(P)$ and
$w^\delta_\b(P^\delta)=w_\b(P)$ of
the normalized holomorphic differentials under the biholomorphic mappings.

Similarly, for the first summand in the right hand side of (\ref{comp1}) we get:
\begin{multline}
\label{comp3}
\sum_{j=1}^\M \left( \Id - \T \right)_{ij} \lambda \int_{\con_k}W(P,P_j)
= \lambda \int_{\con_k} W(P,P_i) \\- 2\pi \i \lambda\sum_{\a,\b=1}^g
\left[ (C\B+D)^{-1}C\right]_{\a\b}
w_\a(P_i) \left[  \E\left(\int_{\con_k}w_\b(P)\right) -\lambda
  \e\left(\int_{\con_k} w_\b(P)\right)\right]=
\lambda \int_{\con_k} W(P,P_i) .
\end{multline}
Here the action of the fields $\E$ and $\e$ are computed similarly to the above: $\E\left(\int_{\con_k}w_\b(P)\right)=0$ and $\e\left(\int_{\con_k} w_\b(P)\right)=0.$

Thus, plugging relations (\ref{comp2}), (\ref{Euler1}), (\ref{e1}) and  (\ref{comp3}) into (\ref{comp1}) and using the expression (\ref{FayW}) for the transformed bidifferential $W$, we get (\ref{comp1}).

\item The integration contours $\con_k$, as well as $W(P,Q)$, remain unchanged, but some of distinguished local parameters change sign, i.e. $Y\neq \Id$.
Then the differentials $W(P,P_j)$ change to $\epsilon_j W(P,P_j)$, which implies the transformation $\Phi\to Y\Phi$
of the matrix $\Phi$.

\end{enumerate}

$\Box$

 Note that  we can rewrite the transformation (\ref{Schlesinger}) in the form
%$\tilde{\Phi} (\lambda) = \left(\Id +\T_{\rm{{\bf \iE}}}-\lambda  \T_{\rm{{\bf e}}} \right) \Phi (\lambda),$
$\widehat{\Phi} (\lambda) =Y \left(\Id +\T_1-\lambda  \T_2 \right) \Phi (\lambda)\,R\,,$
where the matrices $\T_1$ and $ \T_2$ do not depend on $\lambda.$

Therefore, in the case  $R=\Id$ (\ref{Schlesinger})  is nothing but a special type of the
Schlesiger transformation (multiplication from the left by a rational function); this transformation does not change the
monodromy matrices of  $\Phi$. In the case of a non-trivial matrix $R$ the monodromy matrices of $\widehat{\Phi}$
are obtained from monodromy matrices of $\Phi$ via the conjugation by $R^{-1}$.

Let us formulate the following technical lemma:
\begin{lemma}
\label{lemma_detSchles}
The matrix $\Id-\T$ from Theorem \ref{thm_Schlesinger} is non-degenerate. Its inverse is given by $\Id + \T.$
\end{lemma}
{\it Proof.}
The statement of the lemma follows from the relation $\T^2=0,$ which holds due to the following identity:
\begin{equation}
 \label{det1}
 \sum_{j=1}^\iM (\lambda_j-\lambda) w_\a(P_j) w_\b(P_j) = 0 \;\;\; \mbox{ for any } \a,\b = 1,\dots,g.
 \end{equation}
To prove (\ref{det1})  we notice that by virtue of the Rauch variational formulas (\ref{RauchB}) for the Riemann matrix,
 the left hand side of (\ref{det1}) is a multiple of the quantity $\E(\B_{\alpha\beta}) - \lambda \e(\B_{\alpha\beta})$.
%Recall that $\E$ and $\e$ are, respectively, the Euler (\ref{Euler}) and the unit (\ref{unit}) vector fields on the Frobenius manifold.
The constancy of the Riemann matrix $\B$ along the Euler  and the unit
vector fields, $\E(\B_{\a\b})=0$ and $\e(\B_{\a\b})=0,$ is proved as
in (\ref{comp3}) choosing  the contour of integration to be $\con_k=b_\b.$
% (\ref{Euler1}) and (\ref{e1}) using the invariance of the Riemann matrix under the biholomorphic transformations of the surface.
%
$\Box$

This lemma implies the following corollary of Theorem \ref{thm_Schlesinger}, which will be used in the proof of the completeness of the constructed set of solutions to the Fuchsian system (\ref{ls1}), (\ref{ls2}):
\begin{corollary}
\label{corollary_det}

Let  matrix $\Phi$ be a fundamental matrix of solutions to the
system (\ref{ls1}), (\ref{ls2}) for some choice of symplectic basis $({\bf a}_\a, {\bf b}_\a)$
in $H_1(\L)$ and a basis $\{\con_j\}$ in $H_1(\L\setminus f^{-1}(\infty)\;,\;f^{-1}(\l))$.
Then the matrix $\widehat{\Phi}$ corresponding to any other choice of the bases in these homology spaces is
also a fundamental matrix of solutions. In other words, the non-degeneracy of $\Phi$ for some $\l$,
 implies the  non-degeneracy of   $\widehat{\Phi}$.
\end{corollary}

\subsubsection{ Deformation of bidifferential $W$.}
%\begin{remark}
 Note that while the transformation (\ref{symplectic}) of the homology basis is defined by a
symplectic matrix with integer entries, we can construct a bidifferential $\widehat{W}^\C$ as in (\ref{FayW}) with $C$ and $D$ being the corresponding blocks of a symplectic matrix with complex entries. Such a bidifferential $\widehat{W}^\C$ gives also a ``deformation" of the original bidifferential $W.$

Namely, let
\begin{equation}
\left(\begin{array}{rr}
      A & B  \vspace{.2cm} \\
      C & D\\
   \end{array} \right) \in {\rm Sp}(2g,\C)
   \label{SpMatrix}
\end{equation}
and assume the matrix $C\B+D$ to be non-degenerate. Then the bidifferential $\widehat{W}^\C(P,Q),$ $P,Q\in\surf,$ given by
\begin{equation}
\label{WC}
\widehat{W}^\C(P,Q) = W(P,Q) - 2 \pi\i \; w^\iT(P) (C\B+D)^{-1}C  w(Q),
\end{equation}
can be characterized as a unique symmetric bidifferential with a second order pole at the diagonal $P=Q$ with biresidue $1,$ normalized by the conditions:
\begin{equation*}
\sum_{\a=1}^g C_{\b\a}\oint_{{\bf b}_\a} \widehat{W}^\C(P,Q) +
\sum_{\a=1}^g D_{\b\a}\oint_{{\bf a}_\a} \widehat{W}^\C(P,Q) =0\;,
\end{equation*}
where the integration is perfomed with respect to either of the
arguments. (Note that due to the non-degeneracy of the matrix
$C\B+D,$  the vanishing of the above combinations of periods of a
holomorphic differential $v$, namely,
$\sum_{\a=1}^g C_{\b\a}\oint_{{\bf b}_\a} v + \sum_{\a=1}^g D_{\b\a}\oint_{{\bf a}_\a} v =0$
for all $\b=1,\dots,g$ implies $v=0.$)

We note here that the matrices $A$ and $B$ from (\ref{SpMatrix}) do not enter the deformation (\ref{WC}). In other words, the bidifferential $\widehat{W}^\C(P,Q)$ is defined for any pair of matrices $C$ and $D$ which can be completed to a symplectic matrix of the form (\ref{SpMatrix}), namely the matrices $C$ and $D$ should be related by $CD^\iT = D C^\iT.$ This condition is necessary for the bidifferential $\widehat{W}^\C(P,Q)$ to be symmetric with respect to the arguments $P$ and $Q$.

The variational formulas for $\widehat{W}^\C$ have the same form as the Rauch variational formulas (\ref{RauchW}) for the $W.$ The deformed bidifferential $\widehat{W}^\C$ is also invariant with respect to biholomorphic transformations of the Riemann surface.

Thus the matrix $\widehat{\Phi}^\C(\lambda)$ given by (\ref{phisol}),
(\ref{defPhicon}), (\ref{bacyc})-(\ref{gkkp1}) with the $W$ replaced
by its deformation $\widehat{W}^\C$ solves the system (\ref{ls1}),
(\ref{ls2}) with $\q=-1/2$ and the matrix $V$ built from the
entries $V_{ij}=\widehat{W}^\C(P_i,P_j)(\l_i-\l_j)/2.$ The deformed
system is related to the original one by the Schlesinger
transformation of the form (\ref{Schlesinger}), (\ref{T}) with $Y=R=\Id$ and the
matrices $C$ and $D$ having complex-valued entries.

If the matrix $C$ is invertible, the definition (\ref{WC}) yields the bidifferential  ${W}_{{\bf q}}(P,Q) = W(P,Q) - 2 \pi\i \; w^\iT(P) (\B+{{\bf q}})^{-1}C  w(Q),$ where ${{\bf q}}=C^{-1}D.$ This is the deformation of the bidifferential $W$ con\-si\-de\-red in \cite{deform}, where the corresponding deformations of Frobenius structures were built - the Frobenius structures with rotation coefficients $\Gamma_{ij}=W_{{\bf q}}(P_i,P_j)/2.$ Apparently, one can generalize the deformations from \cite{deform} to Frobenius structures with rotation coefficients $\Gamma_{ij}=\widehat{W}^\C(P_i,P_j)/2.$ (Here the ``values'' of $\widehat{W}^\C$ and $W_{\bf q}$ at the points $\{P_i\}$ are defined similarly to (\ref{coefGamma}).)

\section{Monodromy group of the Fuchsian system}

\label{sect_monodromy}

In this section we study the transformations of the solution $\Phi$ (\ref{phisol}) under analytical continuation with respect to $\lambda$ along the paths from $\pi_1({\mathbb C}\setminus\{\lambda_1,\dots,\lambda_\M\}, \lambda_0)$. Since $\Phi(\lambda)$ depends non-linearly on $\lambda$ only through the dependence on $\lambda$ of the integration contours $\gamma_{n,n+1}(\lambda)$, the monodromy transformation in question is given by the corresponding transformation of the integration contours in the space $H_1(\L\setminus f^{-1}(\infty), f^{-1}(\lambda)).$

\subsection{Preliminaries}
\label{sect_monodromy0}

For any set of $\N$ points $Q_1,\dots,Q_\N$ on a
Riemann surface $\L$ one can introduce the surface
braid group $B_\N(\L\,,\,\{Q_j\}_{j=1}^\N)$ (see \cite{Birman}; if $\L$ is
the complex plane, the surface braid group coincides with the
Artin braid group).

For a description of the monodromy group of the Fuchsian system (\ref{ls1})
we  introduce the surface braid group $B_\N(\L\setminus f^{-1}(\infty)\,,\,f^{-1}(\l_0))$.
The corresponding strands end at $\N$ points from
$f^{-1}(\l_0),$ i.e., at $\l_0^{(1)},\dots,\l_0^{(\N)}$.

The lift $f^{-1}(\g)$ of a path $\g\in\pi_1(\CP1\setminus \{\l_1,\dots,\l_{\M},\infty\},\lambda_0)$ from $\CP1$ to
$\L\setminus f^{-1}(\infty) $ consists of $\N$ non-intersecting (other than at the end points) paths on $\L$ which start and
end in the set $\{\l_0^{(1)},\dots,\l_0^{(\N)}\}.$ Therefore,  $f^{-1}(\g)$
naturally defines an element of the  group $B_\N(\L\setminus
f^{-1}(\infty)\,,\,f^{-1}(\l_0))$ (see review \cite{Paris}). We denote this map which takes a loop $\gamma$ to the corresponding surface braid by $\map$.  Obviously, for any two elements
$\gamma$ and $\tilde{\gamma}$ of $\pi_1(\CP1\setminus \{\l_1,\dots,\l_{\M},\infty\},
\lambda_0)$ the element  of the
surface braid group corresponding to $f^{-1}(\gamma\circ \tilde{\gamma})$ coincides with that corresponding to the product
$f^{-1}(\gamma)\circ f^{-1}(\tilde{\gamma})$.
Therefore, we get the following
\begin{proposition}\label{fm1}
The map $\map$ from $\pi_1(\CP1\setminus \{\l_1,\dots,\l_{\M},\infty\},\lambda_0)$ to
$B_\N(\L\setminus f^{-1}(\infty)\,,\,f^{-1}(\l_0))$ defined above is a group homomorphism.
\end{proposition}

There exists also a standard homomorphism from the surface braid
group $B_\N(\L\setminus f^{-1}(\infty)\,,\,f^{-1}(\l_0))$ to the symmetric
group $\symm$ acting on the set of $\N$ points
$\l_0^{(1)},\dots,\l_0^{(\N)}$.
The superposition of this homomorphism with the homomorphism $\map$ from Proposition \ref{fm1} gives  the standard group
homomorphism $\hom$ from $\pi_1(\CP1\setminus \{\l_1,\dots,\l_{\M},\infty\},
\lambda_0)$ to the symmetric group $\symm$; the image of $\pi_1(\CP1\setminus \{\l_1,\dots,\l_{\M},\infty\},
\lambda_0)$ under the homomorphism $\hom$ is called the {\it monodromy
  group of the covering}.

Now, for any Riemann surface $\L$ and a set of $\N$
points $\{Q_n\in\L\}_{n=1}^\N$ one can define a natural action of the surface braid
group $B_\N(\L\,,\,\{Q_n\}_{n=1}^\N)$ on the relative homology group
$H_1(\L\,,\,\{Q_n\}_{n=1}^\N)$.
%(this action is analogous to the action
%of the mapping class group on the first homology group of a compact Riemann
%surface).
Namely, on the space of absolute homologies $H_1(\L)$
(which is a linear  subspace of $H_1(\L\,,\,\{Q_n\}_{n=1}^\N)$) the
 group  $B_\N(\L\,,\,\{Q_n\}_{n=1}^\N)$ acts identically.
On an
element of $H_1(\L\,,\,\{Q_n\}_{n=1}^\N)$ represented by an oriented contour
$\gamma_{mn}$ which starts at the point $Q_m$ and ends at $Q_n$  an element $G\in B_\N(\L\,,\,\{Q_n\}_{n=1}^\N)$ acts in the following way. The element $G$ induces a permutation $(i_1,\dots,i_\N)\in\symm$ of points $Q_1,\dots,Q_\N$ and is defined by $\N$ oriented paths $\{l_n\}$ on $\L$; the path $l_n$ goes from $Q_n$ to $Q_{i_n}$.
%
%Here we speak of contours and of the elements of the corresponding homology groups which they represent interchangeably.
The natural action of $G\in B_\N(\L\,,\,\{Q_n\}_{n=1}^\N)$ on a contour $\gamma_{mn}$  is  defined by
\be
\g_{mn}\to \g_{mn}-l_m+l_n =: \gamma_{i_m\,i_n}.
\label{acG}
\ee
%
%The contour $\mu_m$ connects the points $Q_m$ and $Q_{i_m}$; thus
%$\mu_m=\g_{m\, i_m}+ C_m$, where $C_m\in H_1(\L)$; also
%$\mu_n=\g_{n\, i_n}+ C_n$, where  $C_n\in H_1(\L)$.
%Therefore, the action (\ref{acG}) of $G$ on $\g_{mn}$ has the form: $\g_{mn}\to \g_{i_m\,i_n} + C_{mn}$, where $C_{mn}\in
%H_1(\L)$  (i.e. $C_{mn}$ correspond to  some closed contours in $H_1(\L\,,\,\{Q_n\}_{n=1}^\N)$.
%
In this way to each $G\in B_\N(\L\,,\,\{Q_n\}_{n=1}^\N)$ a linear automorphism of $H_1(\L\,,\,\{Q_n\}_{n=1}^\N)$ is assigned.

\begin{proposition}
This map from  $B_\N(\L\,,\,\{Q_n\}_{n=1}^\N)$ to the group of linear
automorphisms of $H_1(\L\,,\,\{Q_n\}_{n=1}^\N)$ is a group homomorphism.
\end{proposition}

The proof is geometrically obvious: it is easy to see that the action of the product of two elements of $B_\N(\L\,,\,\{Q_n\}_{n=1}^\N)$ on
$H_1(\L\,,\,\{Q_n\}_{n=1}^\N)$ corresponds to the  superposition of the automorphisms corresponding to each of these elements.

Let us now denote by $R$ the homomorphism from the surface braid group  $B_\N(\L\setminus f^{-1}(\infty)\,,\,f^{-1}(\l_0))$ to the group of linear automorphisms of the vector space  $H_1(\L\setminus f^{-1}(\infty)\,;\,f^{-1}(\l_0))$.

The superposition $\F:=R\circ \map$  defines a group homomorphism from  $\pi_1(\CP1\setminus \{\l_1,\dots,\l_{\iM},\infty\}, \lambda_0)$ to ${\rm Aut}[H_1(\L\setminus f^{-1}(\infty)\,;\,f^{-1}(\l_0))]$.

The next theorem states that, essentially, the image of $\pi_1(\CP1\setminus \{\l_1,\dots,\l_{\iM},\infty\},
\lambda_0)$ in ${\rm Aut}[H_1(\L\setminus f^{-1}(\infty)\,;\,f^{-1}(\l_0))]$ under $\F$ coincides with the monodromy
group of the Fuchsian system (\ref{nfls}).

\begin{theorem}
Consider a standard system of generators $\g_1,\dots,\g_{\iM},\g_\infty$ (\ref{genpi1}) in the
fundamental group $\pi_1(\CP1\setminus \{\l_1,\dots,\l_{\iM},\infty\},
\lambda_0)$ based at $\l_0$. 
Let a solution $\Phi(\l)$ to the Fuchsian system (\ref{ls1}) in a neighbourhood $D$
of a base point $\l_0$ be given by (\ref{defPhicon}), (\ref{phisol}),
where the basis $\{\con_j\}$   in the relative homology group $H_1(\L\setminus f^{-1}(\infty)\,;\,f^{-1}(\l))$
is given by (\ref{bacyc}), (\ref{lsss}) and  (\ref{gkkp1}).
Let the automorphisms $\F(\g_k)\in {\rm Aut}[H_1(\L\setminus f^{-1}(\infty)\,;\,f^{-1}(\l_0))]$
(where  the homomorphism $\F$ is defined  before the theorem)
be defined in the basis
$\{\con_j\}$ by the matrices $F_k,$ i.e., $T(\gamma_k)(\con_j) = \sum_{i} (F_k)_{ji}\con_i$
Then the solution $\Phi(\l)$ transforms under the analytical continuation
along the path $\g_k$ as follows:  $\Phi\to \Phi M_k$, where the monodromy matrices $M_k$ are
related to the matrices $F_k$ by:
\be
M_k= (F_k)^{\iT}\;,\hskip0.7cm k=1,\dots,\M,\infty \,.
\label{monmatFt}
\ee
\end{theorem}
{\it Proof.} To prove the theorem one has to remember that the neighbourhood $D$ of $\l_0$ was chosen such that for $\lambda\in D$ the contours $\con_j(\l)$  can be obtained by a smooth deformation from the contours $\con_j(\l_0)$.  Then the statement of the theorem is just a corollary of the definition of the function $\Phi$ (\ref{defPhicon}), (\ref{phisol}) in terms of  integrals of certain meromorphic differentials over the contours  $\con_j(\l),$ as well as of the definitions of monodromy matrices and the homomorphism $\F.$
$\Box$

The transposition in the relation (\ref{monmatFt}) between the matrices $M_k$ and $F_k$ appears since the cycles  $\con_j$ label the {\it columns} of matrix $\Phi$.  Thus the map from $\pi_1(\CP1\setminus \{\l_1,\dots,\l_{\iM},\infty\}, \lambda_0)$ to $GL(\M,\C)$ given by the monodromy map is an {\it anti}-homomorphism (i.e., the monodromy matrices multiply in the order  opposite to the order of multiplication of the corresponding paths in $\pi_1(\CP1\setminus \{\l_1,\dots,\l_{\iM},\infty\}, \lambda_0)$), see (\ref{genpi1}), (\ref{relationmonodromy}).

In our situation,
when all finite branch points are simple and the covering is connected,  the monodromy group of the covering $\X$ (i.e., the image of  $\pi_1(\CP1\setminus \{\l_1,\dots,\l_{\iM},\infty\}, \lambda_0)$ in $\symm$ under the homomorphism $\hom$) coincides with the whole symmetric group $\symm$. Let us denote the permutations corresponding to the loops $\gamma_k$ by $\sigma_k$, i.e., $\sigma_k=\hom(\g_k)$,
$k=1,\dots,\M,\infty$. The permutations satisfy the relation
\begin{equation*}
\sigma_1\sigma_2\dots\sigma_{\iM}\sigma_{\infty}=id.
\end{equation*}

One can make the following  statement about the structure of the monodromy matrices:
\begin{theorem}
The monodromy matrices of the function $\Phi$ defined by (\ref{defPhicon}), (\ref{phisol}) have the following block structure:
\be
M_k= \left(\begin{array}{cc} I & S_k\\
                             0 & \Sigma_k \end{array}\right),
\label{formMk}
\ee
where $I$ is the $(2g+\K-1)\times (2g+\K-1)$ identity matrix; $0$ is the $(\N-1)\times (2g+\K-1)$ matrix with zero entries;
$S_k$ and $\Sigma_k$ are matrices with integer entries of size $ (2g+\K-1)\times (\N-1)$ and $(\N-1)\times (\N-1),$ respectively.
Moreover, the matrix $\Sigma_k$ depends only on the element $\sigma_k$ of the monodromy group of the covering.
\end{theorem}
{\it Proof.}
The diagonal unit block of the size  $(2g+\K-1)\times (2g+\K-1)$ and the zero matrix in the left lower corner of $M_k$ appear since the first
$2g+\K-1$ columns of the matrix $\Phi$ are either linear functions of $\l$ or constant with respect to $\l$.
Therefore, these $2g+\K-1$ columns remain  invariant under the analytical continuation of $\Phi$ along any $\gamma_k$ (this can also be seen
from the fact that the contours $\con_k$, $k=1,\dots,2g+\K-1,$ are independent of $\lambda$ and, therefore, do not change
under $T(\gamma_k)$). The matrices $S_k$ and $\Sigma_k$ define the transformation of the contours $\g_{n,n+1}(\l_0)$,
$n=1,\dots,\N-1$,
under the homomorphism $T(\gamma_k)$. The contour $\g_{n,n+1}(\l_0)$ is mapped under such a transformation to
some contour connecting the points $\l_0^{(i_n)}$ and  $\l_0^{(i_{n+1})}$ (where $(i_1,\dots,i_\N)\in {\bf S}_\N$ is
an element $\hom(\g_k)$ of the monodromy group of the covering $\L$ corresponding to $\g_k$).
This contour can be expressed in $H_1(\L\setminus f^{-1}(\infty)\,;\,f^{-1}(\l_0))$ as a linear combination of the contours  $\{\g_{n,n+1}(\l_0)\}_{n=1}^{\N-1}$,  basis $a$- and $b$-cycles, and the cycles around $\infty^{(s)}$. The coefficients
in front of  $\{\g_{n,n+1}(\l_0)\}$ are given by the matrix $\Sigma_k$; clearly, they depend only on the permutation $\hom(\g_k)$;
thus the matrices $\Sigma_k$ are entirely determined by the monodromy group of the covering $\L$.
The matrices $S_k$, which determine the coefficients in front of the  $a$- and $b$-cycles,
and the cycles around $\infty^{(s)}$, depend also on the choice of a canonical basis of cycles in $H_1(\L)$.
$\Box$

It is thus easy to see that under a change of the basis
$(a_\a,b_\a,l_s)$ in $H_1(\L\setminus f^{-1}(\infty))$ the matrices
$\Sigma_k$ do not change; the matrices $S_k$ transform in an obvious way given
by the next proposition.
We note also that the matrices $\Sigma_k$ satisfy the relation
$$
\Sigma_{\infty}\Sigma_{\iM}\dots\Sigma_1=id.
$$

\begin{proposition}\label{changebacyc}
Let a $(2g+\K-1)\times (2g+\K-1)$ matrix $Q$ define a transformation between a basis $(a_\a,b_\a,l_s)$ in
$H_1(\L\setminus f^{-1}(\infty))$ and  a new basis $(\tilde{a}_\a,\tilde{b}_\a,\tilde{l}_s)$, i.e.,
\be
\left(\begin{array}{c} a_\a\\ b_\a\\
l_s\end{array}\right)=
Q\left(\begin{array}{c}\tilde{a}_\a\\
\tilde{b}_\a\\
\tilde{l}_s\end{array}\right).
\label{changeababt}
\ee
Then the new monodromy matrices (the monodromy matrices of the solution $\Phi$ given by the integrals (\ref{defPhicon}) over the new basis of contours) have the form (\ref{formMk}) with
  the same matrices $\Sigma_k$ and new matrices $S_k$ given by:
\be
\tilde{S}_k= Q^{\iT} S_k\,.
\label{transSk}
\ee
\end{proposition}
The {\it proof} is an immediate corollary of the definition of the matrices
$S_k$; it is also easy to observe that the simultaneous transformation (\ref{transSk}) of
all matrices $S_k$   preserves the relation
(\ref{relationmonodromy}) between the monodromy matrices.
Indeed, the transformed monodromy matrices $\tilde{M}_k$ (\ref{formMk}), (\ref{transSk}) are related to
the matrices $M_k$ by a simultaneous conjugation :
\begin{equation*}
\tilde{M}_k =\left(\begin{array}{cc} Q^{\iT} & 0 \\
                             0 & I \end{array}\right)M_k
\left(\begin{array}{cc} (Q^{\iT})^{-1} & 0 \\
                             0 & I \end{array}\right);
\end{equation*}
the corresponding solutions of the Fuchsian system are related by
\be
\tilde{\Phi} = \Phi \left(\begin{array}{cc} (Q^{\iT})^{-1} & 0 \\
                             0 & I \end{array}\right).
\label{PhiPhit}
\ee
\begin{remark}
{\rm We would like to stress that in Proposition \ref{changebacyc} we only
consider the dependence of $\Phi$ on the change of some of the integration
contours $\con_k$ in (\ref{defPhicon}); the canonical basis of cycles $({\bf a}_\a,{\bf
  b}_\a)$ used in the definition of the bidifferential $W$ (see Section \ref{sect_preliminaries}) is assumed to
remain the same. The dependence of $\Phi$ on the choice of a basis $({\bf a}_\a,{\bf
  b}_\a)$ (i.e., on the normalization of $W$) was discussed in Section \ref{sect_normalization}.}
\end{remark}

\subsection{Monodromy group}

\subsubsection{Spaces of meromorphic functions with simple poles}
Here we describe the group $\Gr$ generated by the monodromy matrices computed in Appendix \ref{sect_simpleinfty}  as a semidirect product of the free group $\mathbb{Z}^{(2g+\N-1)\times(\N-1)}$, where $\N$ is the degree of the covering, and the symmetric group $\symm,$ the monodromy group of the covering.

\begin{figure}[htb]
\centering
\includegraphics[scale=0.5]{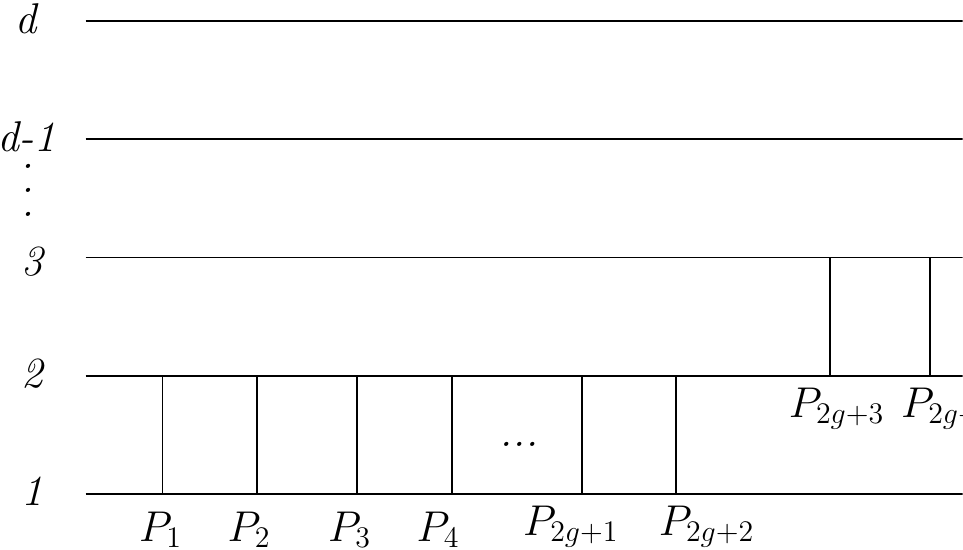}
\caption{A Hurwitz diagram for the space $\H_{g;\N}(1,\dots,1).$ }
\label{fig_diagram}
\end{figure}

Consider a Hurwitz space of coverings represented by Figure \ref{fig_diagram}.
Introduce the group $\tilde{\Gr}$  generated by the monodromy matrices $M_1, M_{2g+3}, M_{2g+5},\dots, M_{\iM-1}$, i.e.,
\begin{equation*}
\tilde{\Gr}:=\langle M_1,\,\{M_{2g+2n-1} \}_{n=2}^{\N-1}\rangle\;.
%\label{Gtilde}
\end{equation*}
These generators have the form (see Appendix \ref{app_monodromy}, (\ref{M1}) and (\ref{M2gp2km1})):
\begin{equation}
\label{eltsG_1}
M_i=\left(
    \begin{array}{cc}
     I & 0 \\
     0 & \Sigma_i
          \end{array}\right), \qquad \mbox{for} \;\; i=1\;\; \mbox{or odd} \;\; i\geq 2g+3\;,
\end{equation}
 where the matrix $\Sigma_i$ corresponds to the element $\sigma_i=h(\gamma_i)$  of the monodromy group of the covering.

As is easy to see, for the coverings from Figure \ref{fig_diagram}, $\tilde{\Gr}$ is isomorphic to the monodromy group of the covering, i.e., to the symmetric group $\symm.$

Denote by $\widehat{\Gr}$ the following group:
\begin{equation}
\label{G2}
\widehat{\Gr}:=\langle \{M_1M_k\}_{ k=2}^{2g+2}\,;\,
\{M_{2g+2n-1}M_{2g+2n}\}_{ n=2}^{\N-1}\rangle\; ,
\end{equation}
and consider its normal closure $\widehat{\Gr}^\Gr$ in $\Gr$. 
Since all elements of $\Gr$ can be represented as products of elements from 
$\widehat{\Gr}$ and $\tilde{\Gr}$, the
monodromy group $\Gr$ is given by a semidirect product $\Gr=\widehat{\Gr}^\Gr \rtimes \tilde{\Gr}.$
\begin{theorem}
\label{thm_group}
The normal closure $\widehat{\Gr}^\Gr$ of the group $\widehat{\Gr}$
(\ref{G2}) in the monodromy group
$\Gr=\langle \{M_k\}_{k=1}^{\iM}\rangle$ is isomorphic to the free group $\mathbb{Z}^{(2g+\N-1)(\N-1)}$. Here $\N$ is the degree and $g$ is the genus of the covering; $\M$ is the number of simple finite branch points.
\end{theorem}
{\it Proof.}
The matrices generating the group $\widehat{\Gr}$ (\ref{G2}) have the form (see Appendix \ref{app_monodromy}):
\begin{equation}
\label{eltsG_2}
M_1M_k = \left(
    \begin{array}{cc}
     I & S_k \\
     0 & I
          \end{array}\right), \qquad
          M_{2g+2n-1}M_{2g+2n}= \left(
    \begin{array}{cc}
     I & S_{2g+2n}\\
     0 & I
          \end{array}\right),
\end{equation}
where again $k=2,\dots 2g+2; \;\;n=2,\dots,\N-1$ and $S_l$ is the block above the diagonal in the monodromy matrix $M_l$, see (\ref{formMk}). We recall that the second diagonal block in $M_lM_k$ (the block $\Sigma$ in (\ref{formMk})) depends only on the permutation $h(\gamma_k\gamma_l)$. The permutations $h(\gamma_k\gamma_1)$ and $h(\gamma_{2g+2n}\gamma_{2g+2n-1})$ are trivial for the coverings from Figure \ref{fig_diagram}, thus the corresponding diagonal blocks are trivial in (\ref{eltsG_2}).

From (\ref{eltsG_2}) and (\ref{eltsG_1}) we get elements of $\widehat{\Gr}^\Gr$ in the form:
\begin{equation}
\label{elts}
M = \left(
    \begin{array}{cc}
     I & S \\
     0 & I
          \end{array}\right)
\end{equation}
with some matrix $S.$ We shall now show that $S$ can be any  matrix with integer entries.

Consider an element of the normal closure $\widehat{\Gr}^\Gr,$ obtained from matrices (\ref{eltsG_2}) by conjugation with one of the generators (\ref{eltsG_1}) of the group $\tilde{\Gr}$. We get
\begin{equation}
\label{eltsG}
M_i \left(
    \begin{array}{cc}
    I & S_k \\
     0 & I
          \end{array}\right)M_i^{-1} = \left(
    \begin{array}{cc}
     I & S_k\Sigma_i \\
     0 & I
          \end{array}\right),
\end{equation}
where $k$ runs through the set $\{2,3,\dots, 2g+2 \} \cup \{ 2g + 2n \}_{n=2}^{\N-1}.$

Now we note that the last $\N-1$ rows in the block $S_\infty$ above the diagonal in the monodromy matrix $M_\infty$ (\ref{Minfty}) form an $(\N-1)\times(\N-1)$ matrix which coincides with the Cartan matrix for $A_{\N-1}.$ Therefore each row of the matrix $M_\infty$  is a coordinate vector of a root of $A_{\N-1}$ with respect to a basis of weight vectors $\{v_i\}$ defined by
%Now we note that the nonzero row in the matrix $S_k$ for $k=2,\dots,2g+2; k=2g+2n\geq 2g+4$ coincides with the corresponding row in %$M_\infty$ (\ref{Minfty}). The last $\N-1$ rows in the block $S_\infty$ above the diagonal in the monodromy matrix $M_\infty$ form an
% $(\N-1)\times(\N-1)$ matrix which coincides with the Cartan matrix for $A_{\N-1}.$ Therefore, each row of the matrix $M_\infty$ (and thus %the nonzero row in $S_k$ in (\ref{eltsG_2}), (\ref{eltsG})) is a coordinate vector of a root of $A_{\N-1}$ with respect to a basis of weight %vectors $\{v_i\}$, defined by
%
\begin{equation*}
\frac{\langle v_i,r_j\rangle}{\langle r_i,r_i\rangle}=\delta_{ij},
\end{equation*}
 where $\{r_i\}$ are the root vectors and $\langle\, ,\,\rangle$
 denotes the scalar product in $\mathbb{R}^\N$. The Weyl group for
 $A_{\N-1}$ is the symmetric group $\symm$, thus the orbit of one
 of the root vectors of $A_{\N-1}$ under the action of  $\symm$  contains all the roots of $A_{\N-1}.$

 Furthermore, as can be seen from  (\ref{M2}), (\ref{Mn}), (\ref{M2gp2k}) and (\ref{Minfty}), the only non-zero row in each of the blocks $S_k$ and $S_{2g+2n}$ in the matrices (\ref{eltsG_2}) is the respective row from $M_\infty,$  i.e., a row of a Cartan matrix for $A_{\N-1}$.

 Therefore the matrices (\ref{eltsG}) for $i=1$ and odd $i\geq 2g+3$
 and for $k=2,\dots 2g+2$ and $k=2g+2n$ with $n=2,\dots,\N-1$ have
 above the diagonal the products $S_k\Sigma,$ where matrices $\Sigma$ represent all generators of the
  group $\symm$ and the nonzero rows in $S_k$ run through all the root vectors of $A_{\N-1}.$ Thus, for each of the above $k$,  the only nonzero row in the product $S_k\Sigma_i$, the $k$th row, runs through all the root vectors of $A_{\N-1},$ i.e., through an integer basis in $\mathbb{R}^{\N-1}.$ Multiplying matrices of the form (\ref{eltsG}), we get matrices of the form (\ref{elts}) with all possible integer blocks $S$ of the size $(2g+\N-1)\times(\N-1)$ above the diagonal, which implies that the normal closure $\widehat{\Gr}^\Gr$ contains the free group $\mathbb{Z}^{(2g+\N-1)(\N-1)}.$ Since from Section \ref{sect_monodromy0} we know that entries of the blocks $S$ in (\ref{elts}) are always integer numbers, we arrive at the isomorphism
$\widehat{\Gr}^\Gr\simeq\mathbb{Z}^{(2g+\N-1)(\N-1)}.$
 $\Box$

To summarize, we repeat that in the case of Hurwitz spaces of
coverings shown in  Fig.\ref{fig_diagram}, Theorem \ref{thm_group}
implies the isomorphism between the monodromy group $\Gr$ of the
solution $\Phi(\lambda)$ (\ref{defPhicon}) to the Fuchsian system
(\ref{ls1}), (\ref{ls2}) and a  semidirect product of the free
abelian group
$ \mathbb{Z}^{(2g+\N-1)(\N-1)}$ with the symmetric group  $\symm$, i.e., $\Gr\simeq \mathbb{Z}^{(2g+\N-1)(\N-1)}  \rtimes \symm.$

\subsubsection{Spaces of meromorphic functions with poles of higher multiplicity}

We consider a covering with ramification over the point at infinity as a limit case of the coverings from Fig.\ref{fig_diagram} when some of the points $P_{2g+2n}$ with $n\geq 1$ tend to the point at infinity without crossing any branch cuts on the covering.

As discussed in Appendix \ref{sect_branchedinfty}, the monodromy matrices corresponding to the ramification points that do not merge in the limit  are obtained from the matrices
 %computed in Section \ref{sect_simpleinfty}
 for simple coverings  (see Appendix \ref{sect_simpleinfty}) by deleting a trivial row and the corresponding trivial column. The monodromy matrices corresponding to the ramification points that are sent to infinity contribute to new monodromy matrix at infinity.

Therefore the reasoning from the proof of Theorem \ref{thm_group} remains valid in the limit: the nonzero rows of the blocks $S_k$ above the diagonal remain unchanged and coincide with rows of the Cartan matrix for $A_{\N-1}$. Since sending a ramification point  to infinity results in deleting one  of the first $2g+\N-1$ rows and one of the first $2g+\N-1$ columns from the monodromy matrices, the dimension of the blocks $S_k$ in the limit is $(2g+\K-1)\times (\N-1)$, where $\K$ is the number of points projecting to $\lambda=\infty$ on the base of the covering arising in the limit.

Thus we get a similar result for the spaces of coverings ramified over the point at infinity: the corresponding monodromy group is isomorphic to the semidirect product: $\Gr\simeq \mathbb{Z}^{(2g+\K-1)(\K-1)} \rtimes \symm.$

For the space of polynomials of degree $\M$ we have $g=0$ and $\K=1$; then the monodromy group of the Fuchsian system
(\ref{ls1}) coincides with the symmetric group $\symm$ - the Weyl group of $A_{\N-1}$ (as well as with the monodromy group of
the covering  $\X$).

If $g=0$ and $\K=2$ (this is the space of rational functions of degree $\N$ with 1 simple pole and one pole of degree $\N-1$),
then $2g+\K-1=1$ and the monodromy group of the Fuchsian system coincides with the Weyl group
$\mathbb{Z} \rtimes \symm$   of the affine Lie algebra
$\widehat{A}_{\N-1}$, i.e. the algebra of formal power series of one variable with coefficients from $A_{\N-1}$.

For arbitrary $g$ and $\K$ the monodromy group is the Weyl group  of the Lie algebra of  formal power series in
$2g+\K-1$ variables with coefficients from $A_{\N-1}$.

\begin{remark}\rm
We recall that the Fuchsian system (\ref{ls1}) we study here is related  to the non-Fuchsian system (\ref{nfls}) studied in \cite{V3} by a Laplace transform (\ref{Laplace}). Therefore the monodromy matrices of our solution $\Phi(\lambda)$ are related to the Stokes matrices for the solution $\Psi(z)$ from \cite{V3}. Explicit formulas establishing this relation can be found in \cite{BJL} for $q=1$ and in \cite{Schaefke} for an arbitrary $q$. However, the formulas in \cite{BJL} relate connection matrices (product of which gives a monodromy matrix) of a solution to the Fuchsian system and Stokes multipliers of the corresponding solution $\Psi(z)$. The Stokes matrix computed in \cite{V3} is a product of certain Stokes multipliers, though the individual Stokes multipliers are not given in \cite{V3}.  Therefore the monodromy matrices for our solution $\Phi(\lambda)$ can be computed using the results of \cite{BJL, Schaefke, V3} by a straightforward though lengthy calculation. Whereas in Appendix A, we present a direct way of computing monodromy matrices of the solution $\Phi$ we construct, covering also 
the case of arbitrary multiplicities of poles of $f$, which was not treated in \cite{V3}. 

Formulas relating the monodromy matrices of a Fuchsian system and the Stokes matrix of the corresponding non-Fuchsian system are given also  in \cite{DubrovinPainleve}, see Lemma 5.3, formulas (5.43) and (5.45). However, this lemma cannot be applied in our case since the matrix  from (5.45) of \cite{DubrovinPainleve}, namely $S+S^\iT$, where $S$ is the Stokes matrix from \cite{V3}, is degenerate. 
\end{remark}

\section{ Action of braid group on  solution to the Fuchsian system}

\label{sect_braid}

\subsection{Braid monodromy group}

The braid group $B_\M$ on $\M$
strands (on the plane)  naturally acts on the set $\{\l_k\}_{k=1}^\M$ and thus on our Hurwitz space.
To each covering $\X=(\L,f)\in \H_{g,\N}$ one can naturally associate a subgroup $B_\M(\X)$ of $B_\M$
such that any element  $\sigma\in B_\M(\X)$ transforms the covering $\X$ into a
covering $\X^{\sigma}$ which is holomorphically equivalent to $\X$ (i.e., $B_\M(\X)$  is the fundamental group of $\H_{g,\N}$ with the base at $\X$). In
particular, for $\N=2$, when the covering $\X$ is hyperelliptic, the
subgroup $B_{\M}(\X)$ coincides with the whole braid group $B_{\M}$.
An equivalence between $\X$ and $\X^\sigma$ is defined by an element of $\symm $; therefore, in the case when the automorphism group of $\X$ is trivial, we get a group homomorphism from
 $B_\M(\X)$ to $\symm$; the image of this homomorphism we call the {\it braid monodromy group of the covering} $\X$.
The action of the braid group on coverings with $\Z_d$ symmetry (all branch points have in this case multiplicity $d-1$) was recently studied in \cite{McMullen}.

In the case when the covering $\X$ admits no automorphisms (this is obviously the case when all branch points are
simple and distinct with the exception of hyperelliptic coverings),  each element $\sigma\in B_\M(\X)$ naturally induces some $Sp(2g,\Z)$
transformation on homologies $H_1(\L,\Z)$; in this way one gets a
group homomorphism
$$h\;:\;\; B_\M(\X)\;\to\;Sp(2g,\Z)\;.$$
The image
$\Gamma(\X)$ of $B_\M(\X)$ under the
homomorphism $h$ is a subgroup of  $Sp(2g,\Z)$. Some partial results
about the subgroup $\Gamma(\X)$ were obtained
(in the simplest case of hyperelliptic coverings) in \cite{Arnold,Mumford}.
In particular, it was proved in \cite{Arnold} that $\Gamma(\X)$ coincides
with the whole group $Sp(2g,\Z)$ for hyperelliptic coverings with $3,4$ and
$6$ branch points, and only in these cases. In \cite{Mumford} it was shown
that the image of the subgroup of pure braids
of $B_{\M}$  in  $Sp(2g,\Z)$
under the homomorphism $h$
coincides with the principal congruence subgroup $\Gamma(2)$.

Let us fix some canonical basis of cycles $\{ a_\a,b_\a\}$ on $\L$.
In this section we shall identify the symplectic basis $({\bf a}_\a,{\bf b}_\a)$ of $H_1(\L)$ with respect to which $W$ is normalized
with the symplectic basis $(a_\a,b_\a)$ which forms a part of the set of the integration contours $\con_j$:
$$
\{a_\a,b_\a\} = \{ {\bf a}_\a,{\bf b}_\a \}\;.
$$

Denote
by $B_{\iM}^{\{a\}} (\X)$ the subgroup of $B_{\iM}(\X)$ whose elements
preserve the $g$-dimensional subspace spanned by the set of $a$-cycles.
The image of this subgroup in $Sp(2g,\Z)$ is a subgroup
$\Gamma(\X,\{a\})$ consisting of matrices $S\in Sp(2g,\Z)$   (\ref{symplectic})
with $C=0:$

\begin{equation}
\left(\begin{array}{r}
      { \hat{b}}  \vspace{.2cm} \\
    {  \hat{a}} \\
   \end{array} \right) = S
   \left(\begin{array}{r}
      { b}  \vspace{.2cm} \\
     { a} \\
   \end{array} \right), \qquad S=
\left(\begin{array}{rr}
      A & B  \vspace{.2cm} \\
      0 & D\\
   \end{array} \right).
    \label{ababt}
   \end{equation}

As before, we consider the space ${\cal H}_{g,\N}^{\{a\}}$ which is the space
of  equivalence classes of
pairs $(\X,\{a\})$, where $\X=(\L,f)$ is a covering
of genus $g$ and degree $\N$ with
simple branch points, and $\{a\}$ is a choice of a subspace of
dimension $g$ in $H_1(\X,\Z)$ spanned by $a$-cycles.     The
subgroup  $B^{\{a\}}_\M (\X)$ coincides with the
fundamental group of the space ${\cal H}_{g,\N}^{\{a\}}$ with the
base point given by the pair $(\X,\{a\})$:
$$
B^{\{a\}}_\M (\X)=\pi_1\left({\cal H}_{g,\N}^{\{a\}},(\X,\{a\})\right).
$$

The role of the subgroup  $B^{\{a\}}_\M (\X)$ in our context is
the following: this subgroup consists of braids which not only map the covering
$\X$ to a holomorphically equivalent covering, but also preserve the canonical
bidifferential $W$.
This follows from the normalization of $W(P,Q)$: $\oint_{a_\a}
W(\cdot,Q)=0$ for all $\a=1,\dots,g$.

Therefore any transformation $\sigma\in B^{\{a\}}_\M (\X)$
preserves the coefficients (\ref{coefGamma}) of the Fuchsian linear
system (\ref{ls1}), (\ref{ls2}). However, the solution $\Phi$ of the
system (\ref{ls1}), (\ref{ls2}) may transform under the action of any
braid $\sigma\in B^{\{a\}}_\M (\X)$ to a new solution  $\Phi^\sigma$ of the same
system, which differs from $\Phi$ by a right monodromy factor
$M^{\sigma}(\{a\})$ independent of $\l$ and $\{\l_j\}$ (but dependent on the choice of the subspace spanned by $a$-cycles):
\be
\Phi^\sigma=\Phi M^\sigma\,, \hskip0.7cm \sigma\in B^{\{a\}}_\M (\X).
\la{PhiPhis}
\ee
One therefore obtains a monodromy representation of the fundamental
group $B^{\{a\}}_\M (\X)$ of the space ${\cal H}_{g,\N}^{\{a\}}$ in
$GL(\M,\Z)$.

The corresponding group, which we call  {\it the braid monodromy group of the
  Fuchsian system,} will be  denoted by ${\bf M}^{\{a\}}$
(the index $\{a\}$ indicates that this group may depend on the
choice of the subspace of $a$-cycles). This group  is
of course  different from the monodromy group $\Gr$ of the Fuchsian
system discussed above in Section \ref{sect_monodromy}.

It seems rather hard to study the groups  ${\bf M}^{\{a\}}$ explicitly
for general coverings: even  description of the subgroup of the
braid group preserving a given covering seems to be not known in
general.

In the case of hyperelliptic coverings every braid from $B_{\M}$ preserves the covering; however
there remains the problem of describing the subgroup of the braid group
which preserves the chosen subspace spanned by $a$-cycles.

Here we restrict ourselves to the simplest case of the space ${\cal
  H}_{1,2}(1,1)$ which consists of two-sheeted coverings of genus $1$
  with four finite branch points $\l_1,\dots,\l_4$ and give an
  explicit description of the corresponding  group ${\bf
  M}^{\{a\}}$. In particular, we shall show that in this case the
  groups ${\bf M}^{\{a\}}$ corresponding to different choices of the
  $a$-cycle are isomorphic to each other. %can be actually identified.

\subsection{Genus one coverings of degree 2}

\begin{figure}[htb]
\centering
\includegraphics[scale=0.5]{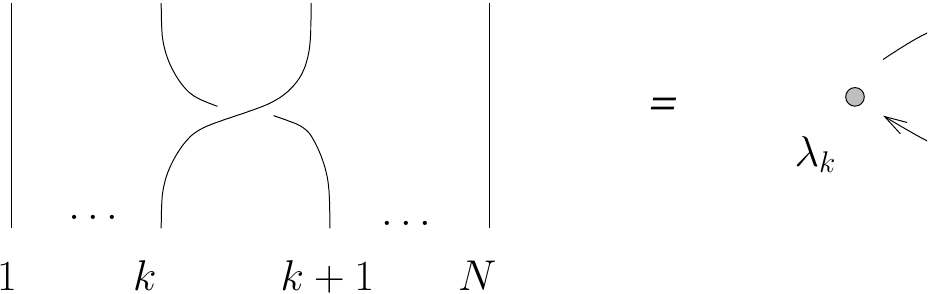}
\caption{ Generators $\sigma_k$ of the braid group $B_\M$ acting on critical values $\lambda_k$.}
\label{gen_braid_group}
\end{figure}

The braid group $B_4$ on 4 strands  has three standard generators: $\sigma_1$
(interchanging $\l_1$ and $\l_2$),  $\sigma_2$
(interchanging $\l_2$ and $\l_3$) and $\sigma_3$
(interchanging $\l_3$ and $\l_4$), see Figure \ref{gen_braid_group}. These generators satisfy the
standard relations $\sigma_1\sigma_2\sigma_1=\sigma_2\sigma_1\sigma_2$,
$\sigma_2\sigma_3\sigma_2=\sigma_3\sigma_2\sigma_3$ and $\sigma_1\sigma_3=\sigma_3\sigma_1$.

Although every element of $B_4$ preserves the covering $\X$, there are two different biholomorphic mappings between
 the initial covering and the transformed one; this ambiguity arises due to the existence of the nontrivial automorphism of $\X$, which interchanges the sheets. However, in our case we have a natural  marking of the sheets of $\X$, namely, the choice of the contour $l_1$ encircling the point $\infty^{(1)}$ on the first sheet of the covering. We therefore have
a natural choice of identification of two coverings - we identify the sheets in a way that the contour $l_1$ stays on the first sheet.
%
%Since any element of $B_4$ preserves the covering $\L$, we get a group
%homomorphism (which we denoted by $h$) from $B_4$ to $Sp(2,\Z)$.
%
%\vskip0.5cm
\begin{figure}[htb]
\centering
\includegraphics[scale=0.5]{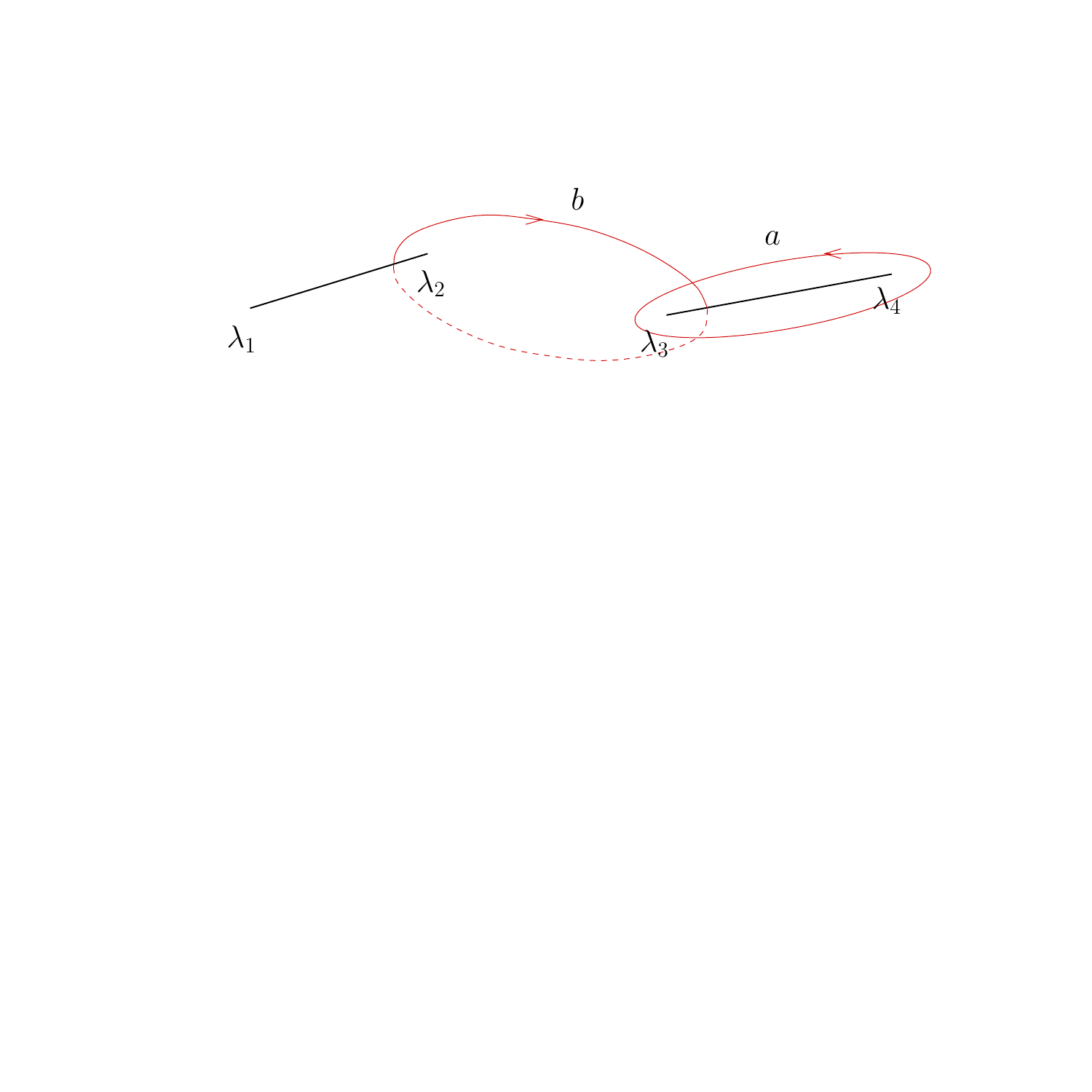}
\caption{ Basis cycles on two-sheeted covering of genus one with four
  branch points; the dash line indicates the first sheet, the solid line - the second sheet.}
\label{cyclesgenus1}
\end{figure}
%\vskip0.5cm
Considering the action of $B_4$ on the homology group of the covering $\X$, we get a group
homomorphism (which we denoted by $h$) from $B_4$ to $Sp(2,\Z)$.

Let us  choose a canonical basis of cycles $(a,b)$ on
$\X$ as shown in Fig.\ref{cyclesgenus1}. Then the generators $\sigma_1$ and $\sigma_3$ act on $\X$ as Dehn half-twists with respect to the cycles $a$ and $-a$, respectively. The generator  $\sigma_2$ acts as a Dehn half-twist along the $b$-cycle.

The Picard-Lefschetz formulas (see for example \cite{DNF}, Th. 24.3) give the following transformations of a contour $l\in H_1(\L)$ under such an action of $\sigma_i$:
\begin{equation}
\label{PL}
\sigma_1: \;\; l \mapsto l+(l\circ a)a; \qquad \sigma_2: \;\; l \mapsto l+(l\circ b)b; \qquad \sigma_3: \;\; l \mapsto l+(l\circ a)a,
\end{equation}
where $(l\circ \gamma)$ stands for the intersection index of two contours $l$ and $\gamma.$

\begin{figure}[htb]
\centering
\includegraphics[scale=0.5]{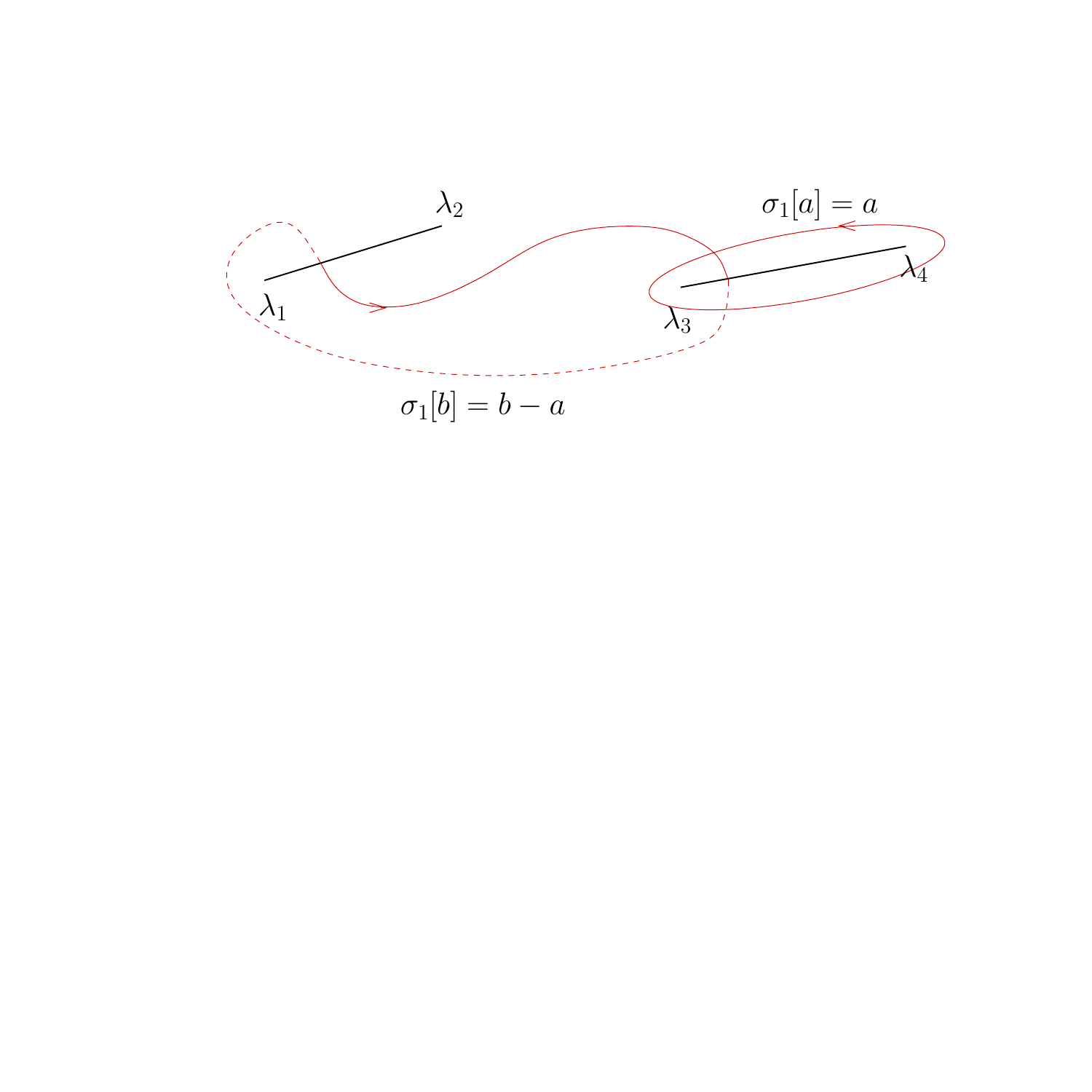}
\caption{ Action of generator $\sigma_1$ on basis cycles.}
\label{sigma1_cycles}
\end{figure}

Thus the symplectic transformations (acting on the column $(b,a)^T$) corresponding to $\sigma_i$ look as follows:
\be
\MA:=h(\sigma_1)=\left(\begin{array}{cc} 1 &  -1 \\
                                0  &  1 \end{array}\right)\;,\hskip0.7cm
\MB:=h(\sigma_2)=\left(\begin{array}{cc} 1 &  0 \\
                                1  &  1 \end{array}\right)\;,\hskip0.7cm
h(\sigma_3)=h(\sigma_1)\;.
\la{hsi0}
\ee
The transformations (\ref{hsi0}) can also be easily deduced by
an appropriate deformation of the branch cuts and basic cycles shown in Figure
\ref{cyclesgenus1}.  For example, in Fig.\ref{sigma1_cycles} the transformation of the $a$- and $b$-cycles under the action of
 $\sigma_1$ is shown.
%
%\vskip0.5cm

>From (\ref{hsi0}) we see that
% $h(\Delta)=-I$. Observe also that
$h(\sigma_1)$ and $h(\sigma_2)$ span the whole group
$\rm{Sp}(2,\Z)$ since
\begin{equation}
\label{Spgenerators}
h(\sigma_2\sigma_1\sigma_2)= \left(\begin{array}{cc} 0 &  -1 \\
                                         1  &  0
\end{array}\right)\;,
\hskip0.7cm
h(\sigma_1^{-1})=\left(\begin{array}{cc} 1 &  1 \\
                                     0  &  1 \end{array}\right)
\end{equation}
are the standard generators of $\rm{Sp}(2,\Z)$.

We are now in a position to formulate the following
\begin{lemma}
The group $B_4^{\{a\}}(\X)$ corresponding to the choice of the basis
cycles shown in Figure \ref{cyclesgenus1} is the subgroup of the braid group $B_4$
generated by the elements $\sigma_1$, $\sigma_3$, $\sigma_2^2\sigma_1\sigma_2^2$ and $\sigma_2^2\sigma_3\sigma_2^2$.
\end{lemma}

{\it Proof.}  The group  $B_4^{\{a\}}(\X)$ coincides with the preimage of the subgroup $U$ of upper triangular matrices in $\rm{Sp}(2,\Z)$
under the homomorphism $h$.
The subgroup $U$ is generated by the matrices
\begin{equation}
h(\sigma_1^{-1})=\left(\begin{array}{rr}
       1& 1  \vspace{.2cm} \\
      0 & 1\\
   \end{array} \right), \qquad \mbox{and}  \qquad h(\sigma_2^2 \sigma_1 \sigma_2^2)=\left(\begin{array}{rr}
       -1& -1  \vspace{.2cm} \\
      0 & -1\\
   \end{array} \right).
   \label{h}
\end{equation}
Let us introduce the notation $\br:=\sigma_2^2 \sigma_1 \sigma_2^2$. Then
\begin{equation}
\label{U}
U=\langle h(\sigma_1), \; h(\br)\rangle\;.
\end{equation}
Due to the equality $h(\sigma_3)=h(\sigma_1)$ (see (\ref{hsi0})) it is sufficient to study only the image of the  braid group ${B}_3$
(generated by $\sigma_1$ and $\sigma_2$) in $\rm{Sp}(2,\Z)$.

Let us denote by $\tilde{h}: {B}_3\to \rm{Sp}(2,\Z)$ the restriction of the homomorphism $h$ to ${B}_3$ and verify that
\begin{equation}
{\tilde{h}^{-1} (U)}= \langle
\sigma_1,\br\rangle .
\label{claim}
\end{equation}
To prove (\ref{claim}) we show that the kernel of $\tilde{h}$ is a subgroup of the group
$\langle \sigma_1,\br\rangle.$ Indeed, consider the composition $F\circ \tilde{h}: {B}_3 \to \rm{PSL}(2,\Z)$, where $F:\rm{Sp}(2,\Z) \to \rm{PSL}_2(\Z)$ is the factorization over plus or minus the identity matrix.
Then the following holds:
\begin{equation}
\label{claim2}
{\rm Ker}(F\circ \tilde{h}) =\langle
 (\sigma_2 \sigma_1 \sigma_2)^2\rangle = Z({B}_3)
\end{equation}
(the center $Z({ B}_3)$ of the braid group ${ B}_3$ is generated by $ (\sigma_2 \sigma_1 \sigma_2)^2$, see \cite{Garside}).
Namely, from (\ref{Spgenerators}) we get the inclusion $\langle (\sigma_2 \sigma_1 \sigma_2)^2\rangle < {\rm Ker}(F\circ \tilde{h})$, where the notation $<$ means that $\langle (\sigma_2 \sigma_1 \sigma_2)^2\rangle$ is a subgroup of ${\rm Ker}(F\circ \tilde{h})$.
Moreover, since ${B}_3$ is a central extension of $\rm{PSL}_2(\Z)$, we have the isomorphism ${\rm{PSL}_2(\Z) }\simeq {B}_3/N$, where $N$ is a subgroup of the center $Z{{(B}}_3)$ of the braid group. On the other hand, groups ${\rm{PSL}}_2(\Z)$  and $ {{B}}_3/ {\rm {Ker}}(F\circ \tilde{h})$ are isomorphic since ${\rm{PSL}}_2(\Z)=F\circ \tilde{h}{{(B}}_3)$. Therefore, ${\rm Ker}(F\circ \tilde{h})$ is isomorphic to a subgroup $N$ of $Z({ B}_3)$ and thus, due to the above inclusion $Z({B}_3) < {\rm Ker}(F\circ \tilde{h})$,  equality (\ref{claim2}) holds.

It is easy to verify that $\tilde{h}((\sigma_2 \sigma_1 \sigma_2)^2)=-I$; thus
\begin{equation}
{\rm Ker}(\tilde{h})= \langle(\sigma_2 \sigma_1 \sigma_2)^4\rangle.
%=\langle(\sigma_2 \sigma_1 \sigma_2)^4\rangle\;.
\label{Kernel}
\end{equation}

We thus have the inclusion ${\rm Ker}(\tilde{h}) < \langle \sigma_1, \br \rangle$, which follows from (\ref{Kernel}) and the relation $(\sigma_2 \sigma_1 \sigma_2)^4=(\sigma_1\br)^2$.  This inclusion and the fact that every upper triangular matrix can be obtained (see (\ref{U})) as an image under $\tilde{h}$ of some braid from the group $\langle \sigma_1, \br \rangle$ implies that every braid mapped to an upper triangular matrix by $\tilde{h}$ belongs to the group $\langle \sigma_1, \br \rangle$, i.e.,  (\ref{claim}) holds.
Now using the relation $\tilde{h}^{-1}(U) = h^{-1}(U)/\{\sigma_1=\sigma_3\}$, we complete the proof of the lemma.

$\Box$

\begin{remark} {\rm If a choice of an $a$-cycle is different from the one shown in Figure \ref{cyclesgenus1}, the fundamental group $\pi_1\left({\cal H}_{1,2}^{\{\tilde{a}\}},(\L,f,\{\tilde{a}\})\right)$ will be a different subgroup $B_4^{\{\tilde{a}\}}$ of the braid group $B_4$. However, by virtue of the result in \cite{Arnold}, for any two cycles $a$ and $\tilde{a}$ there exists a braid $\sigma\in B_4$ such that the transformation $h(\sigma)$ in the homology group $H_1(\L, \Z)$ takes $a$ to $\tilde{a}.$ Then it is straightforward to see that the two subgroups of $B_4$ are related by conjugation, i. e.,  $B_4^{\{\tilde{a}\}} = \sigma B_4^{\{a\}}\sigma^{-1}. $  }
\end{remark}

Now we compute {\it the braid monodromy matrices} for the solution  $\Phi= (\Phi^{\gamma_{1,2}}, \Phi^{l_1}, \Phi^a, \Phi^b)$ (\ref{phisol}), i.e., the transformations of $\Phi$ induced by the braids from the group $B_4^{\{a\}}:=B_4^{\{a\}}(\L,f)$.
%Here the contour $\gamma_{1,2}$ connects the points $\lambda^{(1)}$ and $\lambda^{(2)}$, the two points on  $\L$ projecting to the %point $\lambda$ on the base $\cp;$ the contour $l_1$ encircles the point $\infty^{(1)}$ at infinity on the first sheet of the covering.
Note that only the contours of integration change in the solution $\Phi(\lambda)$ under these transformations, since the bidifferential $W(P,Q)$ remains invariant.

Note that the action induced by the braids from $B_4^{\{a\}}$ on the $a$- and $b$-cycles is only partially described by the matrices $h(\sigma_i)$ (\ref{hsi0}), since these matrices give only the transformation of the contours in the homology space $H_1(\L,\Z).$ The entries of braid monodromy matrices are given by the transformation of the contours in the relative homology space $H_1(\L\setminus
f^{-1}(\infty);\; f^{-1}(\l))$.

In the space $H_1(\L\setminus f^{-1}(\infty);\; f^{-1}(\l))$, the contour going around the points $P_1$ and $P_2$ on the second sheet is equal to $a-l_1$. Therefore the Picard-Lefschetz formulas (\ref{PL}) in this space take the form:
\begin{equation}
\label{PL1}
\sigma_1: \;\; l \mapsto l+(l\circ (a-l_1))(a-l_1); \qquad \sigma_2: \;\; l \mapsto l+(l\circ b)b; \qquad \sigma_3: \;\; l \mapsto l+(l\circ a)a,
\end{equation}

Let  the contour $\gamma_{1,2}$ be chosen as in Appendix
\ref{sect_simpleinfty}, see Figure \ref{fig_hyperelliptic_M1} below. In other words,  for the standard basis $\{ \gamma_k\}_{k=1}^4$ (\ref{genpi1}) in the fundamental group of the punctured sphere $\pi_1(\CP1\setminus \{\l_1,\dots,\l_4,\infty\}, \lambda_0)$ with some base point $\lambda_0,$ we consider ${\rm lift}^{(1)}(\gamma_1)$, the lift  of $\gamma_1$ to the first sheet of the covering. Then $\gamma_{1,2}$ is taken to be a deformation of ${\rm lift}^{(1)}(\gamma_1)$; this deformation takes the end points $\lambda_0^{(1)}$ and
$\lambda_0^{(2)}$ to  the endpoints $\lambda^{(1)}$ and $\lambda^{(2)}$, respectively.
The intersection index of $\gamma_{1,2}$ with $a-l_1$ is $-1$.

Thus the braid group generators act in $H_1(\L\setminus f^{-1}(\infty);\; f^{-1}(\l))$ as follows:
\begin{equation}
\sigma_1: \left\{\begin{array}{l}\gamma_{1,2} \mapsto \gamma_{1,2}-a+l_1 \\ a \mapsto a \\ b \mapsto b-a+l_1 \end{array}\right.  \qquad
\sigma_2: \left\{\begin{array}{l}\gamma_{1,2} \mapsto \gamma_{1,2} \\ a \mapsto a+b \\ b \mapsto b\end{array}\right. \qquad
\sigma_3: \left\{\begin{array}{l}\gamma_{1,2} \mapsto \gamma_{1,2} \\ a \mapsto a \\ b \mapsto b-a \end{array}\right.
\label{sigma_ab}
\end{equation}

The contour $l_1$ stays invariant under our transformations.

Applying (\ref{sigma_ab}) consecutively, we prove the following
\begin{theorem}
\label{thm_braidmonodromies}
Consider the Hurwitz space $\H^{\{a\}}_{1;2}(1,1)$  of two-fold elliptic coverings with the choice of the canonical homology basis as in Fig. \ref{cyclesgenus1}.
Then the braid monodromy group ${\bf{M}}(\{a\})$ of the corresponding solution  $\Phi= (\Phi^{\gamma_{1,2}},
\Phi^{l_1}, \Phi^a, \Phi^b)$ (\ref{phisol}) to the Fuchsian system
(\ref{ls1}), (\ref{ls2}) is generated by the following monodromy matrices:
\begin{align*}
&&M_{\sigma_1}=\left(\begin{array}{rrrr}
       1& 0&0&0   \\
      1&1&0 &1 \\
      -1&0&1&-1\\
      0&0&0&1\\
   \end{array} \right);
&&M_{\sigma_2^2\sigma_1\sigma_2^2}=\left(\begin{array}{rrrr}
       1& 0&0&0   \\
      1&1&2 &1 \\
      -1&0&-1&-1\\
      -2&0&0&-1\\
   \end{array} \right); \\ \\
&&M_{\sigma_3}=\left(\begin{array}{rrrr}
       1& 0&0&0   \\
      0&1&0 &0 \\
      0&0&1&-1\\
      0&0&0&1\\
   \end{array} \right);
&&M_{\sigma_2^2\sigma_3\sigma_2^2}=\left(\begin{array}{rrrr}
       1& 0&0&0   \\
      0&1&0&0 \\
      0&0&-1&-1\\
      0&0&0&-1\\
   \end{array} \right);
\end{align*}
these matrices define the group homomorphism from the subgroup
 $B_4^{\{a\}} = \langle \sigma_1, \; \sigma_3, \;
 \sigma_2^2\sigma_1\sigma_2^2, \; \sigma_2^2\sigma_3\sigma_2^2
\rangle$  of the braid group $B_4$ to $GL(4,\Z)$.
\end{theorem}

\begin{corollary}

The braid monodromy group of  function $\Phi$ associated to the two-fold genus one covering with the homology basis chosen as shown in Fig.\ref{cyclesgenus1}, consists of all matrices of the form
\be
\left(\begin{array}{cccc}
       1 & 0 & 0 & 0   \\
      k+2r & 1 & 2m  & k \\
      2mn-k\epsilon & 0 &  \epsilon & n \\
      2m\epsilon & 0 & 0 & \epsilon \\
   \end{array} \right),
\la{genbrmon}
\ee
where $k,m,n,r$ are arbitrary integers, $\epsilon=\pm1$.

The center of the group is isomorphic to $\Z$. The center is an image of the
center of the braid group which is generated by the element $(\sigma_1\sigma_2\sigma_3\sigma_1\sigma_2\sigma_1)^2$;
 it consists of all matrices of the form
\be
\left(\begin{array}{cccc}
       1 & 0 & 0 & 0   \\
      2r & 1 & 0 & 0 \\
      0 & 0 &  1 & 0 \\
      0 & 0 & 0 & 1 \\
   \end{array} \right)
\la{centerbrgr}
\ee
with $r\in\Z$.
\end{corollary}
{\it Proof.} A straightforward computation shows that matrices of the form (\ref{genbrmon})
form a group and all four generators of the braid monodromy group listed in Theorem
\ref{thm_braidmonodromies} are of this form.

The center of the braid group $B_4$ is generated by the element $ \Delta^2=(\sigma_1\sigma_2\sigma_3\sigma_1\sigma_2\sigma_1)^2$, see \cite{Garside}. Using the braid group relations, $\Delta^2$
can be written in the form $\Delta^2=(\sigma_2^2\sigma_1\sigma_2^2\sigma_3)^2$ which shows that it belongs to the group $B_4^{\{a\}}$. The image of $\Delta^2$ under the homomorphism $h$ is the matrix (\ref{centerbrgr}) with $r=1$.

On the other hand, the general form (\ref{genbrmon}) of a matrix from the braid monodromy group implies that elements in the center of this group are given by the identity matrix plus a matrix with a single nonzero $(21)$ entry.
 Since matrices of this form with an odd $(21)$-entry do not belong to the group of matrices of the form (\ref{genbrmon}), the center consists of the matrices (\ref{centerbrgr}).

% any matrix of the form (\ref{centerbrgr}) lies in the center of the group.

To see that the integers $k,m,n$, as well as $\epsilon$, can be chosen arbitrarily one introduces three elements:
$$
x:= M_{\sigma_3};
$$
$$
y:= M_{\sig_1} M_{\sig_2^2\sig_3\sig_2^2}M_{\sig_2^2\sig_1\sig_2^2}M_{\sig_3}=
\left(\begin{array}{cccc}
        1 & 0 & 0 & 0   \\
        4 & 1 & 2 & 0 \\
        0 & 0 &  1 & 0 \\
        2 & 0 & 0 & 1 \\
   \end{array} \right);
$$
and
$$
z:=M_{\sig_1}M^2_{\sig_2^2\sig_1\sig_2^2}M_{\sig_3}=\left(\begin{array}{cccc}
        1 & 0 & 0 & 0   \\
        -1 & 1 & 0 & -1 \\
        1 & 0 &  1 & 0 \\
        0 & 0 & 0 & 1 \\
   \end{array} \right).
$$
By taking products $x^{-n}y^{m}z^{-k}$
 we get a  matrix of the form (\ref{genbrmon}) with $\epsilon=1$, and arbitrary given $k,m,n$.
On the other hand, the expression
$x^{n+1}y^{m}z^{k+2m}M_{\sig_2^2\sig_3\sig_2^2}$ gives a  matrix of the form (\ref{genbrmon}) with $\epsilon=-1$,
and arbitrary given $k,m,n$.

Now, for any $k,m,n$ and $\epsilon$ one can get an arbitrary value of $r$
by multiplication with an appropriate central element.

$\Box$

\subsection{Action of braid group on monodromy matrices}

Here we discuss the action of the braid group on the set of monodromies of our solution $\Phi$
to system (\ref{ls1}), (\ref{ls2}).

The action of the braid group on the sets of monodromy matrices was used in \cite{DubMaz,Boalch}
to study the algebraic solutions of the Painlev\'e VI equation. Finiteness (up to a simultaneous conjugation)
 of the orbit of the action of the braid group  on the set of monodromies of a given Painlev\'e VI equation
is a necessary condition for algebraicity of the corresponding solution.

Although an analogous result seems to be not explicitly formulated for the Schlesinger systems of an
arbitrary dimension with an arbitrary number of singularities, it is instructive to see how the
braid group acts on the set of monodromies of the function $\Phi$.

Let us briefly recall how $B_\M$ acts on monodromies $\{M_j\}$
of an arbitrary Fuchsian system of the form (\ref{lsintr}). Choose the
 standard set of generators of $\pi_1(\CP1\setminus\{\l_1,\dots,
\l_{\iM},\infty\})$ satisfying relation (\ref{genpi1}); then the monodromy matrices
satisfy (\ref{relationmonodromy}).

The action of the generator $\sigma_k\in B_\M$ on the generators $\gamma_j$ of $\pi_1(\CP1\setminus\{\l_1,\dots,
\l_{\iM},\infty\})$ is as follows: $\sigma_k(\gamma_k)=\gamma_{k+1}$\;;
$\sigma_k(\gamma_{k+1})=\gamma_{k+1}^{-1}\gamma_k\gamma_{k+1}$;
$\sigma_k(\gamma_j)=\gamma_j$ for $j\neq k, \,k+1$.

 Therefore, under the action of $\sigma_k$, the set of matrices $\{M_j\}$ transforms into the set $\{M^{\sigma_k}_j\}$ where
%
%\begin{equation}
%\sigma_k\;:\;\left\{\begin{array}{lll} M_k\; \to \;M_k^{-1} M_{k+1} M_k \\
  %                                     M_{k+1}\; \to \;\;\; M_k   \\
    %                                   M_j \;\;\; \to \;\;\; M_j\;, \hskip0.5cm j\neq k,k+1 \end{array} \right.
\begin{equation}
     M^{\sigma_k}_k =M_k^{-1} M_{k+1} M_k , \qquad
    M^{\sigma_k}_{k+1}= M_k ,  \qquad
    M^{\sigma_k}_j = M_j\;, \hskip0.5cm j\neq k,k+1 .
    \la{actionS_kmon}
\end{equation}

Recall that the subgroup of $B_\M$ consisting of the braids which map the covering $\X$ to an equivalent one was denoted by $B_\M(\X)$.
The index of  $B_\M(\X)$ in $B_\M$ is finite; it is given by the number of inequivalent coverings with simple branch points
 of given degree and given ramification at $\infty$ - the Hurwitz number, which is denoted by $h_{g,\N}(k_1,\dots,k_\K)$.

The following theorem is a simple corollary of  Theorem \ref{thm_Schlesinger}:
\begin{proposition}
\la{monbraidth}
Let $\sigma\in B_\M(\X)$. Suppose the action of $\sigma$ on the chosen basis $\con=(\con_1,\dots,\con_\M)$ in
$H_1(\L\setminus f^{-1}(\infty)\;,\;f^{-1}(\l))$ is defined by a matrix $R_{\sigma}$, i.e., $\sigma: \con \mapsto \con R_\sigma$. Then  $\sigma$ acts
 on monodromy matrices of solution (\ref{defPhicon}) to the system (\ref{ls1}), (\ref{ls2}) by simultaneous conjugation
with the matrix $R_{\sigma}$:
\be
M_k^\sigma= (R_{\sigma})^{-1} M_k R_{\sigma}\;.
\la{changemon}
\ee
\end{proposition}

{\it Proof}. An action of $\sigma$ may result in a transformation of a basis $({\bf a}_\a,{\bf b}_\a)$ in $H_1(\L)$ used to
normalize $W$ and  the basis $\{\con_k\}$ in $H_1(\L\setminus f^{-1}(\infty)\;,\;f^{-1}(\l))$. It may also change signs of some of distinguished local parameters.
A transformation of  $({\bf a}_\a,{\bf b}_\a)$ results in a change of normalization of $W$;
according to Theorem \ref{thm_Schlesinger}, the induced action  on $\Phi$ is given by a Schlesinger transformation with the matrix $\Id-{\bf T}$ (\ref{T}), and, therefore,
does not change the monodromy matrices $M_k$. A change of signs of some of $x_j$ corresponds to multiplication of $\Phi$ from the left by a constant matrix $Y$,
which does not change the monodromy matrices either.

A transformation of the basis  $\{\con_k\}$ leads to the right multiplication of the solution with
the matrix $R_\sigma$ (\ref{Schlesinger}) which implies the transformation (\ref{changemon}) of monodromy matrices.

$\Box$

Let us introduce an equivalence relation $\sim$ on the space of the sets of monodromy matrices: two sets $\{M_k\}$ and $\{\tilde{M}_k\}$
are called equivalent if there exists a matrix $J$ such that $\tilde{M}_k= J M_k J^{-1}$ for all $k$.

Proposition \ref{monbraidth} implies the following immediate corollary:

\begin{corollary}
Consider a covering  $\X\in H_{g,\N}(k_1,\dots,k_\K)$, choose some Lagrangian subspace $\{{\bf a}\}$ of $a$-cycles
and a basis $\{\con_j\}$  in $H_1(\L\setminus f^{-1}(\infty)\;,\;f^{-1}(\l))$; define a function $\Phi$ by (\ref{defPhicon}) and denote the corresponding monodromy matrices by  $\{M_1,\dots,M_\M\}$.
The braid group $B_\M$ acts on the set  $\{M_1,\dots,M_\M\}$ according to (\ref{actionS_kmon}).
Then the number  of inequivalent (modulo the equivalence relation $\sim$) sets of $\M$ matrices obtained by the action of $B_\M$ on the set $\{M_1,\dots,M_\M\}$ equals the Hurwitz number $h_{g,\N}(k_1,\dots,k_\K)$.
\end{corollary}

{\it  Proof.}
Recall that the Hurwitz number equals the number of inequivalent  simple $\N$-sheeted coverings whose  branching at $\infty$ has the type $(k_1,\dots,k_\K)$. According to Theorem \ref{monbraidth}, if
$\sigma\in B_\M(\X)$, then the  initial set of monodromy matrices
is equivalent to the set of monodromies obtained by the action of an element $\sigma$ of the braid group.

Conversely, consider two coverings $\X$ and $\tilde{\X}$ with the same sets of branch points,
denote corresponding solutions of the Fuchsian system by $\Phi$ and $\tilde{\Phi}$
 and  the  sets of monodromy matrices by $\{M_k\}$ and $\{\tilde{M}_k\}$, respectively. Assume that the sets  $\{M_k\}$ and $\{\tilde{M}_k\}$
are equivalent, i.e., there exists a matrix $T$ such that $\tilde{M}_k = T M_k T^{-1}$.
Since the matrices
$$
M_k=  \left(\begin{array}{cc} I & S_k\\
                             0 & \Sigma_k \end{array}\right)\;,\hskip0.7cm
\tilde{M}_k=  \left(\begin{array}{cc} I & \tilde{S}_k\\
                             0 & \tilde{\Sigma}_k \end{array}\right)
$$
have the upper block-triangular structure, the matrix $T$ must have the same upper block-diagonal structure. Namely, denote the lower off-diagonal block
of the matrix $T$ by $T_0$. Then the relation  $\tilde{M}_k T = T M_k $ implies that $T_0= \tilde{\Sigma}_k T_0$ for all $k$, i.e., each non-vanishing
column of $T_0$ is an eigenvector with the eigenvalue $1$  of all matrices $\Sigma_k$ simultaneously. Using the explicit form (\ref{Sigma1}),
(\ref{Sigman}) of the matrices $\Sigma_k$ one easily sees that this is impossible, so $T_0=0$.

Denoting the lower diagonal block of the matrix $T$ by $T_1,$ we see that $\tilde{\Sigma}_k = T_1 \Sigma_k T_1^{-1}$ for all $k$.

Therefore, the $\M$-tuple $(\Sigma_1,\dots,\Sigma_\M)$ of $(\M-1)\times(\M-1)$ matrices  is equivalent to the $\M$-tuple $(\tilde{\Sigma}_1,\dots,\tilde{\Sigma}_\M)$ of $(\M-1)\times(\M-1)$ matrices.
Thus, the corresponding $\M$-tuples of elements of the group $S_\N$ defining the coverings
$\X$ and $\tilde{\X}$ are equivalent, too, and the coverings
$\X$ and $\tilde{\X}$ are isomorphic.

$\Box$

\section{Concluding remarks}

The present work poses a number of interesting questions.
\begin{itemize}

\item
According to the general idea of the work by Dubrovin and Mazzocco \cite{DubMaz}, finiteness of the orbit of the action of the braid group on
the set of monodromy matrices of a Fuchsian system is the necessary condition for the algebraicity of the corresponding solution to the Schlesinger system (see also \cite{Boalch}).
For the solutions to the Fuchsian system constructed here, these numbers are finite and equal to the Hurwitz numbers $h_{g,d}(k_1,\dots,k_m)$. Therefore,
it seems natural to expect that all solutions to the Fuchsian linear system discussed in this paper,
 as well as the corresponding solutions to the Schlesinger system, are algebraic, which would reflect the algebraic nature of the Hurwitz spaces. To find a complete proof of this fact would be an interesting problem.

\item

In another work \cite{DubMaz1} by Dubrovin and Mazzocco the idea of reducibility of Schlesinger systems was developed:
a solution to a Schlesinger system is
called reducible if it can be expressed in terms of solutions to Schlesinger systems with smaller number of singularities or lower matrix dimension.
In particular, it was proved that if all monodromy matrices have the same block-triangular structure, than the solution is reducible.
Since all monodromy matrices of the $N\times N$ Fuchsian systems considered here have the structure of this type,
 their solutions should be expressible in terms of solutions of  lower-dimensional $(d-1)\times (d-1)$ Schlesinger system with the same number of singularities ($N=d-1$ only in the case of the space of polynomials
of degree $d$); the monodromy matrices of this $(d-1)\times (d-1)$-dimensional system are supposed to coincide with matrices $\Sigma_i$ from (\ref{formMk}).

 It is natural to ask how does the solution to the corresponding  $(d-1)\times (d-1)$ Riemann-Hilbert problem look like and what is the corresponding Jimbo-Miwa tau-function. Is it different from the Bergman tau-function?

\item

In this paper we solve the Fuchsian systems corresponding to the value $q=-1/2$ from the one-parametric family of Fuchsian systems arising from the Frobenius structures on Hurwitz spaces, while the system used by Dubrovin in \cite{2D} has $q=1/2$. In principle, one could get a solution to Dubrovin's system by a simple differentiation of the solution $\Phi$ to our system; however, since generically some of the columns of our matrix $\Phi$ are constants, in this way one does not get a complete set of solutions to the system with $q=1/2$. Therefore, there arises a problem of finding the missing set of vector functions satisfying Dubrovin's system.

\item

Two solutions to the Fuchsian system are equivalent up to a multiplication with constant factors from both sides if the corresponding coverings are equivalent as elements of the Hurwitz space ${\cal H}_{g,d}^{\{ {\bf a}\}}(k_1,\dots,k_m)$.
The  fundamental group of this space is a subgroup of the braid group which preserves  the covering together with the Lagrangian subspace spanned by the $a$-cycles in the homologies. In this paper we described this subgroup
in the simplest case of two-sheeted coverings of genus one. An extension of this result to hyperelliptic and more general coverings is an interesting problem.

\end{itemize}

{\bf Acknowledgments.} We thank M. Bertola, V. Kac, A. Kokotov and P. Zograf for fruitful discussions.
The authors thank the Max Planck Institute for Mathematics in Bonn, where
the main part of this work was done, for warm hospitality, support and
excellent working conditions. The work of DK was partially supported
by NSERC, NATEQ and Concordia University Research Chair grant. The work of VS was supported by the EPSRC Postdoctoral Fellowship and NSERC.

\newpage
\setcounter{section}{0}
\renewcommand{\thesection}{\Alph{section}}

\section{Explicit form of monodromy matrices}

\label{app_monodromy}
%\renewcommand{\theequation}{A-\arabic{equation}}

%\subsection{Explicit form of monodromy matrices}
%\label{sect_monodromy}

Here we compute the explicit form of monodromy matrices of our solution of the Fuchsian system (\ref{lsintr}).

\subsection{Spaces of meromorphic functions with simple poles}
\label{sect_simpleinfty}

Consider the  Hurwitz space $\H_{g;\N}(1,\dots,1)$ of functions with $\N$ simple poles and  simple critical points on a
Riemann surface of genus $g.$
The branched covering  $\L$ corresponding to such a function has
$\M=2g+2\N-2$ finite branch points $\l_j$ and no branching over $\l=\infty$;
the covering $\L$ can be defined by a choice of  $\M$ generators of the fundamental group $\pi_1({\mathbb C}\setminus \{\l_1,\dots,\l_{\iM}\},\lambda_0)$ of the base of the covering and a set of elements of the symmetric
group $\symm$ assigned to these generators.
For an explicit computation of monodromy matrices of the solution (\ref{defPhicon}), (\ref{phisol}) to the Fuchsian system (\ref{ls1}), (\ref{ls2}) it is useful to
represent
the branched covering $\L$ in a standard form. For that purpose we
make use of Clebsch's result (\cite{Clebsch}, see \cite{Eisenbud} for
the modern exposition) stating that one can always choose
generators  $\{\g_j\}$ of $\pi_1({\mathbb C}\setminus \{\l_1,\dots,\l_{\iM}\},\lambda_0)$  satisfying (\ref{genpi1}) in such a way that the loop $\g_j$ encircles
only the point $\l_j$ and the set of the corresponding elements $\sigma_k\in \symm$ of
the monodromy group of the covering has the form:
\be
\sigma_1,\dots,\sigma_{\iM}= (1,2), (1,2),\dots,(1,2),(1,2),(2,3),(2,3),(3,4),(3,4),\dots,(\N-1,\N),(\N-1,\N)\;,
\label{genmongroup}
\ee
where the first transposition $(1,2)$ occurs $2g+2$ times at the
beginning and the other transpositions $(j,j+1),$ $j\geq 2,$ each occur twice, in order.
Such a  covering  can be visualized as a hyperelliptic Riemann surface
of genus $g$ with $\N-2$ Riemann spheres attached to it, see the Hurwitz diagram from Figure \ref{fig_diagram}.

\begin{figure}[htb]
\centering
\includegraphics[scale=0.5]{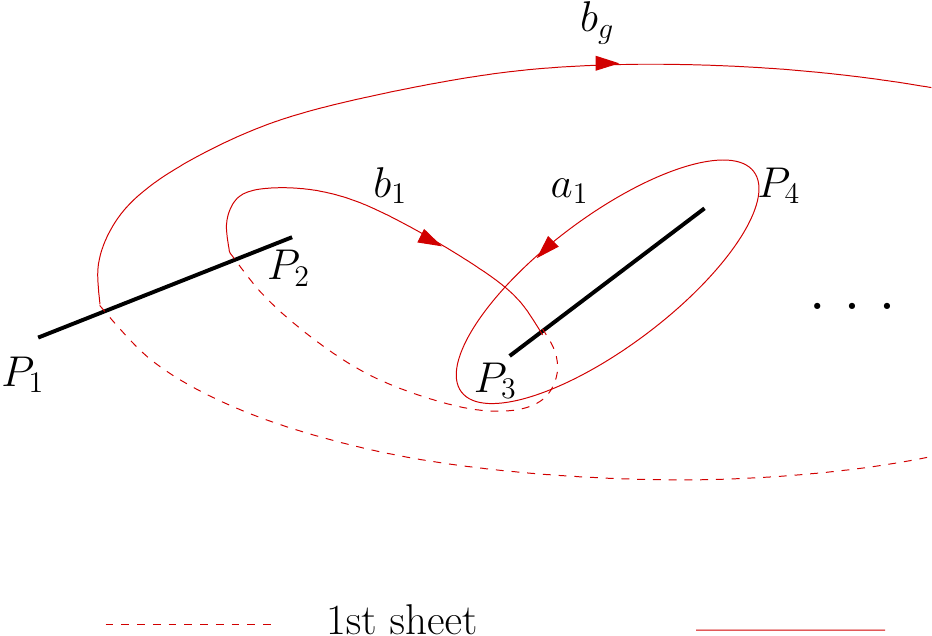}
\caption{Canonical homology basis for a hyperelliptic curve.}
\label{fig_hyperelliptic_ab}
\end{figure}

Assume the canonical homology basis  to be chosen on the ``hyperelliptic part'' of the Riemann surface in the standard way, i.e., the cycle ${a}_\alpha$ encircles the ramification points $P_{2\a+1},$ $P_{2\a+2}$ on the second sheet, and the cycle ${ b}_\a$ goes around the points $P_2$ and $P_{2\a+1},$ see Figure \ref{fig_hyperelliptic_ab}.
Assume also that the branch cuts are chosen to connect the points $P_{2k-1}$ and $P_{2k}$ for $k=1,\dots,g+\N-1.$

It is convenient also to consider a basis of contours in the space $H_1(\L\setminus f^{-1}(\infty)\,;\, f^{-1}(\l))$ different from the basis
(\ref{bacyc})-(\ref{gkkp1}). To describe the new basis, assume the contour  $\gamma_{1,2}(\l)$ (see (\ref{gkkp1})) to go around the point $P_1$ when passing from the first sheet to the second, i.e, $\gamma_{1,2} = {\rm lift}^{(1)}(\gamma_1)$, the lift of the generator $\gamma_1\in\pi_1({\mathbb C}\setminus \{\l_1,\dots,\l_{\iM}\},\lambda_0)$.  Analogously, for $n=2,\dots,\N-1$ we assume $\gamma_{n,n+1}(\l) ={\rm lift}^{(n)}(\gamma_{2g+2n-1})$, i.e., $\gamma_{n,n+1}(\l)$ passes from $n$th to $(n+1)$st sheet of the covering  by going around the point $P_{2g+2n-1}$.

Recall from Section \ref{sect_monodromy0} that there is a natural homomorphism $\F:\pi_1({\mathbb C}\setminus \{\l_1,\dots,\l_{\iM}\}, \lambda) \to {\rm Aut}[H_1(\L\setminus f^{-1}(\infty)\,;\,f^{-1}(\l))]$. Let us denote the images of the generators $\gamma_k$ of the fundamental group under $\F$ by $\F_{\lambda_k}.$

Then $\F_{\lambda_j}[\gamma_{n,n+1}]$ is the transformation of the contour $\gamma_{n,n+1}(\l)$  as $\lambda$ goes around $\l_j$ on the base of the covering. Let us take the basis in $H_1(\L\setminus f^{-1}(\infty)\,;\,f^{-1}(\l))$ formed by the following $2g+2\N-2$ paths on the surface $\L$:
\begin{eqnarray}
\label{Scont1}
& \Scont_j:=\frac{1}{2} \left( \gamma_{1,2} + \F_{\lambda_j}[\gamma_{1,2}] \right),  &\mbox{for} \;\; j=2,\dots, 2g+2;\\
%&& S_{2g+4}:= \gamma_{23}+\F_{[\lambda_{2g+4}]}(\gamma_{23})
\label{Scont2}
& \Scont_{2g+2n}:= \frac{1}{2} \left( \gamma_{n,n+1}+\F_{\lambda_{2g+2n}}[\gamma_{n,n+1}] \right),   &\mbox{for} \;\; n=2,\dots, \N-1;\\
\label{Scont3}
& \gamma_{n,n+1}(\lambda), &\mbox{for} \;\; n=1,\dots, \N-1.
\end{eqnarray}
The factor of $1/2$ is introduced for computational convenience in what follows.
Note that the paths $2\Scont_k$ (\ref{Scont1})-(\ref{Scont2}) are closed contours on the surface.

In this section we compute monodromy matrices for the fundamental matrix solution to our Fuchsian system (\ref{ls1}), (\ref{ls2}), associated to the Hurwitz space $\H_{g;\N}(1,\dots,1)$, whose columns are the vectors (\ref{defPhicon}) with integration contours $\con$ given by the  basis (\ref{Scont1})-(\ref{Scont3}).

The monodromy matrices of the solution $\Phi(\lambda)$ (\ref{defPhicon}) with integration paths (\ref{Scont1})-(\ref{Scont3}) in the given order still have the structure (\ref{formMk}), where the number $\K$ of pre-images of the point at infinity equals $\N$, the degree of the covering. The monodromy matrices are determined by the transformations of the contours (\ref{Scont1}) -
(\ref{Scont3}) in  $H_1(\L\setminus
f^{-1}(\infty);\; f^{-1}(\l))$ which occur as the point $\lambda$
describes the loops $\gamma_k$ on the base of the covering (i.e., under the automorphisms $\F_{\lambda_k}$). The first
$2g+\N-1$ columns of the matrix $\Phi$ remain unchanged under these
transformations.
%The matrices $S_k$ in (\ref{formMk}) are thus vectors of the length $2g+1$ and the matrices $T_k$ are scalars.

 We now look at the transformations of the last $\N-1$ columns of the matrix $\Phi(\l)$ given by the integrals (\ref{defPhicon}) over the contours $\gamma_{n,n+1}(\l)$ with $n=1,\dots,\N-1$ and find the corresponding $S_k$ and $\Sigma_k$ (see (\ref{formMk})) for $k=1,\dots,2g+2\N-2,\infty.$

{\bf Monodromy matrix $M_1$.}

\begin{figure}[htb]
\centering
\includegraphics[scale=0.4]{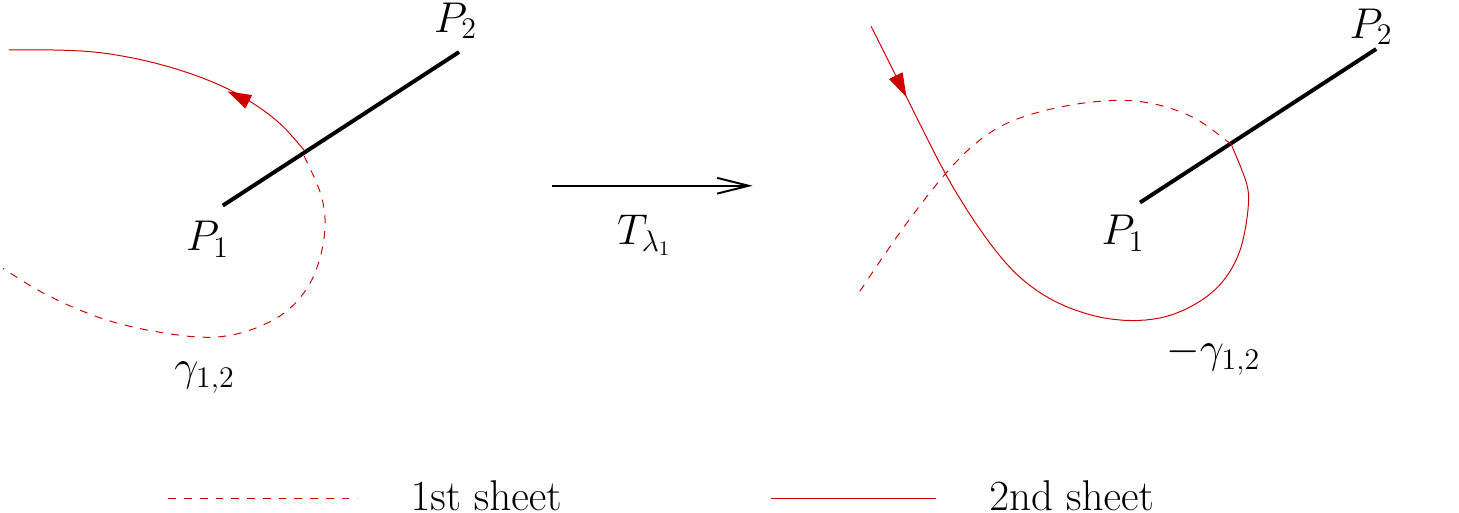}
\caption{The transformation of the contour $\gamma_{1,2}$ corresponding to the monodromy matrix $M_1.$}
\label{fig_hyperelliptic_M1}
\end{figure}

When $\lambda$ travels along the loop $\gamma_1$ on the base, the contour $\gamma_{1,2}$ transforms to $-\gamma_{1,2},$ as shown in Figure \ref{fig_hyperelliptic_M1}. Note that the sum of the two contours in Figure \ref{fig_hyperelliptic_M1} is the closed contour encircling the point $P_1;$ this contour is trivial in the space $H_1(\L\setminus f^{-1}(\infty);\; f^{-1}(\l)).$

As is easy to see, the contour $\gamma_{2,3}$ becomes the sum $\gamma_{2,3}+\gamma_{1,2}$ under the automorphism $\F_{\l_1}$. The other contours $\gamma_{n,n+1}$ with $n>2$ do not change. Thus, the monodromy matrix $M_1$ has the form
\begin{equation}
M_1= \left(
    \begin{array}{ccc}
     I_{2g+\N-1}  & 0 \\
     0 & \Sigma_1
          \end{array}\right),
\label{M1}
\end{equation}
 where $\Sigma_1$ is the following $\N-1\times \N-1$ matrix
\begin{equation}
\Sigma_1= \left(
    \begin{array}{ccccc}
     -1  & 1&0 & \dots & 0\\
     0 & 1&0 & \dots & 0\\
     0&0&1 & \dots & 0\\
     \vdots&\vdots&\vdots&\ddots & \vdots\\
     0&0&0&\dots&1
          \end{array}\right).
\label{Sigma1}
\end{equation}

{\bf Monodromy matrix $M_2$.}

\begin{figure}[htb]
\centering
\includegraphics[scale=0.6]{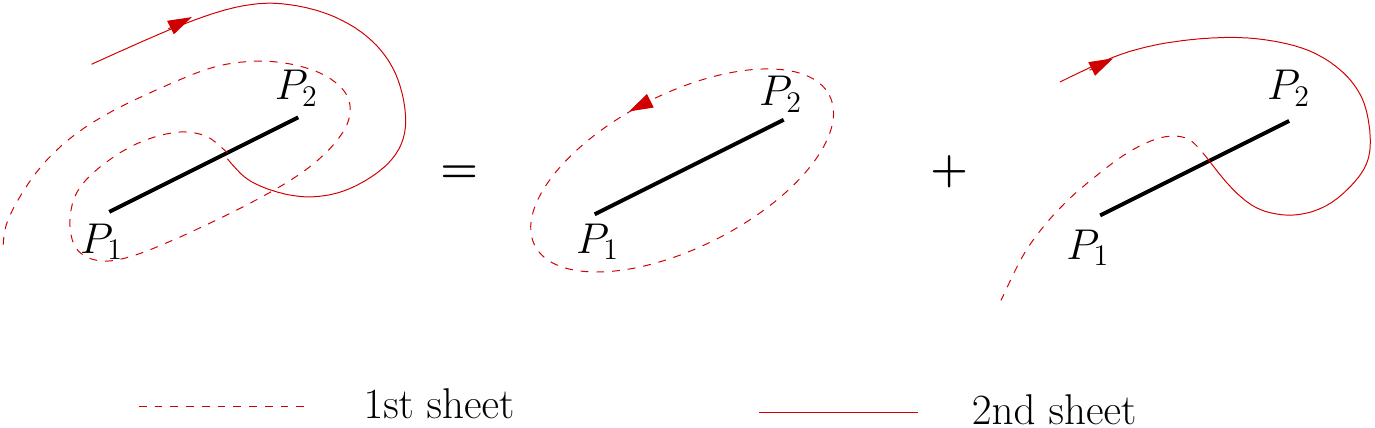}
\caption{The transformation of the contour $\gamma_{1,2}$ corresponding to monodromy around $\l_2.$}
\label{fig_hyperelliptic_M2}
\end{figure}

\begin{figure}[htb]
\centering
\includegraphics[scale=0.6]{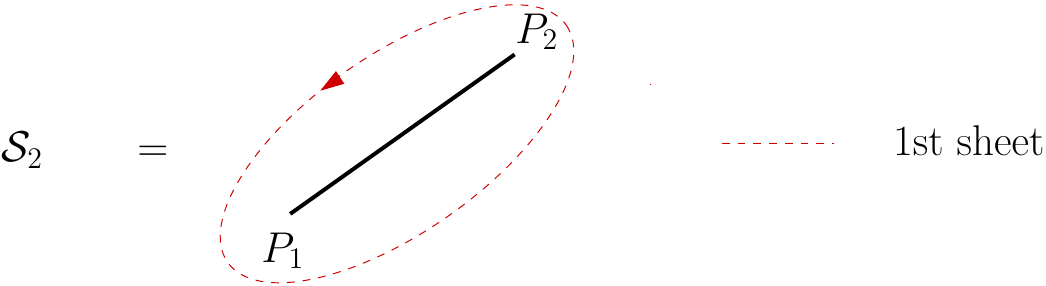}
\caption{The contour $\Scont_2$}
\label{fig_S2}
\end{figure}

The image of the contour $\gamma_{1,2}$ under  $\F_{\lambda_2} \in {\rm Aut}[H_1(\L\setminus f^{-1}(\infty)\,;\,f^{-1}(\l))]$ is shown in Figure \ref{fig_hyperelliptic_M2}.
Let us denote the closed contour encircling the branch cut $[P_1,P_2]$ counter-clockwise on the first sheet, the closed contour from Figure \ref{fig_hyperelliptic_M2}, by $\A_{12}$.
 The sum of the non-closed contour in the right hand side in Figure \ref{fig_hyperelliptic_M2} and $\gamma_{1,2}$ (the contour $-\gamma_{1,2}$ from Figure \ref{fig_hyperelliptic_M1} with inverse orientation) is a closed contour encircling clockwise the branch cut $[P_1,P_2]$ on the second sheet, i.e., again the contour $\A_{12}$.
By (\ref{Scont1}) we have $ \F_{\lambda_2}[\gamma_{1,2}]= 2\Scont_2 - \gamma_{1,2} $. Thus, we get $\Scont_2=\A_{12}$, see Figure \ref{fig_S2}.

\begin{figure}[htb]
\centering
\includegraphics[scale=0.6]{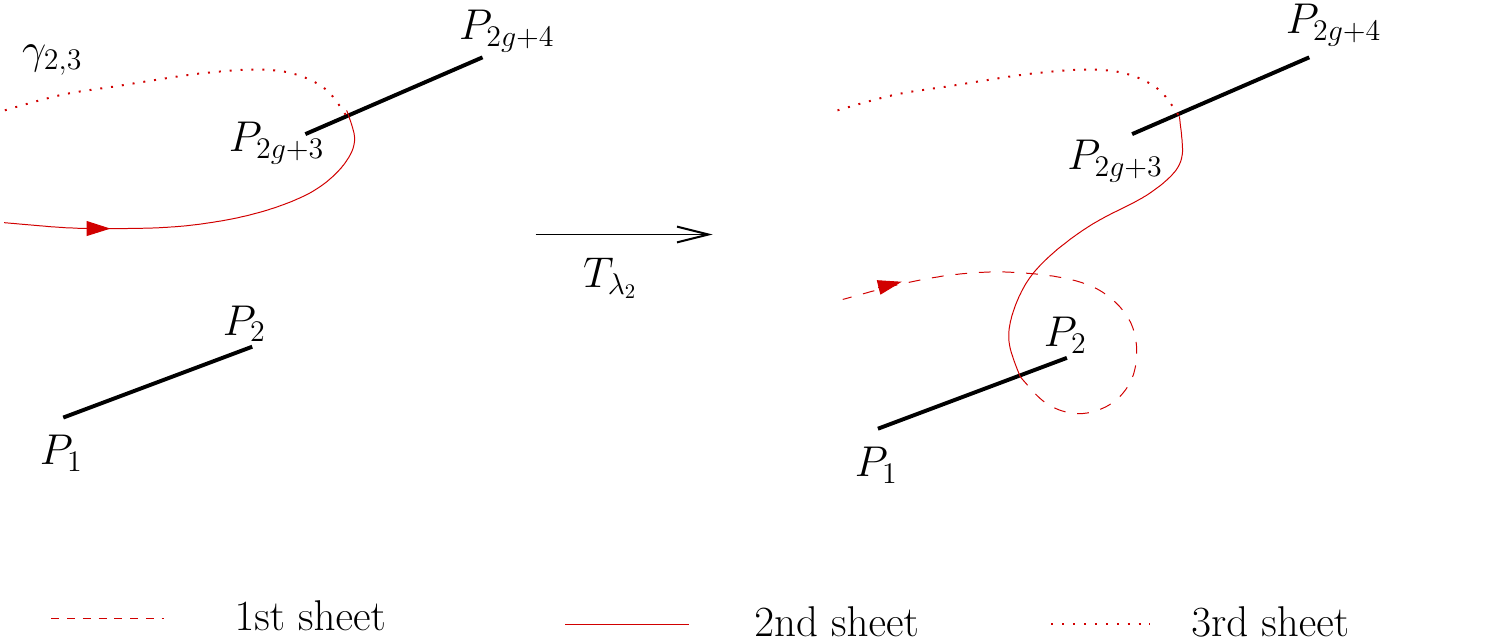}
\caption{The transformation of the contour $\gamma_{2,3}$ corresponding to monodromy around $\l_2.$}
\label{fig_gamma23}
\end{figure}

The transformation of the contour $\gamma_{2,3}$ under $\F_{\lambda_2}$  is shown in Figure \ref{fig_gamma23}.
>From the figure we see that the sum of the contour $\F_{\l_2}[\gamma_{2,3}]$  and the  non-closed contour from the right hand side of Figure \ref{fig_hyperelliptic_M2} gives $\gamma_{2,3}$. In other words, $\F_{\l_2}[\gamma_{2,3}] = \gamma_{2,3} + \gamma_{1,2} - \Scont_2$.

The paths $\gamma_{n,n+1}$ with $n>2$ remain unchanged when $\lambda$ goes around $\lambda_2.$ Thus, the monodromy matrix $M_2$ has the form:
\begin{equation}
M_2= \left(
    \begin{array}{ccc}
     I_{2g+\N-1}  &   S_2\\
     0 & \Sigma_1
          \end{array}\right),
\label{M2}
\end{equation}
where
$$
S_2=\left(\begin{array}{ccccc}
                2  & -1 & 0 &\dots & 0 \\
                0  &  0 & 0 & \dots & 0 \\
                        \dots & \dots & \dots & \dots &\dots\\
                        0&0&0&\dots&0
                      \end{array}\right)
$$
Similarly, we find other monodromy matrices corresponding to the ramification points $\l_{1},\dots,\l_{2g+2}$.

{\bf Monodromy matrix $M_{\lambda_{2k+1}}$ for $1\leq k\leq g$.}

\begin{figure}[htb]
\centering
\includegraphics[scale=0.6]{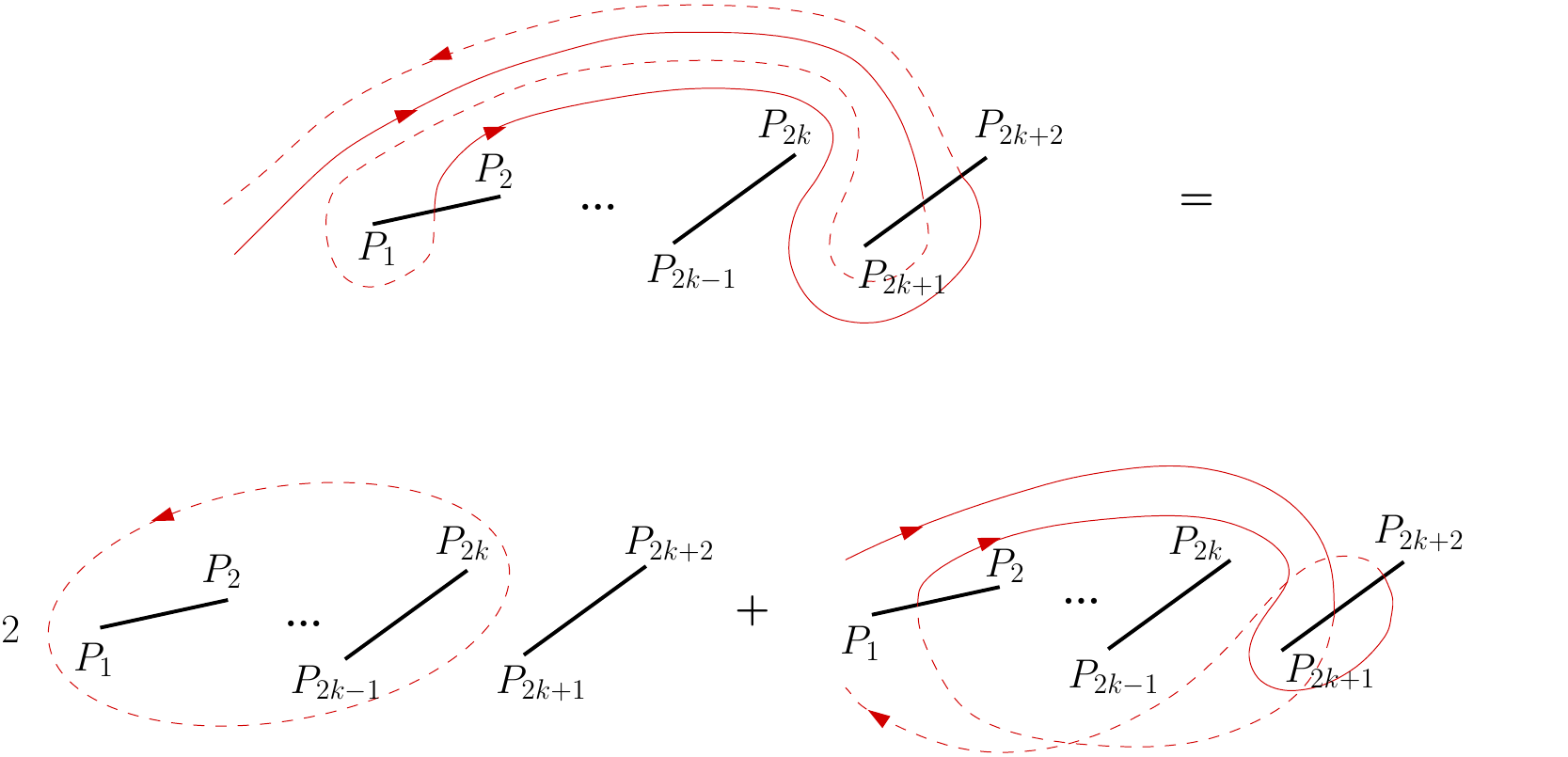}
%[width=8cm]{hyperelliptic_M2kp1_1.eps}\subfigure{\includegraphics[scale=0.5][width=10cm]{hyperelliptic_M2kp1_2.eps}}}
\caption{The transformation of the contour $\gamma_{1,2}$ corresponding to monodromy around $\l_{2k+1,} k>1.$}
\label{fig_hyperelliptic_M2kp1}
\end{figure}

\begin{figure}[htb]
\centering
\includegraphics[scale=0.6]{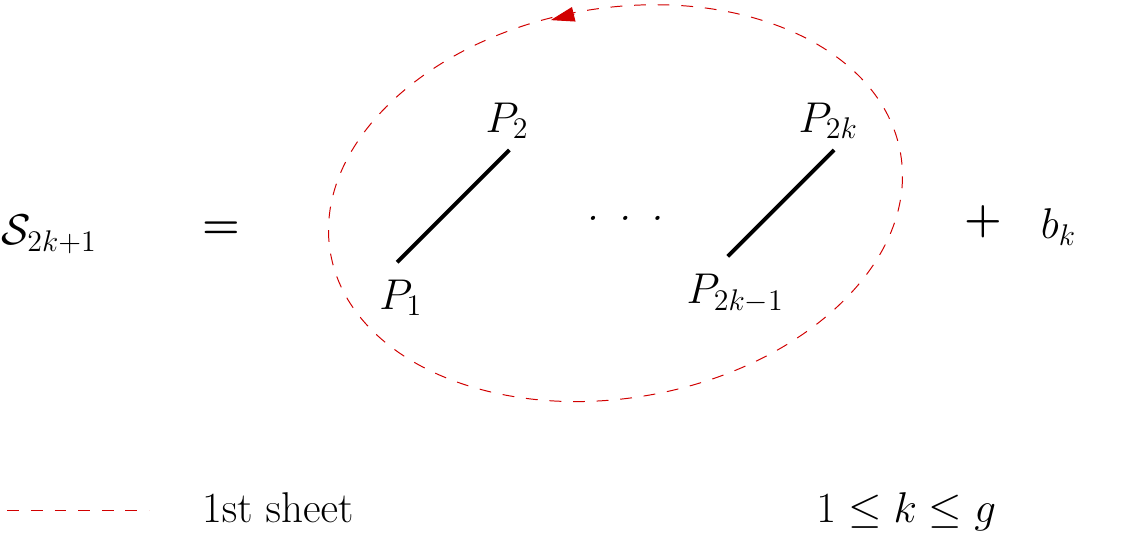}
\caption{The basis contour $\Scont_{2k+1}.$}
\label{fig_S2kp1}
\end{figure}

As the endpoints of the contour $\gamma_{1,2}$ go counterclockwise around  the point $\l_{2k+1},$ $1\leq k \leq g,$
  the contour $\gamma_{1,2}$ transforms to the contour shown in Figure \ref{fig_hyperelliptic_M2kp1}. As before, the paths on the first sheet are drawn with dash line and  solid line corresponds to the second sheet.
The sum of the original contour $\gamma_{1,2}$ and the non-closed
component in the right hand side in Figure
\ref{fig_hyperelliptic_M2kp1} gives a closed contour equivalent to $2{
  b}_k.$ Since, due to (\ref{Scont1}),
$ \F_{\lambda_{2k+1}}[\gamma_{1,2}] = 2 \Scont_{2k+1} -  \gamma_{1,2}$, the closed contour in Figure \ref{fig_hyperelliptic_M2kp1}  encircling the first $k$ branch cuts counter-clockwise on the first sheet is equivalent to $\Scont_{2k+1} - b_k$, see Figure \ref{fig_S2kp1}.

\begin{figure}[htb]
\centering
\includegraphics[scale=0.6]{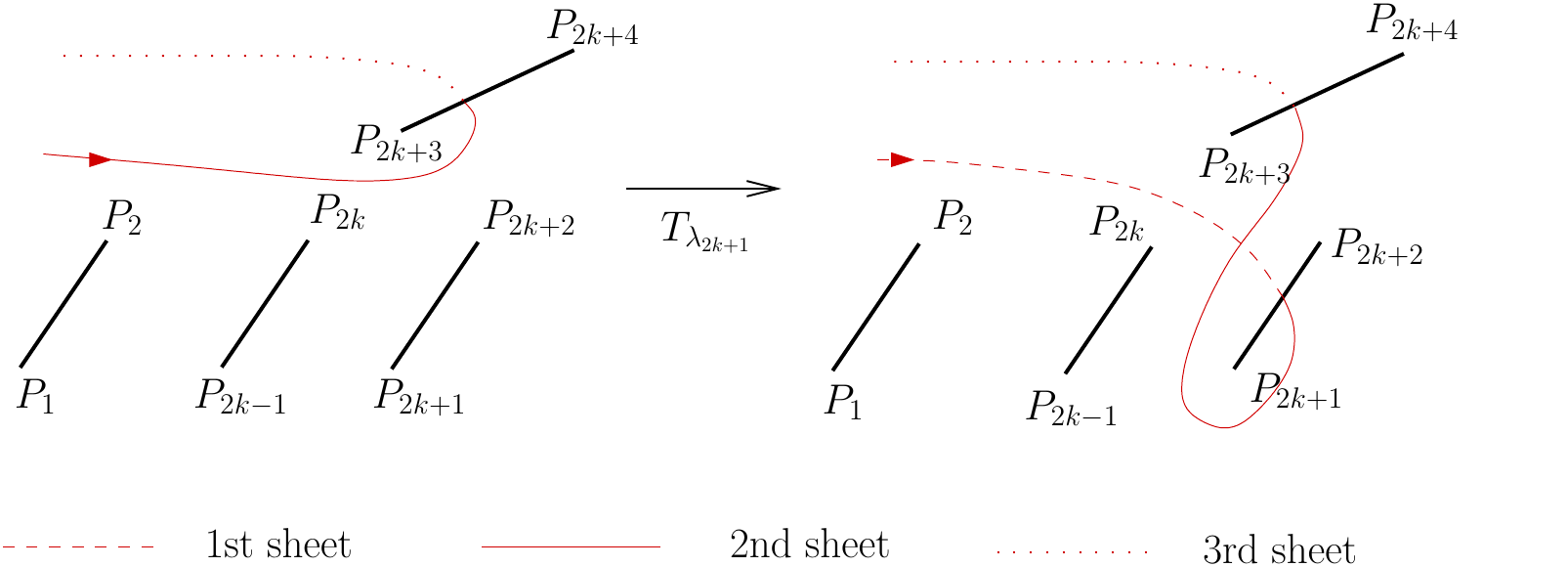}
\caption{The transformation of the contour $\gamma_{2,3}$ corresponding to the monodromy around $\l_{2k+1}.$}
\label{fig_T2kp1_g23}
\end{figure}

The transformation of the contour $\gamma_{2,3}$ under  $\F_{\lambda_{2k+1}}\in{\rm Aut}[H_1(\L\setminus f^{-1}(\infty)\,;\,f^{-1}(\l))]$ is shown in Figure \ref{fig_T2kp1_g23}.
Note that after subtracting $\gamma_{1,2}$ and $\gamma_{2,3}$ from the contour in the right hand side of Figure \ref{fig_T2kp1_g23}, we get the contour equivalent to $-\Scont_{2k+1}$ (see Figure \ref{fig_S2kp1}). Therefore, $\F_{\lambda_{2k+1}}[\gamma_{2,3}] = \gamma_{1,2} + \gamma_{2,3} - \Scont_{2k+1}.$

The remaining paths $\gamma_{n, n+1}$ with $n>2$ do not change under $\F_{\lambda_{2k+1}}$.

{\bf Monodromy matrix $M_{\lambda_{2k+2}},$ $1\leq k \leq g\,.$}

\begin{figure}[htb]
\centering
\includegraphics[scale=0.6]{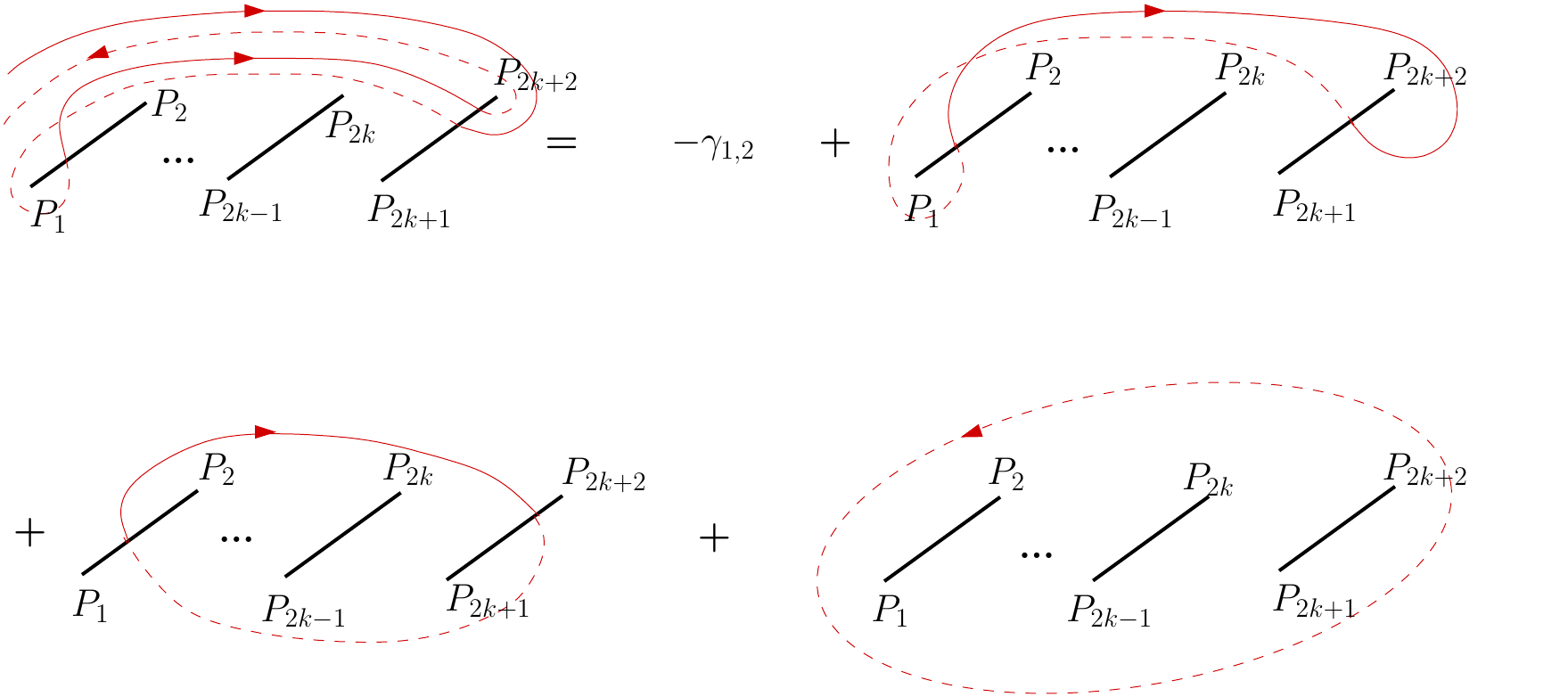}
\caption{The transformation of the contour $\gamma_{1,2}$ corresponding to monodromy around $\l_{2k+2}.$ The parts of the contours lying on the first sheet are drawn with dash line, the ones on the second sheet with solid line.}
\label{fig_M2kp2_gamma12}
\end{figure}

\begin{figure}[htb]
\centering
\includegraphics[scale=0.6]{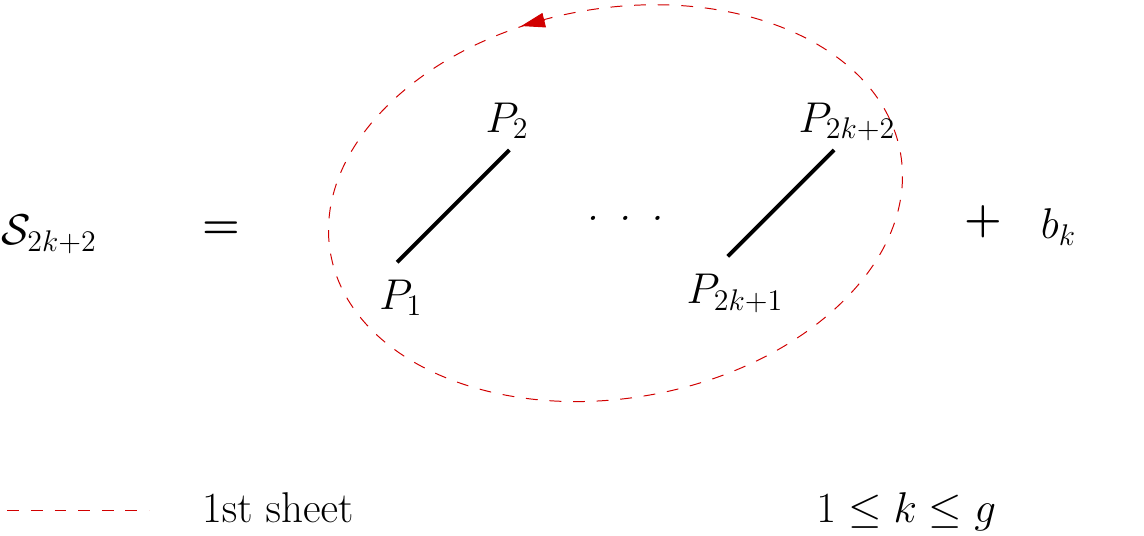}
\caption{The basis contour $\Scont_{2k+2}$}
\label{fig_S2kp2}
\end{figure}

Analogously, the image $ \F_{\lambda_{2k+2}}[\gamma_{1,2}]$ of $\gamma_{1,2}$ is shown in Figure \ref{fig_M2kp2_gamma12}. As is easy to see, the first closed contour from the right hand side in Figure \ref{fig_M2kp2_gamma12} is equivalent to the sum of the other two closed contours. On the other hand, from (\ref{Scont1}), we have $ \F_{\lambda_{2k+2}}[\gamma_{1,2}]=2\Scont_{2k+2} - \gamma_{1,2}$. Thus, the basis contour $\Scont_{2k+2}$ has the form given by Figure \ref{fig_S2kp2}.

\begin{figure}[htb]
\centering
\includegraphics[scale=0.6]{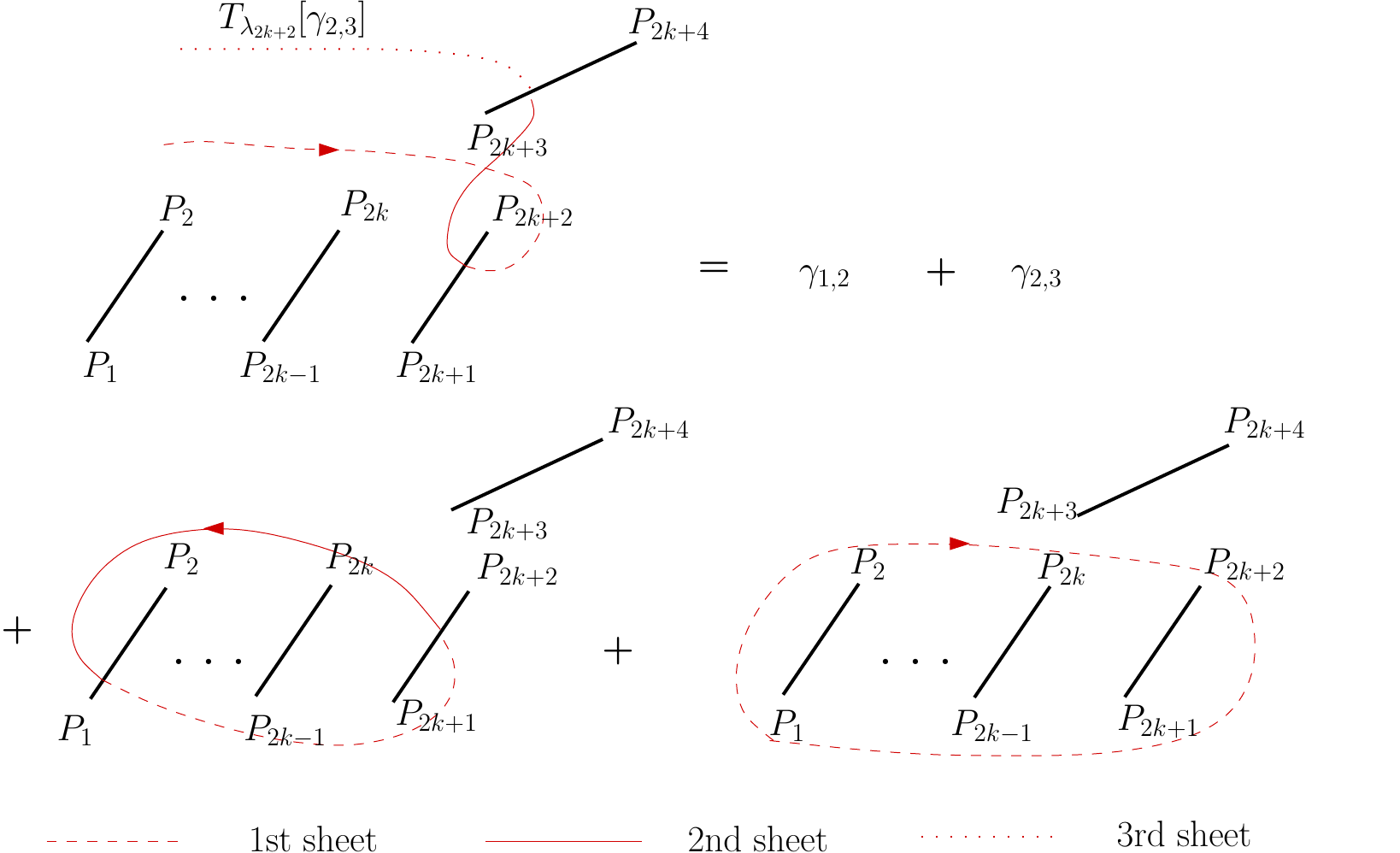}
\caption{The transformation of the contour $\gamma_{2,3}$ corresponding to monodromy around $\l_{2k+2}.$ The correspondence between style of the lines and sheets of the covering is as in Figure \ref{fig_T2kp1_g23}.}
\label{fig_M2kp2_gamma23}
\end{figure}

Similarly, from Figure \ref{fig_M2kp2_gamma23} we see that $\F_{\lambda_{2k+2}}[\gamma_{2,3}] = \gamma_{1,2}+\gamma_{2,3} - \Scont_{2k+2}.$
The contours $\gamma_{n,n+1}$ with $n>2$ do not change under the automorphism $\F_{\lambda_{2k+2}}$.
Hence, we obtain the monodromy matrices $M_n$ in the form:
\begin{equation}
M_n= \left(
    \begin{array}{ccc}
     I_{2g+\N-1}  & S_n \\
     0 & \Sigma_1
          \end{array}\right), \qquad 2\leq n\leq 2g+2.
\label{Mn}
\end{equation}
where
$$
S_n=
 \left(\begin{array}{ccccc}
                            0  &  0 & 0 & \dots & 0 \\
                        \dots & \dots & \dots & \dots &\dots\\
                        0  &  0 & 0 & \dots & 0 \\
                2  & -1 & 0 &\dots & 0 \\
                0  &  0 & 0 & \dots & 0 \\
                \dots & \dots & \dots & \dots &\dots\\
                0&0&0&\dots&0
                      \end{array}\right)
$$
Here,   $\Sigma_1$ is the $(\N-1)\times (\N-1)$ matrix given by (\ref{Sigma1}), and the nontrivial row in the block above the diagonal is the  $(n-1)$st one, i.e., the row corresponding to the integration contour $\Scont_n.$

{\bf Monodromy matrix $M_{\lambda_{2g+2k-1}},$ $2\leq k\leq \N-1\,.$}

The automorphism $\F_{\lambda_{2g+2k-1}}$ transforms the three contours $\gamma_{k-1,k}(\lambda)$, $\gamma_{k,k+1}(\lambda)$, $\gamma_{k+1,k+2}(\lambda)$ - those passing through the $k$th and $(k+1)$st sheets.
The contour $\gamma_{k, k+1}$ transforms to $-\gamma_{k,k+1}$ similarly to Figure \ref{fig_hyperelliptic_M1}.
The contours $\gamma_{k-1,k}$ and $\gamma_{k+1,k+2}$ transform to $ \gamma_{k-1,k}+\gamma_{k,k+1}$ and $\gamma_{k,k+1}+\gamma_{k+1,k+2}$, respectively.
Therefore, the monodromy matrix  has the form:
\begin{equation}
M_{2g+2k-1}= \left(
    \begin{array}{ccc}
     I_{2g+\N-1}  & 0  \\
     0 & \Sigma_{2g+2k-1}
          \end{array}\right), \qquad 2\leq k\leq \N-1.
\label{M2gp2km1}
\end{equation}
Here $\Sigma_{2g+2k-1}$ is the $(\N-1)\times (\N-1)$ matrix, corresponding to the permutation of the sheets of the covering, associated to the branch point $\lambda_{2g+2k-1}$ - the matrix is given by
\begin{equation}
\Sigma_{2g+2k-1}= \left(
    \begin{array}{cccccccc}
        1& \dots & 0 & 0 & 0  &\dots & 0 \\
        \vdots&\ddots&\vdots &\vdots&\vdots&&\vdots\\
        0 & \dots & 1 & 0 & 0 & \dots & 0\\
     0 & \dots & 1 & -1  & 1& \dots & 0\\
     0 & \dots & 0 & 0 & 1 & \dots & 0\\
%    0 & \dots& 0&0& 0 & 1  & \dots & 0\\
     \vdots&  &\vdots&\vdots &\vdots&\ddots & \vdots\\
     0&\dots& 0&0&0&\dots&1
          \end{array}\right), \qquad 2\leq k\leq \N-1,
\label{Sigman}
\end{equation}
where the nontrivial $3\times 3$ diagonal block is formed by the rows from $(k-1)$st to $(k+1)$st and the respective columns.

{\bf Monodromy matrix $M_{\lambda_{2g+2k}},$ $2\leq k\leq \N-1\,.$}

\begin{figure}[htb]
\centering
\includegraphics[scale=0.6]{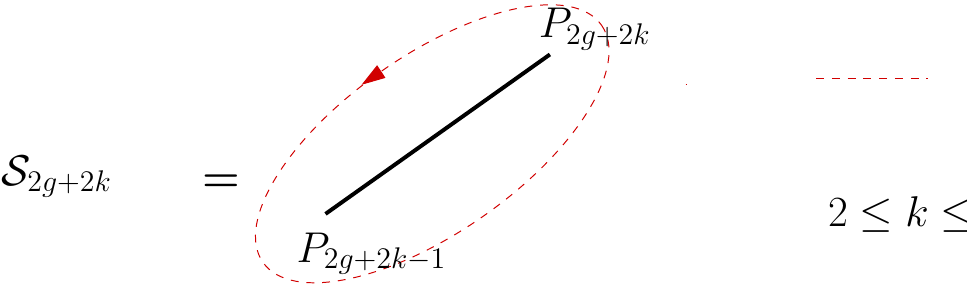}
\caption{The basis contour $\Scont_{2g+2k}$.}
\label{fig_S2gp2k}
\end{figure}

\begin{figure}[htb]
\centering
\includegraphics[scale=0.6]{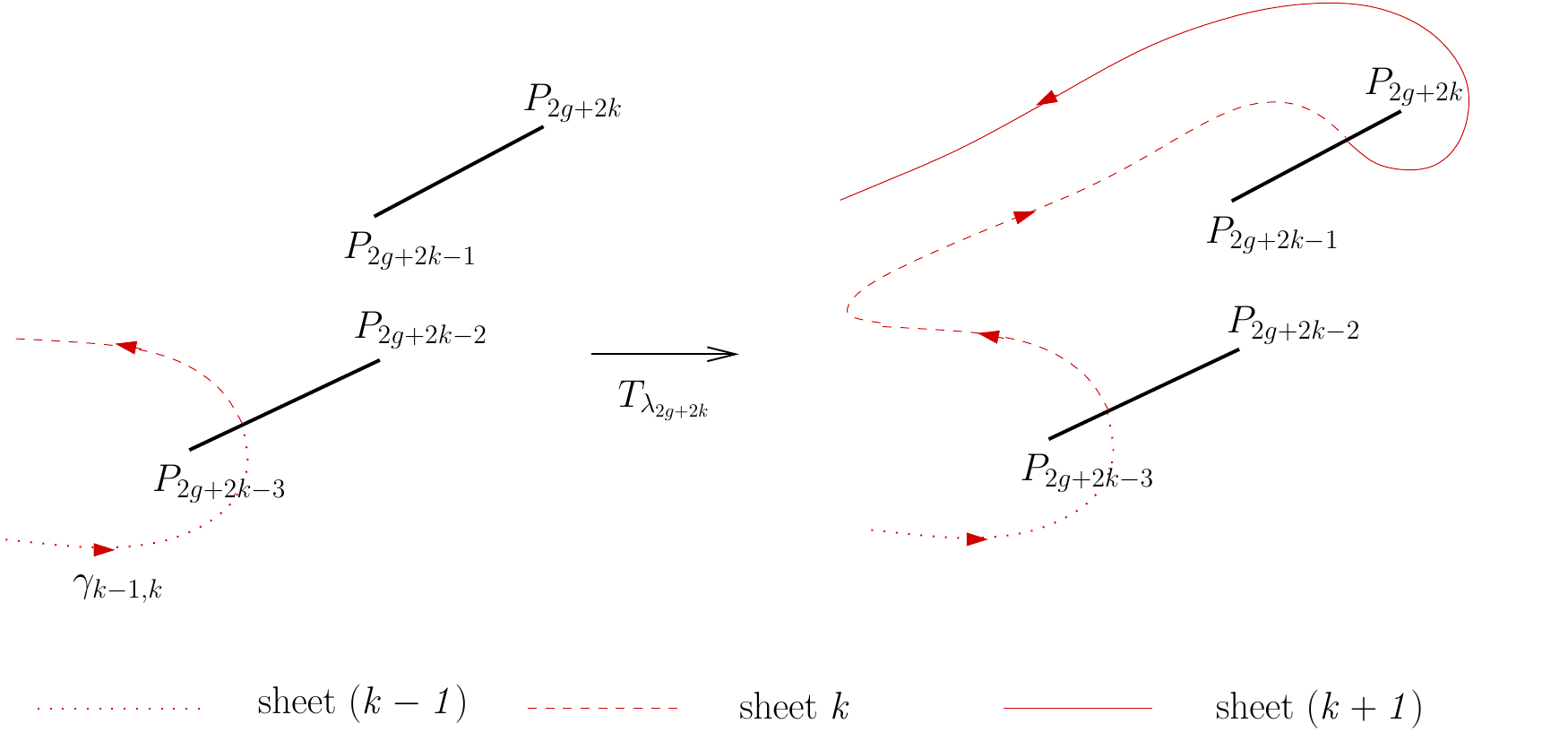}
\caption{The transformation of the contour $\gamma_{k-1,k}$ corresponding to the monodromy around $\l_{2g+2k}.$}
\label{fig_stairs_M2k1}
\end{figure}

\begin{figure}[htb]
\centering
\includegraphics[scale=0.6]{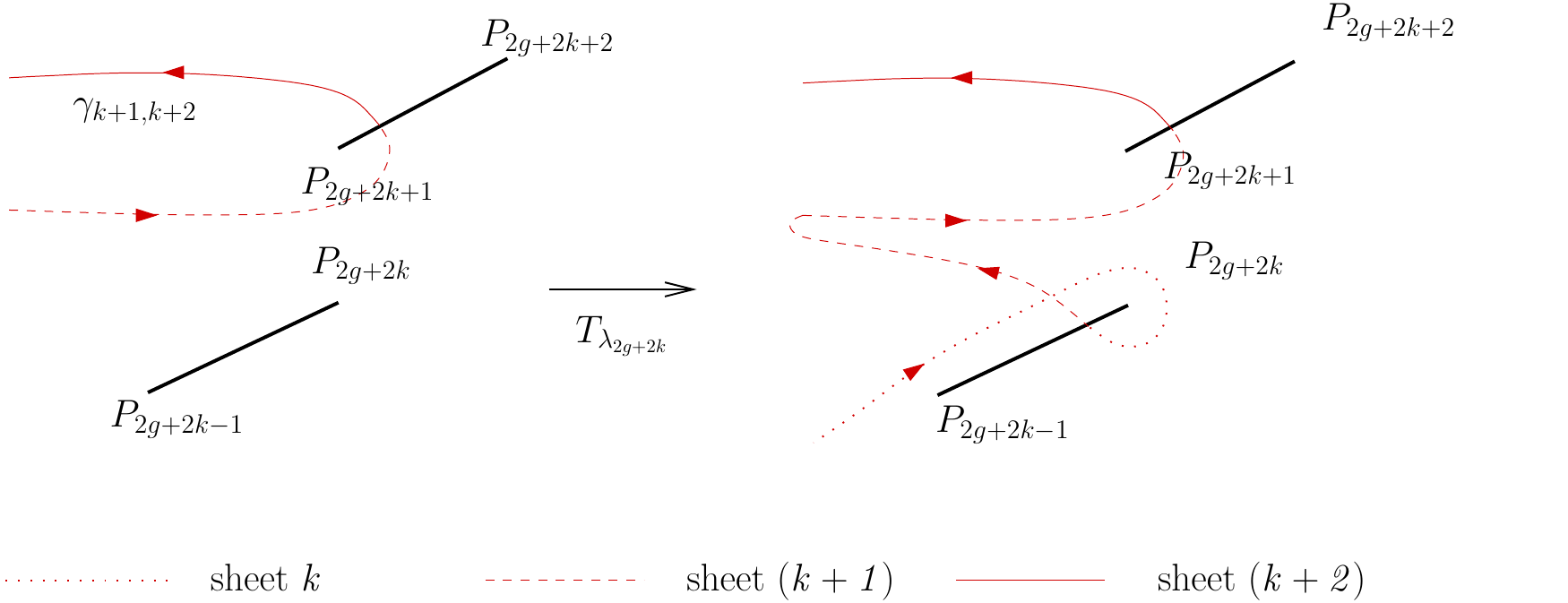}
\caption{The transformation of the contour $\gamma_{k+1,k+2}$ corresponding to the monodromy around $\l_{2g+2k}.$}
\label{fig_stairs_M2k2}
\end{figure}

The same columns of the matrix $\Phi(\l)$ transform, when $\lambda$ describes the loop $\gamma_{2g+2k},$ $k\geq 2,$ on the base of the covering.
The transformation of $\gamma_{k,k+1}$ under $\F_{\lambda_{2g+2k}}$ is analogous to that in Figure \ref{fig_hyperelliptic_M2}, where the ramification points are $P_{2k-1}$ and $P_{2k}$ instead of $P_1$ and $P_2,$ respectively. Since from \ref{Scont2} we have $\F_{\lambda_{2g+2k}}[\gamma_{k,k+1}]= 2\Scont_{2g+2k} - \gamma_{k,k+1},$ we see, similarly to Figure \ref{fig_S2}, that the basis contour $\Scont_{2g+2k}$ is equivalent to the closed contour encircling the branch cut $[P_{2g+2k-1},P_{2g+2k}]$ on the $k$th sheet, see Figure \ref{fig_S2gp2k}.
The transformations of $\gamma_{k-1,k}$ and $\gamma_{k+1,k+2}$ under $\F_{\lambda_{2g+2k}}$ are shown in Figures \ref{fig_stairs_M2k1} and \ref{fig_stairs_M2k2}, respectively.
The sum of the contour in the right hand side of Figure \ref{fig_stairs_M2k1} and the contour $-\gamma_{k,k+1}$ is equivalent to $\gamma_{k-1,k}$ plus the contour encircling the branch cut $[P_{2g+2k-1},P_{2g+2k}]$ clockwise on the $k$th sheet (we use the triviality of the contour encircling one ramification point). Therefore, we get $\F_{\lambda_{2g+2k}}[\gamma_{k-1,k}] =\gamma_{k-1,k}+\gamma_{k,k+1} - \Scont_{2g+2k}.$
Analogously, adding $-\gamma_{k,k+1}$ to the contour in the right hand side of Figure \ref{fig_stairs_M2k2} we get $\gamma_{k+1,k+2}$ plus a closed contour around the branch cut $[P_{2g+2k-1},P_{2g+2k}]$ oriented clockwise on the $k$th sheet. Thus the contour $\gamma_{k+1,k+2}$ transforms to $\gamma_{k+1,k+2}+\gamma_{k,k+1} - \Scont_{2g+2k}$ as $\l$ describes the loop $\gamma_{2g+2k}$ around $\l_{2g+2k}.$
The monodromy matrix thus has the form:
\begin{equation}
M_{2g+2k}= \left(
    \begin{array}{ccc}
     I_{2g+\N-1}  &  S_{2g+2k} \\
     0 & \Sigma_{2g+2k-1}
          \end{array}\right), \qquad 2\leq k\leq \N-1.
\label{M2gp2k}
\end{equation}
where
$$
S_{2g+2k}=\left(\begin{array}{ccccccc}
                        0&\dots&    0  &  0 & 0 & \dots & 0 \\
                    0&\dots&    \dots & \dots & \dots & \dots &\dots\\
                    0&\dots&    0  &  0 & 0 & \dots & 0 \\
0&\dots&                -1  & 2 & -1 &\dots & 0 \\
        0&\dots&        0  &  0 & 0 & \dots & 0 \\
                \dots&\dots&\dots & \dots & \dots & \dots &\dots\\
0&\dots&                0&0&0&\dots&0
                      \end{array}\right)
$$
The nontrivial row in the block above the diagonal is the $(2g+k)$th one, i.e., the row corresponding to the integration contour $\Scont_{2g+2k};$ the nonzero columns being the $(k-1)$st, $k$th and $(k+1)$st. The $(\N-1)\times (\N-1)$ matrix $\Sigma_{2g+2k-1}$ is  given by (\ref{Sigman}).

{\bf Monodromy matrix $M_\infty.$}

%As $\lambda$ goes around $\infty$ on the base of the covering, the contour $\gamma_{1,2}$ transforms to
As $\lambda$ goes counterclockwise around the point at infinity, each contour $\gamma_{k,k+1}$ transforms to  $\gamma_{k,k+1} - l_{k} + l_{k+1},$ where the contour $l_s$ (\ref{lsss}) is a closed contour around the point $\infty^{(s)}$.
% and $l_\iN = - \sum_{i=1}^{\iN-1} l_i.$
We need to express these contours in terms of our basis (\ref{Scont1})-(\ref{Scont3}). As is easy to see from Figures \ref{fig_S2}, \ref{fig_S2kp1}, \ref{fig_S2kp2} and \ref{fig_S2gp2k}, the following relations hold in $H_1(\L\setminus f^{-1}(\infty)\,;\, f^{-1}(\l))$:
\begin{eqnarray}
&& l_1= -\Scont_{2g+2} + b_g =   \sum_{n=1}^g (\Scont_{2n+1} - \Scont_{2n}) -\Scont_{2g+2} ,\\
&& l_2 = -\Scont_{2g+4} + \Scont_{2g+2} - b_g = - \sum_{n=1}^g (\Scont_{2n+1} - \Scont_{2n})  + \Scont_{2g+2}  -\Scont_{2g+4},\\
&& l_{k} = \Scont_{2g+2k-2} - \Scont_{2g+2k}, \qquad 2 \leq k\leq \N-1,\\
&& l_\N=\Scont_{2g+2\N-2}.
\end{eqnarray}

Thus, the monodromy matrix has the form:
\begin{equation}
M_\infty= \left(
    \begin{array}{ccc}
     I_{2g+\N-1}  & S_{\infty} \\
     0 & I_{\N-1}
          \end{array}\right),
\label{Minfty}
\end{equation}
where
$$
S_{\infty}=\left(\begin{array}{rrrrrrrrrrr}
                        2&-1&0&\dots&   \dots&\dots& \dots& \dots & \dots & 0 \\
                    -2&1&0&\dots&   \dots & \dots& \dots & \dots & \dots &0\\
                    \vdots&\vdots&\vdots&\dots&\dots& \dots&\dots&\dots& \dots & \vdots \\
                                          2&-1&0&\dots& \dots&\dots&\dots& \dots &\dots& 0 \\
                    -2&1&0&\dots&\dots&\dots&\dots& \dots& \dots & 0 \\
                        2&-1&0&\dots&\dots&\dots&\dots&\dots& \dots&0\\
                                   -1&2&-1&0&0&0&\dots&\dots& \dots&0\\
                                 0&-1&2&-1&0&0&\dots&\dots& \dots&0\\
                                 0&0&-1&2&-1&0&\dots&\dots& \dots&0\\
                                     \vdots&\dots &\dots&\dots&\dots& \dots & \dots & \dots & \dots &\dots\\
                            0   &\dots&\dots&\dots&\dots&\dots& \dots&-1&2&-1\\
                               0&\dots&\dots&\dots&\dots&\dots& \dots&0&-1&2
                      \end{array} \right).
$$
One can check that the monodromy matrices (\ref{M1}), (\ref{M2}), (\ref{Mn}), (\ref{M2gp2km1}), (\ref{M2gp2k}), (\ref{Minfty}) satisfy the relation $M_\infty M_{\iM} \dots M_1=I,$ see (\ref{relationmonodromy}).

All the above monodromy matrices except for $M_\infty$ are
rank one perturbations of
 the identity matrix, according to the general theory \cite{DubrovinPainleve}.
Note also that the lower $(\N-1)\times (\N-1)$ block of the matrix
$S_{\infty}$ is equal to the Cartan matrix of the group $A_{\N-1}$.

\subsection{Spaces of meromorphic functions with poles of higher multiplicity}
\label{sect_branchedinfty}

Consider the coverings with branching over the point at infinity as the limits of simple coverings when some of the branch points tend to infinity. In the covering represented by the Hurwitz diagram in Figure \ref{fig_diagram}, let one of the points $P_{2g+2k},$ $1\leq k\leq \N-1,$ tend to infinity without leaving the sheets it belongs to, i.e., without crossing any branch cuts. The dimension of the Hurwitz space is thus reduced by one. Then in the space $H_1(\L\setminus \tilde{f}^{-1}(\infty)\,;\, \tilde{f}^{-1}(\l))$ which corresponds to the covering $\tilde{f}:\L\to\cp$ arising in the limit, we take the basis consisting of the contours (\ref{Scont1})-(\ref{Scont2}) less the contour $\Scont_{2g+2k}$ corresponding to the ramification point taken to infinity.  Then the matrix $\Phi(\lambda)$ (\ref{defPhicon}) with integration contours given by this basis in the space $H_1(\L\setminus \tilde{f}^{-1}(\infty)\,;\, \tilde{f}^{-1}(\l))$, solves the Fuchsian problem (\ref{ls1}), (\ref{ls2}) associated to the Hurwitz space of the coverings $\tilde{f}:\surf\to\cp$ arising in the limit considered.

 As we have seen in the previous section, a basis contour $\Scont_n$ only plays a role in the corresponding monodromy transformation $\F_{\lambda_n}$. Therefore, in the limit $P_{2g+2k}\to \infty$ with $1\leq k\leq \N-1$ the monodromies of the solution $\Phi(\lambda)$ around the remaining finite branch points are obtained from the respective monodromy matrices computed in Section \ref{sect_simpleinfty} by deleting the $(2g+k)$th row and column, which are trivial and which correspond to the disappearing in the limit integration contour $\Scont_{2g+2k}$. In other words, the monodromy matrices $M_n,$ $n\neq \infty,$ still have the form (\ref{M1}) - (\ref{M2gp2k}) computed in Section \ref{sect_simpleinfty}. The monodromy matrix $M_\infty$ can be obtained from relation (\ref{relationmonodromy}).

By repeating the procedure for some of the remaining ramification points $P_{2g+2k}$ one can arrive at a covering with any number (not exceeding $\N$) of points projecting to $\lambda=\infty$ on the base.

\subsection{Space of polynomials}
\label{sect_polynomials}

\begin{figure}[htb]
\centering
\includegraphics[scale=0.6]{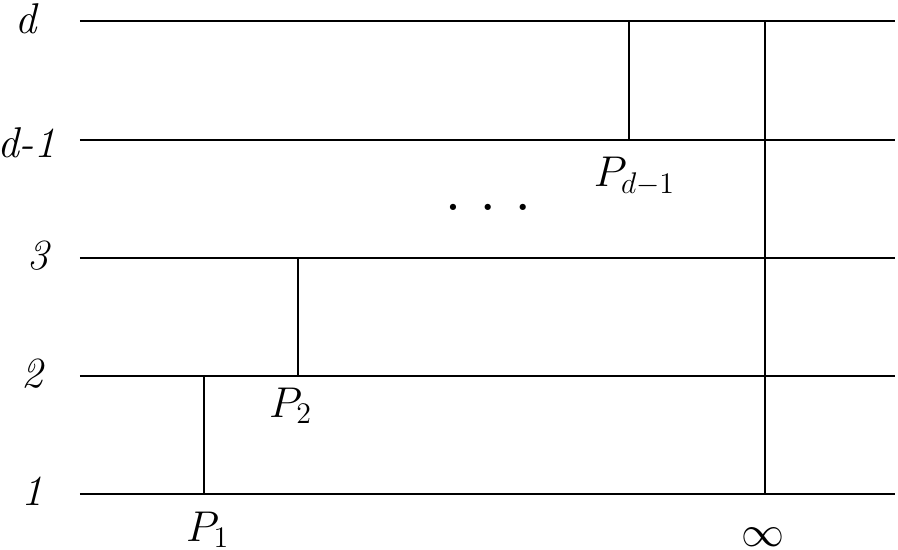}
\caption{A Hurwitz diagram for the space $\H_{0;\iN}(\N).$ }
\label{fig_diagram_poly}
\end{figure}

Here we consider a partial case of the Hurwitz spaces from Section \ref{sect_branchedinfty}, the Hurwitz space $\H_{0;\iN}(\N)$ with a degenerate ramification over $\lambda=\infty$ where all $\N$ sheets are glued together, represented by the Hurwitz diagram from Figure \ref{fig_diagram_poly}.
This space can be regarded as a space of polynomials of degree $\N$ on $\cp.$

As before, we assume the generator $\gamma_k$ of the fundamental group
$\pi_1(\C\setminus\{\lambda_1,\dots,\lambda_{\N-1}\},\lambda_0)$ to encircle only one branch point,
namely  $\lambda_k$.

The $k$th column of the solution $\Phi$  corresponding to this Hurwitz space is given by the integral (\ref{defPhicon}) over the contour $\gamma_{k,k+1}$ (\ref{gkkp1}) for $k=1,\dots,\N-1$ (the contour $\gamma_{k,k+1}$ is again defined as the lift of the loop $\gamma_k$ to the $k$th sheet, it starts on the $k$th and ends on the $(k+1)$st sheet).
Then,  as $\lambda$ describes the loop $\gamma_k$ on the base of the covering, the contours
$\gamma_{1,2},\dots,\gamma_{\iN-1,\iN}$
change as follows: $\gamma_{k-1,k}$ becomes $\gamma_{k-1,k}+\gamma_{k,k+1};$ the contour $\gamma_{k,k+1}$ turns into its negative $-\gamma_{k,k+1}$ (as in Figure \ref{fig_hyperelliptic_M1}), and $\gamma_{k+1,k+2}$ becomes $\gamma_{k,k+1}+\gamma_{k+1,k+2}$.  Here we assume that if one of the indexes becomes $0$ or $\N$, the contour equals 0.

All other contours $\gamma_{j,j+1}$ for $j\neq  k-1, k, k+1$ remain unchanged.

 Thus the monodromy matrices have the following form:
\begin{equation}
\label{poly_Mk}
M_k= \left(
    \begin{array}{ccc}
     I_{k-\itwo} & 0 & 0 \\
     0 & M & 0\\
     0 & 0 & I_{\iN-k-\itwo}
          \end{array}\right),\qquad 1< k < \N-1,
\end{equation}
where the block $M$ is
\begin{equation*}
M= \left(
    \begin{array}{crc}
     1 & 0 & 0 \\
     1 & -1 & 1\\
     0 & 0 & 1
          \end{array}\right).
\end{equation*}
Note that the matrices $M_k$ coincide with the blocks $\Sigma_k$ from (\ref{formMk}) in this case.
The monodromies at $\l_1$ and $\l_{\iN-1}$ are given by
\begin{equation*}
M_1= \left(
    \begin{array}{ccc}
     -1 & 1 & 0 \\
     0 & 1 & 0\\
     0 & 0 & I_{\iN-\ithree}
          \end{array}\right),\qquad
M_{\iN-1}= \left(
    \begin{array}{ccc}
     I_{\N-\ithree} & 0 & 0 \\
     0 & 1 & 0\\
     0 & 1 & -1
          \end{array}\right).
\end{equation*}

 To compute the monodromy at $\lambda=\infty$ we note that since the covering surface is of genus zero and since the preimage $f^{-1}(\infty)$ consists of just one point, all closed contours on the covering are trivial in the relative homology space $H_1(\L\setminus f^{-1}(\infty)\;,\;f^{-1}(\l)).$ Therefore, the non-closed contours from this space can be  characterized by their end points, i.e., any contour connecting points from $f^{-1}(\l)$ on the $k$th and $(k+1)$th sheet is equivalent to $\gamma_{k,k+1}$ up to orientation. Then it is easy to see that the monodromy matrix corresponding to the loop $\gamma_\infty$ based at $\l_0$ and going around $\l=\infty$ counterclockwise %from Figure \ref{fig_fundamental}
 has the form:
\begin{equation}
\label{poly_Minfty}
M_\infty= \left(
    \begin{array}{ccccccc}
     0  & 0 & \dots & 0 &-1 \\
         1& 0 &\dots&0&-1\\
         0&1&\dots&0&-1\\
         \vdots&\vdots&\ddots&\vdots&\vdots\\
         0 &0& \dots&1& -1%\\
%    0 & 0 & I_{\iN-k-\itwo}
          \end{array}\right).
\end{equation}

\section{Completeness of the set of solutions to the Fuchsian system}

%\section{Completeness of the set of solutions to the Fuchsian system}
%\label{sect_completeness}

\label{app_completeness}

Here we are going to prove the completeness of the set of solutions to system (\ref{ls1}), (\ref{ls2}) given by  formula
(\ref{defPhicon}) with the integration contours (\ref{bacyc}) - (\ref{gkkp1}) forming a basis in $H_1(\L\setminus f^{-1}(\infty);\;f^{-1}(\l))$.%, (\ref{lsss}).

The whole section will be devoted to the proof of the following theorem:
\begin{theorem}
\label{thm_det}
The determinant of the matrix function $\Phi$  defined by (\ref{defPhicon}), (\ref{phisol}) is given by:
\be
{\rm det}\,\Phi = C\prod_{j=1}^{\iM} (\l-\l_j)^{1/2},
\label{det00}
\ee
where $C\neq 0$ is a constant independent of $\l$ and $\{\l_j\}$.
\end{theorem}
{\it Proof.}
Since  the function $\Phi$ satisfies linear system
 (\ref{ls1}) with $\q=-1/2,$ we have:
\begin{equation*}
\f{d}{d\l}\log {\rm det}\Phi = {\rm tr}\left\{-\sum_{j=1}^{\iM}
\f{E_j(V - \frac{1}{2} I)}{\l-\l_j}\right\}=
\f{1}{2}\sum_{j=1}^{\M}\f{1}{\l-\l_j}\;,
\end{equation*}
where we used the relation ${\rm tr}\, V=0$.
Analogously, from  (\ref{ls2}) we get
\begin{equation*}
\f{d}{d\l_j}\log {\rm det}\Phi =-\f{1}{\l-\l_j}\;.
\end{equation*}
Therefore, ${\rm det} \, \Phi$ has the form (\ref{det00}) with some
constant $C$. What remains to check is that  $C$ is not equal to $0,$ i.e.,  the columns of the matrix $\Phi(\lambda)$ form a complete set of linearly independent solutions to  (\ref{ls1}), (\ref{ls2}).

For simplicity, we restrict ourselves to the space of coverings with no
branching at infinity, i.e., $\K=\N$.  According to the Riemann-Hurwitz formula we have
in this case $\M=2g+2\N-2.$

Let us choose
generators of the fundamental group $\pi_1({\mathbb C}\setminus\{\lambda_1,\dots,\lambda_\M\},\lambda_0)$ in such a way that the corresponding generators
of the monodromy group of the covering are given by
(\ref{genmongroup}).

The branch cuts  can then be chosen to connect the
branch points $P_{2k+1}$ and $P_{2k+2}$, $k=0,\dots,g+\N-1$.
%For convenience let us denote the branch points $P_{2g+2+k}$ by $Q_k$;
%then the full set of branch branch points looks as follows:
%\be
%\{P_k\}_{k=1}^{\M}:=\{P_k\}_{k=1}^{2g+2}\cup \{Q_k\}_{k=1}^{2\N-4}
%\label{PPQ}
%\ee
The branch cuts $[P_1,P_2],\dots,[P_{2g+1},P_{2g+2}]$ connect the
sheets number $1$ and $2$; the branch cut $[P_{2g+3},P_{2g+4}]$ connects sheets
number $2$ and $3$ etc; the branch cut $[P_{\iM-1},P_{\iM}]$ connects
sheets number $\N-1$ and $\N.$ %(see Figure ???).
In this way we realize the
branch covering $\X$ as a hyperelliptic Riemann surface of genus $g$
with $\N-2$ Riemann spheres attached to it.

Due to Corollary \ref{corollary_det} and relations
(\ref{changeababt}), (\ref{PhiPhit}), the completeness of the set of
our solutions to the system (\ref{ls1}), (\ref{ls2})  depends  neither
on the choice of a symplectic basis $({\bf a}_\a,{\bf b}_\a)$ used in the
normalization of the bidifferential $W,$ nor on the choice of a symplectic
basis $(a_\a,b_\a)$ in (\ref{bacyc}) used as integration contours in (\ref{defPhicon}). Therefore, we shall verify the
completeness choosing these two bases to our convenience. First, we
choose them to coincide: $(a_\a,b_\a)=({\bf a}_\a,{\bf
  b}_\a) $.
Second, we choose these contours to lie on the ``hyperelliptic part''
of the covering as shown in Figure \ref{fig_hyperelliptic_ab}:  the
cycle $a_\alpha$
encircles the ramification points $P_{2\a+1},$ $P_{2\a+2}$ on the
$2$nd sheet,
and the cycle $b_\a$ goes around the points $P_2$ and $P_{2\a+1}$.

 %Also we assume for simplicity that there is no
%branching over the point at infinity,
%i.e. that  on the covering $\L$ there are  $\N$ simple points at infinity: $\infty^{(1)},\dots, \infty^{(\N)}.$

%The proof is inductive and consists of two steps. First, we shall
%prove that $C\neq 0$ for any covering of
%genus $0$ by induction over the number of
%sheets $\N.$ Second, for a fixed number of sheets we shall perform induction over the genus $g.$

Our proof of the non-vanishing of the constant $C$ will be inductive: first we
check that $C\neq 0$ for any covering with $\N=2$ (i.e., a hyperelliptic
covering) of any genus. Second, we check that $C$ remains
non-vanishing when we attach any number of
Riemann spheres to the 2-sheeted covering keeping the genus of the covering unchanged.

\subsection{Example: two sheets, two branch points}
\label{sect_examples}

%In this section we consider a few interesting partial cases, when all
%ingredients of our general construction can be computed more
%explicitly.
%These partial cases correspond to the following Hurwitz spaces: the space
%of degree two rational functions; the space of degree three rational
%functions with simple poles; the space of polynomials of degree 3, and the
%space of degree two functions on an elliptic curve with one double pole.

%\subsection{Two sheets, two branch points}

%%%%%%%%%%%%%%%%%

In this section
we discuss the simplest  case of rational functions $f$ of degree two with
simple poles, whose equivalence classes form the Hurwitz space
$\H_{0,2}(1,1)$.  Up to a M\"obius  transformation in the $\gamma $-plane, any
degree two rational function with critical values $\l_1$ and $\l_2$
is equivalent to the function
\be
f(\gamma)=
\f{\l_1-\l_2}{4}\left(\gamma+\f{1}{\gamma}\right)+\f{\l_1+\l_2}{2}.
\la{rafundegtwo}
\ee

The function $f(\gamma)$ (\ref{rafundegtwo}) defines a two-sheeted genus zero branched covering $\X$
of the Riemann sphere with two branch points $\l_1$ and $\l_2$; this covering is the Riemann surface of the function
$\sqrt{(\l-\l_1)(\l-\l_2)}$.
For simplicity, in this section we identify the ramification points $P_{1,2}$
with the corresponding branch points $\l_{1,2}$.

The uniformization map, i.e., the map from the covering $\X$ to the Riemann sphere,
is given by  the function
\be
h(\l)=\f{2}{\l_1-\l_2}\left\{\l-\f{\l_1+\l_2}{2}+\sqrt{(\l-\l_1)(\l-\l_2)}\right\}\,;
\la{hlamb}
\ee
the value of $\l$  together with the sign of
the square root  $\sqrt{(\l-\l_1)(\l-\l_2)}$ determines the point
$P\in \X$. The functions $f$ (\ref{rafundegtwo}) and $h$ (\ref{hlamb}) are related by  $f\circ h\; (\l)= \lambda$.
In terms of the function $h$ the bidifferential $W$ has the form:
\begin{equation}
W(\l,\mu)=\f{d h(\l)\,d h(\mu)}{(h(\l)-h(\mu))^2}\,.
\label{WPQ0}
\end{equation}

%%%%%%%%%%%%%%%%%%%%5

The relative homology group  $H_1(\L\setminus f^{-1}(\infty)\,;\,f^{-1}(\l))$
is in this case two-dimensional; a  basis in this group can be chosen
to consist of a closed contour $\con_1:=l_1$ around $\infty^{(1)}$
(\ref{lsss}),
and a contour $\con_2:=\gamma_{1,2}(\l)$ (\ref{gkkp1}) connecting in
some way the points
$\l^{(1)}$ and $\l^{(2)}$; we shall choose $\gamma_{1,2}(\l)$ to
consist of two segments: the first segment lies on the
first sheet and  connects the points $\l^{(1)}$
with the branch point $\l_1$; the second interval lies on the second
sheet and connects the points $\l_1$ and   $\l^{(2)}$.

If one of the arguments of the $W$  coincides with a branch
point
(see (\ref{defW(P,P_j)})), we get from (\ref{hlamb}), (\ref{WPQ0}) and (\ref{defW(P,P_j)}):
\begin{equation}
W(\lambda,\lambda_1)= \f{\sqrt{\l_1-\l_2}}{2}\f{d\l}{(\l-\l_1)^{3/2}(\l-\l_2)^{1/2}}\,;
\label{Wl1}
\end{equation}
\begin{equation*}
W(\lambda,\lambda_2)= \f{\sqrt{\l_2-\l_1}}{2}\f{d\l}{(\l-\l_2)^{3/2}(\l-\l_1)^{1/2}}\,
\label{Wl2}
\end{equation*}
%where $\l:=f(P)$.
(the choice of the sign in these formulas corresponds to the choice of the signs of distinguished local parameters near $\lambda_1$ and
$\lambda_2$, i.e., the factors $\epsilon_1$ and  $\epsilon_2$ from Theorem \ref{thm_Schlesinger}).

Therefore, according to (\ref{Philssp}), for the first column of the matrix $\Phi$ we get:
\begin{equation}
\Phi^{(\con_1)}_1=2\pi \i\, W(\infty^{(1)},\l_1)=- 2\pi \i \, \f{\sqrt{\l_1-\l_2}}{2},
\label{const221}
\end{equation}
\begin{equation}
\Phi^{(\con_1)}_2=2\pi \i\, W(\infty^{(1)},\l_2)=- 2\pi \i \, \f{\sqrt{\l_2-\l_1}}{2}.
\label{const222}
\end{equation}

%Integrating  expressions (\ref{Wl1}) and (\ref{Wl2}), we get
%
%\begin{equation*}
%\int_{P_2}^P W(.,P_1) = -\f{1}{\sqrt{\l_1-\l_2}}\sqrt{\f{\l-\l_2}{\l-\l_1}}\,;
%\end{equation*}
%
%\begin{equation*}
%\int_{P_1}^P W(.,P_2) = -\f{1}{\sqrt{\l_2-\l_1}}\sqrt{\f{\l-\l_1}{\l-\l_2}}\,.
%\end{equation*}
%

Integration over the contour $\g_{12}(\l)$ gives the following expressions for the second column of
the
matrix $\Phi$:
%
%\begin{equation*}
%\Phi^{(\con_2)}_1\equiv \int_{\l^{(1)}}^{\l^{(2)}} d\l(P)\int^P W(\mu,\l_1)
%\end{equation*}
%
\begin{equation}
\Phi^{(\con_2)}_1= - \f{2}{\sqrt{\l_1-\l_2}}
\left\{\sqrt{(\l-\l_1)(\l-\l_2)}+\f{1}{2}(\l_1-\l_2)\log h(\l)\right\}\,,
\label{phi11}
\end{equation}
\begin{equation}
\Phi^{(\con_2)}_2= - \f{2}{\sqrt{\l_2-\l_1}}
\left\{\sqrt{(\l-\l_1)(\l-\l_2)}+\f{1}{2}(\l_2-\l_1)\log h(\l)\right\}\,.
\label{phi22}
\end{equation}

Computing the determinant of the matrix function  $\Phi$ (\ref{const221}) - (\ref{phi22}), we get
\begin{equation*}
{\rm det}\Phi= \pm 8\pi \sqrt{(\l-\l_1)(\l-\l_2)}.
%\label{det22}
\end{equation*}

%The monodromy matrices $M_1$, $M_2$ and $M_{\infty}$ are as follows:
%\begin{equation}
%\label{monodromies22}
%M_1 = \left(
 %  \begin{array}{cc}
%    1  &0 \\
   %      0&-1\\
 %         \end{array}\right) , \qquad
%
%          M_2 = \left(
%   \begin{array}{cc}
%    1  &-2 \\
 %        0&-1\\
 %         \end{array}\right) , \qquad
%
    %
 %         M_\infty = \left(
 %  \begin{array}{cc}
%    1  &-2 \\
   %      0&1\\
   %       \end{array}\right) .
%\end{equation}

\subsection{Completeness for the case of  two simple poles  ($\N=2$) }
\label{sect_CompletenessN2}

We start by proving a few auxiliary facts related to degeneration of
hyperelliptic Riemann surfaces.
Consider a hyperelliptic Riemann surface $\X_g$ defined by the equation
\begin{equation*}
\nu^2=\Pi_{2g+2}(\l):=\prod_{k=1}^{2g+2} (\l-\l_k).
%\label{hyperell}
\end{equation*}
We are going to study the behaviour of the bidifferential $W$ under the
degeneration of one of the branch cuts: we put $\l_0:=\l_{2g+1}$
and consider the limit $\l_{2g+2}\to\l_0.$

As a result of the degeneration of the covering $\X_g$ there arises
the hyperelliptic Riemann surface $\X_{g-1}$ of genus $g-1$ defined by the equation
\begin{equation}
\nu^2=\Pi_{2g}(\l):=\prod_{k=1}^{2g} (\l-\l_k).
\label{hyperellgm1}
\end{equation}

Due to the choice of a canonical basis of cycles $\{a_\a, b_\a\}_{\a=1}^g$ on $\X_g$ as shown in Figure
\ref{fig_hyperelliptic_ab}, the cycles  $\{a_\a,b_\a\}_{\a=1}^{g-1}$
 in the limit $\l_{2g+2}\to\l_{2g+1}$ provide  a canonical basis of cycles on $\X_{g-1}$.

Let us denote by $W_g(P,Q)$ the canonical meromorphic bidifferential $W$ on the covering $\X_g$ of genus $g$.
Consider the behaviour of  $W_g(P,Q)$ in the limit $\l_{2g+2}\to\l_{2g+1} \equiv \l_0$.
%When  $\l_{2g+2}\to\l_{2g+1}$ and  $f(P)$ and $f(Q)$ remain
%independent of $\l_{2g+1}$ and $\l_{2g+2}$, the bidifferential   $W_{g}(P,Q)$ does not gain
%any singularity
%at the double point  $\l_0$ on $\L_{g-1}$
According to the general theory (see \cite{Fay73}), no second or higher order pole of $W_g$ at $P_0$ arises under such degeneration.
Since all $a$-periods of $W_{g}(P,Q)$ with respect to both of its arguments
vanish, and in the limit
 the $a_g$ period becomes the residue at $P_0$, the first order pole at $P_0$ also does not arise in the limit, and
the bidifferential $W_{g}(P,Q)$
does not gain any singularity at $P_0$ on $\X_{g-1}$. At all other points, the singularity
structure of
$W_{g}(P,Q)$ under the degeneration coincides with that of $W_{g-1}(P,Q).$
Therefore, if  $f(P)$ and $f(Q)$ remain independent of $\l_{2g+2}$ and
lie outside of a fixed neighbourhood of $\l_0$, we have
as $\l_{2g+2}\to\l_0\,:$
\begin{equation}
W_{g}(P,Q)= W_{g-1}(P,Q)+ o(1)\;.
\label{Wgp1Wg}
\end{equation}
The analysis becomes more subtle if one of the arguments of $W$
coincides with $P_{2g+1}$ or $P_{2g+2}:$
\begin{lemma}
\label{asWPP1gg}
Let $f(P)$   lie outside of a fixed neighbourhood of $\l_0:=\l_{2g+1}$ and be independent of $\l_{2g+2}$. Then
\begin{equation}
\label{limWPP1}
W_{g}(P,P_{2g+2})=\f{\sqrt{\l_{2g+2}-\l_{0}}}{2}\{W_{g-1}(P,P_0)- W_{g-1}(P,P_0^*)+o(1)\},
\end{equation}
\begin{equation}
\label{limWPP2}
W_{g}(P,P_{2g+1})=\f{\sqrt{\l_{0}-\l_{2g+2}}}{2}\{W_{g-1}(P,P_0)- W_{g-1}(P,P_0^*)+o(1)\},
\end{equation}
as $\l_{2g+2}\to\l_{0}$, where  $P_0$ and $P_0^*$ are the points
on the  2nd and 1st sheets of $\L_{g-1}$, respectively, projecting to $\l_0$ on the $\l$-plane.
\end{lemma}
{\it Proof.} The proof of this lemma can be obtained analogously to (\cite{Fay73},
p.51, 52) using the Rauch variational formulas. Consider, for example, (\ref{limWPP1}).
In the hyperelliptic case considered here, the asymptotics (\ref{limWPP1})
can alternatively be derived from an explicit formula for
$W_{g}(P,P_{2g+2}).$ Namely, the differential $W_g(P,P_{2g+2})$ can be
written as follows:
\begin{equation}
W_g(P,P_{2g+2})=W^0(P)-\sum_{\a=1}^g\left\{\oint_{a_\a}W^0 \right\}w_\a(P),
\label{WgW0}
\end{equation}
where
\begin{equation}
W^0(P):=\f{1}{\l-\l_{2g+2}}\f{\sqrt{\Pi_{2g}(\l_{2g+2})}\sqrt{\l_{2g+2}-\l_{0}}}{2\sqrt{\Pi_{2g+2}(\l)}}d\l
\label{W0P}
\end{equation}
(with $\l=f(P)$) is a non-normalized meromorphic differential having the same singular
part as $W_g(P,P_{2g+2})$; a linear combination of holomorphic
differentials in (\ref{WgW0}) ensures the vanishing of all $a$-periods of
the right hand side.

In the limit $\l_{2g+2}\to\l_0$ we have
\begin{equation}
\f{W^0(P)}{\sqrt{\l_{2g+2}-\l_{0}}}\to \f{d\l}{2(\l-\l_0)^2}\f{\sqrt{\Pi_{2g}(\l_{0})}}{\sqrt{\Pi_{2g}(\l)}}.
\label{W0lim}
\end{equation}
The holomorphic terms in (\ref{WgW0}) guarantee the vanishing of all
periods of the differential
$(\l_{2g+2}-\l_{0})^{-1/2}{W_g(P,P_{2g+2})}$, as
well as the vanishing of the residues at $P_0$ and $P_0^*$ of this differential  in the limit considered. The coefficient in front
of $(\l-\l_0)^{-2}$ in the expansion at $P_0$ and $P_0^*$ of the
differential in the limit  coincides with that in (\ref{W0lim}); therefore, taking into account the
normalization condition $\oint_{a_k}W=0,$ $k=1,\dots,g,$ we arrive at (\ref{limWPP1}).
$\Box$

Below we use also the following

\begin{lemma}
\label{lemma_limits}
In the limit $\l_{2g+2}\to\l_{2g+1}:=\l_0,$ the following asymptotics
hold true:
\begin{equation}
\oint_{a_{g}} f(P) W_g(P,P_{2g+2}) = \pi \i (\l_{2g+2}-\l_{0})^{1/2} (1+ o(1)),
\label{asaWPP}
\end{equation}
\begin{equation}
2{\pi \i} \, w_{g}(P_{2g+2})= (\l_{2g+2}-\l_{0})^{-1/2}(2+ o(1)),
\label{aswg}
\end{equation}
\begin{equation}
\oint_{b_{g}} f(P) W_g(P,P_{2g+2}) = (\l_{2g+2}-\l_{0})^{-1/2}(2\l_0 + o(1))).
\label{asbWPP}
\end{equation}
and
$$
\oint_{a_{g}} f(P) W_g(P,P_{2g+1}) = \pi \i (\l_{0}-\l_{2g+2})^{1/2} (1+ o(1)),
$$
$$
2{\pi \i} \, w_{g}(P_{2g+1})= (\l_{0}-\l_{2g+2})^{-1/2}(2+ o(1)),
$$
$$
\oint_{b_{g}} f(P) W_g(P,P_{2g+1}) = (\l_{0}-\l_{2g+2})^{-1/2}(2\l_0 + o(1))).
$$
\end{lemma}
{\it Proof.}
We  prove only the set of formulas involving $P_{2g+2}$.
To prove (\ref{asaWPP}) we make use of the asymptotics (\ref{limWPP1}),
which  implies as $\l_{2g+2}\to\l_{0}$
\begin{equation*}
\f{1}{\pi \i(\l_{2g+2}-\l_{0})^{1/2}}\oint_{a_{g}} f(P) W_g(P,P_{2g+2}) \to
   \underset{P=P_0} {\rm res} \{ f(P) W_{g-1}(P,P_0)\}=1\;,
\end{equation*}
which yields (\ref{asaWPP}).

To prove (\ref{aswg}), let us write the differential $w_g$ in the
form:
\begin{equation}
w_g(P)=\f{1}{2\pi
  \i}\f{d\l}{\sqrt{(\l-\l_{0})(\l-\l_{2g+2})}}\f{Q_{g-1}(\l)}{\sqrt{\Pi_{2g}(\l)}},
\qquad \l=f(P),
\label{wgP}
\end{equation}
where $Q_{g-1}(\l)$ is a polynomial of degree $g-1$ with coefficients
depending on $\{\l_k\}$.
In the limit $\l_{2g+2}\to\l_0,$ the
differential $w_g$ becomes the normalized Abelian differential of the
third kind with poles at $P_0$ and $P_0^*$ and residues $+1$ and $-1$, respectively (this follows from the normalization $\oint_{a_g}w_\alpha=\delta_{\alpha g}$).
Therefore, if we first take the limit $\l_{2g+2}\to\l_0$, and then put
$\l=\l_0$, we get $Q_{g-1}(\l_0)=\sqrt{\Pi_{2g}(\l_0)}$.
Since from (\ref{wgP}) we have
\begin{equation*}
w_g(P_{2g+2})=\f{1}{\pi \i}\f{1}{\sqrt{\l_{2g+2}-\l_{2g+1}}}\f{Q_{g-1}(\l_{2g+2})}{\sqrt{\Pi_{2g}(\l_{2g+2})}}\;,
\end{equation*}
in the limit  $\l_{2g+2}\to\l_0$ we arrive at (\ref{aswg}).

The asymptotics (\ref{asbWPP}) can be deduced from (\ref{aswg}) and (\ref{asaWPP}) by
noticing that the integral  $\oint_{b_{g}} ( f(P)-\l_0) W(P,P_{2g+2})$
remains finite in the limit $\l_{2g+2}\to\l_{0}$. One should also use
the relation $2\pi \i \, w_{g}(P_{2g+2})=\oint_{b_{g}} W(P,P_{2g+2})$.
$\Box$

Now we are in a position to prove the following
\begin{proposition}
\label{propgen0}
The constant $C$ in (\ref{det00}) is non-vanishing for $\N=2,$ i.e., for all hyperelliptic
coverings of genus $g$ (with no branching at $\infty$).
\end{proposition}
{\it Proof.} For $\N=2$ the number of ramification points is $\M=2g+2.$
We prove the  proposition  by induction over the genus of the covering. That is we reduce
the computation of the determinant of the $(2g+2)\times (2g+2)$-dimensional matrix  $\Phi_{g}$
to the computation of the determinant of the $2g\times 2g$-dimensional matrix $\Phi_{g-1}$ arising from $\Phi_{g}$ in the limit $\l_{2g+2}\to\l_{2g+1} \equiv \l_0.$ The base case of the induction, the determinant of the $2\times 2$ matrix-solution $\Phi_0$ corresponding to the genus zero two-fold coverings with two ramification points,  was computed in Section \ref{sect_examples}: ${\rm det}\Phi_0= \pm 8\pi \sqrt{(\l-\l_1)(\l-\l_2)} $.

Consider the $2g\times 2g$ matrix  obtained from $\Phi_g$ by crossing out two columns  and two rows. The crossed out columns are number $2g-1$ and $2g$ -- those given by the integrals (\ref{defPhicon}) over the contours $a_g$ and $b_g$ degenerating in the limit. The two rows correspond to the ramification points $P_{2g+1}$ and $P_{2g+2}$. The resulting matrix, due to (\ref{Wgp1Wg}), tends in the limit $\l_{2g+2}\to\l_{2g+1}$
to a solution $\Phi_{g-1}$ given by (\ref{defPhicon}) to the Riemann-Hilbert problem associated to the hyperelliptic curve  (\ref{hyperellgm1}) of genus $g-1.$  According to the assumption
of our induction, ${\rm det}\,\Phi_{g-1}(\l) \neq 0$ for $\l\in\C\setminus\{\l_1,\dots,\l_\iM\}.$
%it contains a complete set of solutions,
%i.e. corresponding constant $C_{g-1}$ from (\ref{det00}) is non-zero..

Due to Lemma \ref{asWPP1gg}, $W_{g}(P,P_{2g+2})$ and
$W_{g}(P,P_{2g+1})$ tend to $0$ as  $\l_{2g+2}\to\l_{2g+1} \equiv \lambda_0$ if $f(P)$ is independent of
$\l_{2g+1}$ and $\l_{2g+2}$. Therefore, the entries of the deleted $(2g+1)$st and $(2g+2)$nd rows of the matrix $\Phi_{g}$ not belonging to the deleted columns tend to $0$ as $\l_{2g+2}\to\l_{2g+1}$.

%The deleted diagonal $2\times 2$ block consists of the integrals (\ref{defPhicon})
%over the contours $a_g$ and $b_g$ of the differentials $W(P,P_j)$ and $f(P)W(P,P_j)$ with $j=2g+1,2g+2.$

Thus in our limit, ${\rm det}\,\Phi_{g}$ tends to the product of ${\rm
det}\,\Phi_{g-1}$ and the determinant of the $2\times 2$ block at the intersection of the deleted rows and columns:
\begin{equation*}
{\rm det}\Phi_{g}\to {\rm det}\, {\bf A}\;{\rm det}\Phi_{g-1},
\end{equation*}
where
\begin{equation*}
{\bf A} =\lim_{\l_{2g+2}\to\l_0}\left(\begin{array}{cc}
\oint_{a_{g}}f(P) W(P,P_{2g+1})  &  2\pi \i \, w_{g}(P_{2g+1})\l-\oint_{b_{g}}f(P)W(P,P_{2g+1})\\
\oint_{a_{g}}f(P) W(P,P_{2g+2})  &  2\pi \i \,
w_{g}(P_{2g+2})\l-\oint_{b_{g}}f(P)W(P,P_{2g+2})\end{array}\right).
\end{equation*}
Using Lemma \ref{lemma_limits}, we find the behaviour of ${\rm det}\,{\bf A}$  in the limit:
\begin{equation*}
{\rm det}\left(\begin{array}{cc}
 \pi \i (\l_{2g+1}-\l_{2g+2})^{1/2} &  2 (\l_{2g+1}-\l_{2g+2})^{-1/2}(\l-\l_0)  \\
 \pi \i (\l_{2g+2}-\l_{2g+1})^{1/2} &  2 (\l_{2g+2}-\l_{2g+1})^{-1/2}(\l-\l_0)   \end{array}\right)=\pm
4\pi(\l-\l_0).
\end{equation*}
The corresponding constants in (\ref{det00}) are thus related by $C_{g}=\pm 4\pi C_{g-1}$ and
$C_{g}\neq 0$ if $C_{g-1}\neq 0.$
$\Box$

\subsection{Completeness for an arbitrary number of simple poles}
\label{sect_Completeness_simple}

We are going to prove the completeness of our set of solutions for the
Hurwitz diagram shown in Fig.\ref{fig_diagram}; i.e., the permutations
associated to branch points are given by (\ref{genmongroup}). Then the
completeness for an arbitrary set of elementary permutations follows
from the well-known fact that the space $\H_{g;\N}(1,\dots,1)$ is
connected and independence of the constant $C$ from (\ref{det00})  of $\l$ and $\{\l_j\}$.

For the coverings defined by (\ref{genmongroup}) we shall perform an
 induction over the number of sheets without changing
the genus of the covering $\X$ (in this section we denote
it by $\X_\N$); on each step we detach one sheet by a degeneration of one branch cut.
 Put $P_0:=P_{\iM-1}$ (and $\l_0:=\l_{\iM-1}$) and take the limit $P_{\iM}\to
P_0$. In this limit the $\N$th sheet of $\X_\N$ detaches and the $\N$-sheeted covering
 splits into a $(\N-1)$-sheeted covering $\X_{\N-1}$
of the same genus with the ramification points $\{P_k\}_{k=1}^{\M-2}$,
and the Riemann sphere, which we denote by $\X_1$. Denote the
bidifferential $W$ on $\X_\N$ by $W_\N$, on $\X_{\N-1}$ by $W_{\N-1}$ and
on $\X_1$ by $W_1$ (note that $W_1(\l,\mu)=(\l-\mu)^{-2} d\l d\mu $).
The points in the set $f^{-1}(\l_0)$ on the covering  we denote by $\l_0^{(k)}$ (the upper index indicates the sheet number).

Let us prove a few auxiliary facts about this type of degeneration. First, we determine the behaviour of  the bidifferential
 $W_{\N}(P,Q)$  in our limit. Assuming that $f(P)$ and $f(Q)$ are independent of $\l_{\iM}$ and
 $\l_{\iM-1}$ we have the following obvious asymptotics (see \cite{Fay73}):
\be
W_{\N}(P,Q)\to W_{\N-1}(P,Q)\;,\hskip0.7cm P,Q  \in \X_{\N-1};
\label{WNWNm1}
\ee
\begin{equation*}
W_{\N}(P,Q)\to W_1(P,Q)\equiv \f{d\mu(P)\,d\mu(Q)}{(\mu(P)-\mu(Q))^2}\;,\hskip0.7cm P,Q  \in \X_1,
%\label{WNW1}
\end{equation*}
where $\mu$ is a coordinate on the Riemann sphere $\X_1;$ and
\begin{equation*}
W_{\N}(P,Q)\to 0\;,\hskip0.7cm P\in \X_{\N-1}\,,\;\; Q\in \X_1.
%\label{WNLNL1}
\end{equation*}

The next lemma is less trivial.

%Consider now the behaviour of $W(P,P_{\M-1})$ and $W(P,P_{\M})$ in the limit
%
\begin{lemma}
\label{lemWNWNWN}
There are the following asymptotic expansions as $P_{\M}\to P_{\M-1}=P_0$:
\be
W_{\N}(P,P_0)=\f{\sqrt{\l_0-\l_{\M}}}{2}\{W_{\N-1}(P,\l_0^{(\N-1)})+ O(\l_0-\l_{\M})\}, \qquad P\in \X_{\N-1},
\label{aswln}
\ee
%
%\be
%W_{N}(P,P_{\M})=\f{\sqrt{\l_{\M}-\l_{\M-1}}}{2}(W_{N-1} (P,P_0)+ O(\l_{\M-1}-\l_{\M}))
%\label{aswln2}
%\ee
%where $P\in \X_{\N-1}$
and
\begin{equation*}
 W_{\N-1}(P,\l_0^{(\N-1)}) :=\f{W_{\N-1}(P,Q)}{d f_0(Q)}\Big|_{Q=\l_0^{(\N-1)}} \,,
\end{equation*}
where $f_0$ is the meromorphic function on $\L_{\N-1}$ which defines the covering $\X_{g-1}$;
\be
W_{\N}(P,P_0)=\f{\sqrt{\l_0-\l_{\M}}}{2}\{W_1 (P,\l_0^{(\N)})+ O(\l_0-\l_{\M})\}, \qquad P\in \X_1;
\label{aswl1}
\ee
%\be
%W_{\N+1}(P,P_{\M})=\f{\sqrt{\l_{\M}-\l_{\M-1}}}{2}(W_1 (\l,P_0)+ O(\l_{\M-1}-\l_{\M}))
%\label{aswl2}
%\ee
  %$P\in \L_1$
  and
\begin{equation*}
W_1(P,\l_0^{(\N)}):=\f{d\mu(P)}{(\mu(P)-\l_0)^2}\,,
\end{equation*}
$\mu$ being the coordinate on the Riemann sphere $\X_1.$
\end{lemma}
{\it Proof.} Following \cite{Fay73}, Chapter 3, consider a domain
$\Omega\subset\X_\N$, which contains the segment $[P_0,P_{\M}]$ on both $1$st
and $2$nd sheets, and can be
conformally mapped  to an annulus by the map
$$
h(\l)=\f{1}{\l_0-\l_\M}\left\{\l-\f{\l_0+\l_{\M}}{2}+\sqrt{(\l-\l_0)(\l-\l_{\M})}\right\}\;;
$$
the union of two banks of the branch cut $[P_0,P_{\M}]$  is mapped by the function $h(\l)$ to the unit circle.
The Laurent series for $W_\N(P,P_0)$  in the coordinate
$h(\l)$ in a neighbourhood of the unit circle can be written as follows in terms
of the coordinate $\l$ within the domain  $\Omega$ \cite{Fay73}:
\be
W_\N(P,P_0)=
\f{1}{\sqrt{(\l-\l_0)(\l-\l_{\M})}}\sum_{k=-1}^{\infty}
  a_k(\tau) (\l-\l_0)^k \,d\l+ \sum_{k=0}^{\infty}
  b_k(\tau) (\l-\l_0)^k \,d\l \,,
\label{WlMlM}
\ee
where $\l=f(P)$; $\tau=\sqrt{\l_{\M}-\l_0}$; coefficients $a_k(\tau)$
and $b_k(\tau)$ are holomorphic at $\tau=0$. The first sum in
(\ref{WlMlM}) starts from $k=-1$ since $W_\N(P,P_0)$ has a quadratic
pole at $P_0$. Since the singular part of $W_\N(P,P_0)$ at
$P=P_0$ has the form $(\l-\l_0)^{-1}d\sqrt{\l-\l_0}$,
we have $a_{-1}(\tau)=\sqrt{\l_0-\l_{\M}}/2$.
The term in the second sum in
(\ref{WlMlM}) corresponding to $k=-1$ is absent since the residue of
$W_\N(P,P_0)$ at $P=P_0$ equals zero.

Therefore, the differential
\be
\lim_{\l_{\M}\to\l_0} \f{2}{\sqrt{\l_0-\l_{\M}}}W_\N(P,P_0), \qquad P\in \Omega,
\label{limWPPMm1}
\ee
has a singular part of the form
\begin{equation*}
\f{d\l}{(\l-\l_0)^2}\,, \qquad \l=f(P),
\end{equation*}
in  neighbourhoods of  $\l_0^{(\N)}$ and $\l_0^{(\N-1)}.$ The term
containing the first order pole must vanish since the integral of (\ref{limWPPMm1}) over the (homologous to zero) contour on $\X_\N$ encircling
the branch cut $[P_0, P_{\M}]$  is  zero; thus  the residues of (\ref{limWPPMm1}) at $\l_0^{(\N)}$ and
$\l_0^{(\N-1)}$ vanish.

The differential (\ref{limWPPMm1}) does not have any other singularities
 on either $\X_{\N-1}$ or $\X_1$; this differential  has
all vanishing $a$-periods on $\X_{\N-1}$. Therefore, we arrive at
(\ref{aswln}), (\ref{aswl1}).
$\Box$
\begin{lemma}
\label{gen0intW}
There are the following asymptotic expansions as $\l_{\M}\to\l_{\M-1}\equiv\l_0$:
\be
\sqrt{\l_0-\l_{\M}}\int_{P}^Q W_{\N}(R,P_0) =  2  + O(\l_{\M}-\l_0) ;
\label{asiwln}
\ee
\be
\sqrt{\l_0-\l_{\M}}\int_{P}^Q f(R) W_{\N}(R,P_0) = 2\l_0+ O(\l_{\M}-\l_0)\,,
\label{asiwln2}
\ee
where $P\in\X_1$, $Q\in\X_{\N-1}$; $f(P)$ and $f(Q)$ are assumed to be
independent of  $\l_{\M}$.
\end{lemma}
{\it Proof.}
The proof is similar to the proof of the previous lemma. Consider
(\ref{asiwln}).
The integral of $W_\N(R,S)$ with respect to $R$ between the points $P$ and $Q$ is
an abelian differential of the third kind in $S$ with simple poles at
$S=P$ and $S=Q$ and residues $-1$ and $1$, respectively. We denote
this differential by $W_\N^{P,Q}(S):=\int_P^Q W_\N(\cdot,S)$.
%As $\l_{\M}\to\l_0$, the differential $W_\N^{P,Q}(S)$ gains simple poles at $\l_0^{(1)}$ and $\l_0^{(2)}.$
Since the sum of the residues of the differential
$W_1(S):=\lim_{\l_{\M}\to\l_0} W_\N^{P,Q}(S)$ on $\X_1$ must vanish, we
conclude that $W_1(S)$ has  two simple poles on $\X_1:$ the pole with the residue
$-1$ at $S=P$, inherited from  $W_\N^{P,Q}(S)$, and a new pole at $\l_0^{(\N)},$
arising as a result of the degeneration, with the residue
$+1$ (the absence of higher order terms of $W_1(S)$  at $\l_0^{(\N)}$ follows from the
expansion (\ref{WlMlM}) for $W(P,P_0)$). Similarly, on $\X_{\N-1},$ the
differential $W_\N^{P,Q}(S)$ tends to the normalized abelian
differential of the third kind with simple poles at $S=\l_0^{(\N-1)}$
and $S=Q$ and residues $-1$ and $+1$, respectively.

Let us now write down an analog of the expansion (\ref{WlMlM})
for  $W_\N^{P,Q}(S)$, when $S\in \Omega$:
\be
\label{asWNPQS}
W_\N^{P,Q}(S)= \f{1}{\sqrt{(\l-\l_0)(\l-\l_{\M})}}\sum_{k=0}^{\infty}
  c_k(\tau) (\l-\l_0)^k \,d\l+ \sum_{k=0}^{\infty}
  d_k(\tau) (\l-\l_0)^k \,d\l \,,
\ee
where $\l=f(S)$; as before, $\tau:=\sqrt{\l_0-\l_{\M}}$; the coefficients $c_k(\tau)$ and $d_k(\tau)$ are holomorphic at
$\tau=0$. Both sums in (\ref{asWNPQS}) start from $k=0$ since the
differential $W_\N^{P,Q}(S)$ is holomorphic at $S=P_0\equiv P_{\M-1}$ and $S=P_{\M}$.
 Since in our limit the differential $W_\N^{P,Q}(S)$ gains simple poles
 at $S=\l_0^{(\N-1)}$ and $S=\l_0^{(\N)}$ with residues $-1$ and $+1$,
 respectively, we conclude that  $c_0= 1+ o(\tau)$ as $\tau\to 0$.
Now, taking $S=P_{0}$, and evaluating $W_\N^{P,Q}$ at $P_0$ with respect to the local parameter $\sqrt{\l-\l_0}$ similarly to (\ref{defW(P,P_j)}), we arrive at (\ref{asiwln}).

The asymptotics (\ref{asiwln2}) easily follows from
(\ref{asiwln}) since the integral $\int_{P}^Q (f(R)-\l_0)W_{\N}(R,P_0)$ behaves as $o(1)$ in our limit.
$\Box$

We note that all the asymptotics computed in the above lemmas
are symmetric under the interchange of $\l_{\M}$ and $\l_{\M-1}$.

By the assumption of the induction, the constant $C$, we denote it by $C_{\N-1}$, in relation (\ref{det00}) corresponding to the branch
covering $\X_{\N-1}$ is non-vanishing. One needs to prove the
non-vanishing of the constant $C_\N$ corresponding to the covering $\X_\N$.

Denote the function $\Phi$ (\ref{defPhicon}) corresponding to the $\N$-sheeted covering
$\X_\N$ by $\Phi_\N$, and the function $\Phi$ corresponding to the $(\N-1)$-sheeted
covering $\X_{\N-1}$ by $\Phi_{\N-1}$. The columns of $\Phi_\N$
given by the integrals over the contours $l_\N$ encircling $\infty^{(\N)}$, and the
contour $\gamma_{\N,\N-1}(\l)$ have, according to (\ref{defPhicon}) and (\ref{Philssp}),
 the form:

\begin{equation*}
\Phi_k^{(\gamma_{\N,\N-1}(\l))}= -\int_{\l^{(\N)}}^{\l^{(\N-1)}}  f(P) W(P,P_k) +\l
\int_{\l^{(\N)}}^{\l^{(\N-1)}}  W(P,P_k)\;,
%\label{phi10}
\end{equation*}
and
\begin{equation*}
\Phi_k^{(l_\N)}= -2\pi \i \, W(\infty^{(\N)}, P_k).
%\label{phi20}
\end{equation*}

The contours $l_\N$ and  $\gamma_{\N,\N-1}(\l)$ are
 absent from the set of integration contours determining $\Phi_{\N-1}$. The rows corresponding to $P_{\M-1}$ and
$P_{\M}$ are also missing in $\Phi_{\N-1}$. The $2\times 2$ block on
the intersection of these rows and columns in the matrix $\Phi_\N$ looks as follows:
\begin{equation*}
{\bf B}=
\left(\begin{array}{cc}
-\int_{\l^{(\N)}}^{\l^{(\N-1)}}  f(P) W(P,P_{\M-1})
+\l \int_{\l^{(\N)}}^{\l^{(\N-1)}}  W(P,P_{\M-1})             & -2\pi \i \, W(P_{\M-1},\infty^{(\N)}) \\
-\int_{\l^{(\N)}}^{\l^{(\N-1)}}  f(P) W(P,P_{\M})
+\l \int_{\l^{(\N)}}^{\l^{(\N-1)}}  W(P,P_{\M})             & -2\pi \i \, W(P_{\M},\infty^{(\N)})
                        \end{array}\right) .
%\label{defANN}
\end{equation*}
According to (\ref{WNWNm1}), the $(2\M-2)\times(2\M-2)$ minor in the matrix
$\Phi_\N$ obtained by deleting  these two rows and two columns tends to
$\Phi_{\N-1}$ in our limit.
Since all
other entries of the two rows of $\Phi_\N$
corresponding to $P_{\M-1}$ and $P_{\M}$, tend to $0$ as
$P_{\M}\to  P_0=P_{\M-1}$, we see that in this
limit ${\rm det}\,\Phi_\N\to {\rm det} \, {\bf B} \,{\rm det}\,\Phi_{\N-1}$.

Now, due to Lemmas \ref{lemWNWNWN} and \ref{gen0intW}, in this limit
\begin{equation*}
{\rm det} \, {\bf B} \to \left(\begin{array}{cc} -2\f{\l-\l_0}{\sqrt{\l_{\M-1}-\l_{\M}}} &
-\f{\sqrt{\l_{\M-1}-\l_{\M}}}{2}\\
-2\f{\l-\l_0}{\sqrt{\l_{\M}-\l_{\M-1}}} & -\f{\sqrt{\l_{\M}-\l_{\M-1}}}{2} \end{array}\right)=
\left\{\sqrt{\f{\l_{\M}-\l_{\M-1}}{\l_{\M-1}-\l_{\M}}}-\sqrt{\f{\l_{\M-1}-\l_{\M}}{\l_{\M}-\l_{\M-1}}} \right\}
(\l-\l_0) =\pm 2 \i (\l-\l_0) ,
\end{equation*}
where $\l_0=f(P_0)$; therefore, $ C_{\N}=\pm 2\i C_{\N-1},$
i.e., $C_{\N-1}\neq 0$ implies  $C_{\N}\neq 0$.
$\Box$

The proof that the set  of solutions
(\ref{phisol}), (\ref{defPhicon}) is complete for any Hurwitz
space $H_{g,\N}(k_1,\dots,k_\K)$ is also inductive
and can be obtained by sending some of the branch points $\lambda_j$ to $\infty$.


\begin{thebibliography}{99}

\addcontentsline{toc}{section}{Bibliography}

\bibitem{Arnold} Arnold, V.I., Remark on the branching of
  hyperelliptic integrals as functions of the parameters,
  Funkts. Analiz i Ego Prilozheniya, {\bf 2} No.3, 1-3 (1968)

\bibitem{BJL} Balser, W., Jurkat, W. B., Lutz, D. A., On the reduction of connection problems for differential equations with an irregular singular point to ones with only regular singularities, SIAM J. Math. Anal. {\bf 12}, 691-721 (1981)


\bibitem{BBIEM} Belokolos, E.,  Bobenko, A., Its, A.,  Enolskij,
  V. and  Matveev,V., Algebro-geometrical Approach to the Nonlinear
  Integrable Systems,
Springer (1994)

\bibitem{BertolaHarnad}
Bertola, M.,  Eynard, B.,   Harnad, J.,
Differential systems for biorthogonal polynomials appearing in
2-matrix models and the
associated Riemann-Hilbert problem
Comm. Math. Phys. {\bf 243}  193-240  (2003)

\bibitem{Birman}  Birman, J.S. Mapping class groups of
  surfaces. Braids  13-43, Contemp. Math., {\bf 78}, Amer. Math. Soc., Providence, RI, (1988)

\bibitem{Boalch} Boalch, P., From Klein to Painleve via Fourier, Laplace and Jimbo,
Proc. London Math. Soc. (3) {\bf 90}  167-208 (2005)


\bibitem{Bolibruch1}
 Bolibruch, A. A. On orders of movable poles of the Schlesinger
 equation. J. Dynam. Control Systems 6, no. 1, 57--73 (2000)

\bibitem{Bolibruch} Bolibruch, A. A., On the Tau Function for the
  Schlesinger Equation of Isomonodromic Deformations,  {
    Mathematical Notes} ({ Matematicheskije Zametki}), {\bf 74}
  no. 2,  177-184 (2003), in Russian

\bibitem{Clebsch} Clebsch, A., Zur Theorie der algebraischen
  Funktionen, Math. Ann., {\bf 29} 171-186 (1887)

\bibitem{DIZ} Deift, P. A., Its, A. R., Zhou, X., A Riemann-Hilbert approach to asymptotic problems arising in the theory of random matrix models, and also in the theory of integrable statistical mechanics. { Ann. of Math. (2)} {\bf 146} no. 1, 149--235 (1997)

%\bibitem{DVV} Dijkgraaf, R., Verlinde, E., Verlinde, H.,  Notes on topological string theory and $2$D quantum gravity. String theory and quantum gravity %(Trieste, 1990), 91--156, World Sci. Publ., River Edge, NJ (1991)

\bibitem{Doran} Doran, C.F., Algebraic and geometric isomonodromic
  deformations, Journal of Differential Geometry, {\bf 59},  33-85 (2001)


\bibitem{Drinfeld} Drinfeld, V., Quasi-Hopf algebras and Knizhnik-Zamolodchikov equations. In: Problems of Modern Quantum Field Theory (Alushta, 1989). Berlin: Springer (1989), 1-13 (Res.Rep.Phys.)

\bibitem{2D} Dubrovin, B., Geometry of $2D$ topological field theories, {\it Integrable Systems and Quantum Groups,} Montecatini Terme (1993), Lecture Notes in Math. {\bf 1620}, Springer, Berlin (1996)

\bibitem{DubrovinPainleve} Dubrovin, B., Painlev\'e transcendents in two-dimensional topological field theory, The Painlev\'e property, 287--412, CRM Ser. Math. Phys., Springer, New York (1999)

\bibitem{DubMaz}Dubrovin, B.,  Mazzocco, M.,
Monodromy of certain Painlev\'e-VI transcendents and reflection groups,
Invent.Math.
{\bf 141} 55-147  (2000)

   \bibitem{DubMaz1} Dubrovin, B.,  Mazzocco, M.,
On the reduction and classical solutions of the Schlesinger equations, Differential equations and quantum groups,
 A.Bolibruch Memorial Volume, IRMA lectures in Mathematics and Theoretical Physics {\bf 9}, 157-187,
EMS (2007)


\bibitem{DNF} Dubrovin, B.A., Novikov, S.P., Fomenko, A.T., Modern
  geometry: methods and applications. Part II, The geometry and
  topology of manifolds,  Graduate Texts in Mathematics, vol. 104, Springer (1985)

\bibitem{Eisenbud}
Eisenbud, D., Elkies, N., Harris, J., Speiser, R., On the Hurwitz
scheme and its monodromy, Compositio Mathematica {\bf 77} No.1, 95-117 (1991)

\bibitem{EKK} Eynard, B.,  Kokotov, A., Korotkin, D. Genus one
  contribution to free energy in hermitian two-matrix model,
  { Nucl.Phys.} {\bf B694}, 443-472 (2004)

\bibitem{FaddeevTakhtadjan} Faddeev, L. D., Takhtajan, L. A.,
 Hamiltonian methods in the theory of
 solitons. Springer, 592pp  (1987)

\bibitem{Fay73} Fay, John D., Theta-functions on Riemann surfaces,
  Lect.Notes in Math., {\bf 352}, Springer (1973)

\bibitem{Fay92}Fay, John D., Kernel functions, analytic torsion, and moduli spaces, Memoirs of the AMS, {\bf 96} no. 464, AMS (1992)

\bibitem{Garside} Garside, F. A., The braid group and other groups, Quart. J. Math. Oxford {\bf 20}, 235-254 (1969)

\bibitem{Harnad} Harnad, J., Quantum isomonodromic deformations and Knizhnik-Zamolodchikov equation, hep-th/9406078,
in ``Symmetries and integrability of difference equations'', CRM Conference and Lecture Note Series {\bf 9} (1996)

\bibitem{HarMum} Harris, J., Mumford, D.,
On the Kodaira dimension of the moduli space of curves
  Invent.Math. {\bf 67} No.1 23-86 (1982)


\bibitem{Kassel} Kassel, C., Quantum groups, Springer, Berlin (1995)

%\bibitem{Its}  Its, A. R., The Riemann-Hilbert problem and integrable systems. {\it Notices Amer. Math. Soc.} {\bf 50}, no. 11, 1389--1400 (2003)

%\bibitem{JMU} Jimbo, M., Miwa, M., Ueno, K., Monodromy preserving deformations of linear ordinary differential equations with rational coefficients, I, {\it %Phys. D} {\bf 2}, 306-352 (1981)





\bibitem{KokKor} Kokotov, A., Korotkin, D., A new hierarchy of integrable systems associated to Hurwitz spaces,  ``Philosophical Transactions of Royal Society A 10.1098/rsta.2007.2061 (2007)

\bibitem {KokKorB} Kokotov, A., Korotkin, D., Isomonodromic tau-function of Hurwitz Frobenius manifolds and its applications.
{ Int. Math. Res. Not.} {\bf 2006}, no. 2, 1-34 (2006)

\bibitem {KokKorG} Kokotov, A., Korotkin, D., On $G-$function of Frobenius manifolds related to Hurwitz spaces, IMRN, no 7, p. 343-360 (2004)


\bibitem{KokKorJDG} Kokotov, A., Korotkin, D.,  Tau-functions on spaces of Abelian  differentials and
higher genus generalizations of Ray-Singer formula, { J.Diff.Geom.}, {\bf 82} 35-100 (2009)

\bibitem{KKZ} Kokotov, A., Korotkin, D., Zograf, P., Isomonodromic tau-function and admissible covers (2009),
 in preparation

\bibitem{DimaRH}  Korotkin, D., Solution of matrix Riemann-Hilbert problems with quasi-permutation monodromy matrices, { Math. Ann.} {\bf 329}, no. 2, 335--364 (2004)


\bibitem{McMullen} McMullen, C.,  Braid groups and Hodge theory, preprint, June 2009

\bibitem{Mumford} Mumford, D., Tata lectures on theta,  Progress in Math., {\bf 43} Birkh\"auser (1984)





%\bibitem{Malgrange} Malgrange, B., Sur les D\'eformation Isomonodromiques,  in Math\'ematique et Physique (E.N.S. S\'eminaire 1979-1982), 401--426, %{\it Progr. Math.}, {\bf 37}, Birkh\"auser Boston, Boston (1983)

%\bibitem{Manin}  Manin, Yu., Frobenius manifolds, quantum cohomology, and moduli spaces, American Mathematical Society (1999)

%\bibitem{Mumford} Mumford, D., Tata Lectures on Theta, Progress in Math., v. 28, 43 Birkh\"auser (1983,84)

\bibitem{Paris} Paris, L., Braid groups and Artin groups,
 in: Handbook on Teichm\"uller theory (A. Papadopoulos, ed.), Volume II, EMS Publishing House, Z\"urich (2008)
arXiv:0711.2372


\bibitem{Rauch}Rauch, H. E., Weierstrass points, branch points, and moduli of Riemann surfaces, { Comm. Pure Appl. Math.} {\bf 12}, 543-560 (1959)

\bibitem{Reshet}Reshetikhin, N., The Knizhnik-Zamolodchikov system as a deformation of the isomonodromy problem, Lett.Math.Phys., {\bf 26} 167-177   (1992)

%\bibitem{doubles}  Shramchenko, V.,  ``Real doubles" of Hurwitz Frobenius manifolds,  {\it Commun. Math. Phys.} {\bf 256}, 635--680 (2005)

\bibitem{Schaefke} Sch\"afke, R., \"Uber das globale analytische Verhalten der L\"osungen der \"uber die Laplacetransformation 
zusammenh\"angenden Differentialgleichungen $tx^\prime=(A +tB)x$ und $(s-B)v^\prime=(p-A)v$, Doctoral 
Dissertation, University of Essen (1979)

\bibitem{deform}   Shramchenko, V., Deformations of Hurwitz Frobenius structures, { Int. Math. Res. Not.}  {\bf 2005} no.6, 339--387 (2005)

\bibitem{V3} Shramchenko, V., Riemann-Hilbert problem associated to Frobenius manifold structures on  Hurwitz spaces:
irregular singularity, { Duke Math. J.}  {\bf 144}, no. 1, 1-52 (2008)

\bibitem{Zorich} Zorich A., Flat Surfaces,  { Frontiers in number
  theory, physics, and geometry.} I, 437--583, Springer, Berlin
  (2006)

%\bibitem{Witten} Witten, E., On the structure of the topological phase of two-dimensional gravity, Nucl. Phys. {\bf B 340}, 281-332 (1990)

\end{thebibliography}
\end{document}